\documentclass[aps,prx,singlecolumn,superscriptaddress,longbibliography]{revtex4-2}
\usepackage{setspace}
\usepackage{array}[=2016-10-06]
\usepackage[english]{babel}
\usepackage[utf8]{inputenc}
\usepackage{makecell}
\usepackage{amsmath,amssymb,mathtools}
\usepackage{multirow}
\usepackage{float}
\usepackage{mathdots}
\usepackage[version=4]{mhchem}
\usepackage{siunitx}
\usepackage[section]{placeins}
\usepackage{graphicx}
\usepackage{pgfplots}
\usepackage{tikz}
\usepackage{todonotes}
\usetikzlibrary{positioning}
\usepackage{booktabs}
\usepackage{subfig}
\pgfplotsset{compat = newest}
\usepackage{tabularx}
\usepackage{url}
\usepackage{booktabs}

\newcommand{\pg}[1]{{\color{black}  #1}}
\newcommand{\an}[1]{{\color{black}  #1}}

\newtheorem{theorem}{Theorem}
\newtheorem{proposition}{Proposition}
\newtheorem{definition}{Definition}

\begin{document}
\flushbottom

\title{Symmetry, Invariant Manifolds and Flow Reversals in Active Nematic Turbulence}

\author{Ángel Naranjo}
\affiliation{Department of Mechanical and Materials Engineering, University of Nebraska--Lincoln, Lincoln, NE 68588, USA}

\author{Rumayel Pallock}
\affiliation{Department of Mechanical and Materials Engineering, University of Nebraska--Lincoln, Lincoln, NE 68588, USA}

\author{Caleb Wagner}
\affiliation{Parsons Corporation, USA
}

\author{Piyush Grover}
\email{Corresponding author. Email: piyush.grover@unl.edu}
\affiliation{Department of Mechanical and Materials Engineering, University of Nebraska--Lincoln, Lincoln, NE 68588, USA}

\date{\today}

\begin{abstract}

\pg{We investigate the role of symmetry, exact coherent structures (ECSs), and their invariant manifolds in organizing flow reversals and turbulent dynamics in channel-confined two-dimensional active nematics. In compact periodic channels, we use equivariant bifurcation theory to trace the origin of dynamically relevant ECSs through symmetry-constrained local and global bifurcations. At low activity, we identify periodic and relative-periodic families created through local and global bifurcations. The invariant manifolds of these ECSs organize potential heteroclinic pathways between left- and right-flowing nearly uniaxial states. These give rise to several symmetry-dictated reversal mechanisms, with and without vortex-lattice intermediate states. In the high activity turbulent regime, this ECS skeleton persists as a dynamical scaffold for the chaotic attractor. We use symmetry signatures to trace the origin of ECSs embedded in turbulence to their preturbulent branches. We find evidence that typical turbulent trajectories repeatedly shadow these ECSs and their unstable manifolds, resulting in near-heteroclinic transitions and persistent two-way reversals. Our results reinforce the view that active nematic turbulence is organized by a low-dimensional, symmetry-governed network of invariant solutions and their manifolds, and identify the dynamical mechanisms underlying spontaneous flow reversals.}
\end{abstract}
\maketitle

\section{Introduction}
Active fluids serve as a paradigmatic class of non-equilibrium systems displaying a rich phenomenology, including spontaneous coherent flows, dynamical vortex patterns, and chaotic hydrodynamics (`active turbulence') \citep{ramaswamy2019active,kurzthaler2023out,alert2025active}. Of particular recent interest among active fluids are active nematics (ANs), which are suspensions of rod-like units that convert energy into mechanical work at the individual level \cite{Doostmohammadi2017}. Examples of ANs include suspensions of elongated bacteria, kinesin-microtubule mixtures, and synthetic rod-like materials activated by chemical, optical or acoustic stimuli. These systems have attracted significant attention because they closely mirror biological processes from subcellular to multicellular scales and offer promising avenues for engineering applications.

Several continuum descriptions of active nematics have been derived using a variety of microscopic, phenomenological and data-driven techniques \cite{Marchetti2013,varghese2021active,golden2023physically,joshi2022data}. The dynamic behavior of the continuum models is usually studied via a mix of numerical simulations \citep{shendruk2017dancing,hillebrand2025discontinuous}, and linear stability analysis of the analytically derivable steady `base' states \citep{Simha2002,duclos2018spontaneous,varghese2021active}. These base steady states are usually the simplest equilibria, or in certain cases, traveling wave solutions. Some works extend the analysis to predict periodic states \citep{giomi2012banding,mori2023viscoelastic}, and/or apply weakly nonlinear analysis \citep{ohm2022weakly,lavi2025nonlinear} to predict first few bifurcations from the base states \citep{Walton2020}. Other recent works have explored the role of nonlinearity and non-reciprocity in characterizing vortical and traveling flow patterns in general active systems \citep{martinez2019selection,you2020nonreciprocity,brauns2024nonreciprocal}. There have also been recent theoretical and computational efforts to derive coarse-grained statistical descriptions \citep{alert2025active,alert2020universal,Giomi2015,linkmann2020condensate} of active turbulence, focusing on energy spectra and spatial organization of statistically steady states, rather than deterministic dynamics.

While these studies provide invaluable insight into certain deterministic and statistical aspects of motion of active fluids, our understanding of the complexities of active turbulence in confinement remains limited \citep{thampi2022channel,henshaw2023dynamic}. In this context, a deterministic, dynamical systems based approach to studying active turbulence was initiated in our prior works~\cite{Wagner2022, Wagner2023}. The basic premise is that the infinite-dimensional phase space of an active fluid is populated by \emph{Exact Coherent Structures} (ECS), which are exact solutions of the physical dynamics with distinct and regular spatiotemporal structure; examples include equilibria, periodic orbits \an{(POs)}, relative periodic orbits \an{(RPOs)}, and traveling waves. The ECSs are connected by dynamical pathways in phase space, built of intersections of invariant manifolds. While original ideas go back to \citep{hopf1948mathematical,ruelle1971nature}, recent computational and algorithmic advances have resulted in successful application of ECS-based framework to partially resolve the long-standing problem of transition to turbulence in \emph{passive, high-Reynolds number} fluid flows \citep{graham2020exact}. Specifically, convincing numerical and experimental evidence that turbulent trajectories track certain `dominant' ECSs has been obtained for such \emph{passive, inertial} systems \citep{Crowley2022}.

A priori, there are multiple reasons to expect that ECS framework can be equally useful in understanding \emph{active} turbulence by revealing the underlying, low-dimensional skeleton of the dynamics.  First, robust motifs such as defect creation/annihilation events and vortical flows recur across experiments and simulations. Second, the lack of energy cascade in active turbulence \citep{alert2025active} implies a narrower band of dynamically relevant scales as compared to inertial turbulence. By identifying the ECSs and \pg{potential} heteroclinic connections that scaffold the turbulent attractor, one can: (i) clarify which nonlinear mechanisms are essential for active turbulence, (ii) provide a reduced description in terms of transitions between a small number of organizing states, and (iii) potentially control or design active flows via targeted perturbations that steer the system between selected coherent structures. \pg{We use `turbulent regime' to denote the activity range beyond the transition value for which the dynamics are chaotic and no stable regular ECS attractors are found. Finite, isolated chaotic windows at lower activity are considered part of the preturbulent regime.} 

In our prior work \citep{Wagner2022}, we analyzed the dynamics of the 2D AN periodic channel flow at a single, preturbulent activity level, identifying several dominant ECSs and the associated heteroclinic network. In the follow-up work \citep{Wagner2023}, we extended this effort by computing a larger collection of ECSs spanning both preturbulent and turbulent activity levels in the same system, and showed numerically that typical trajectories in the preturbulent regime repeatedly shadow various ECSs before ultimately decaying into an attractor. However, the sheer number of ECSs present in the turbulent regime made it impossible to unambiguously detect shadowing by turbulent trajectories, leaving open the question of how ECSs and their invariant manifolds structure active turbulence. Moreover, the question of how the ECS repertoire is organized and how the different ECSs are related to one another was left unresolved. 

Our present work is based on the key observation that, for a given geometry and boundary conditions, many continuum models of active nematics share the same symmetry group and the same action of this group on the relevant fields. Equivariant bifurcation theory then implies that the \emph{types} of primary and secondary bifurcations, and the symmetries of their branches are dictated largely by the system symmetry and isotropy structure rather than by microscopic details of the model \citep{Golubitsky2002,Marzorati2023}. Thus, a symmetry-based classification of ECSs and their bifurcations in one representative active nematic model can provide a template that is expected to be broadly applicable across different continuum descriptions (e.g. for microtubule–kinesin mixtures, bacterial suspensions, or actomyosin layers) that share the same symmetry class.

\an{In the present work, we use equivariant bifurcation theory to systematically trace the origins of dynamically relevant ECSs such as dancing disclinations, oscillatory nearly unidirectional states, and sustained vortices, and to provide a symmetry-based organization of this ECS zoo within \an{small periodic units (SPUs). Motivated by the minimal flow unit concept \citep{smith2005low}, the SPUs are chosen to retain the dynamically relevant structures and sustain active turbulence while remaining sufficiently small to permit a systematic ECS-based analysis.} Guided by this framework, we investigate a robust and striking phenomenon reported in active nematic channel flow (and more generally active fluid) simulations and experiments: spontaneous flow reversals between nearly unidirectional flow states in active turbulence. We identify a phase space skeleton, consisting of ECSs and \pg{potential heteroclinic connections between them}, that captures the essential mechanisms behind flow reversals in the preturbulent regime. In the smaller of the two \an{SPU}s that we analyze, we extend this picture into the turbulent regime, where we uncover the corresponding organizing skeleton consisting of ECSs whose origin can be traced back to the preturbulent regime using their symmetry signatures. Finally, we provide direct numerical evidence that typical turbulent trajectories shadow these ECSs and their invariant manifolds in active turbulence in this SPU. 

In summary, our results show that key features of active nematic turbulence in 2D channels can be understood in terms of a low-dimensional, symmetry-governed network of invariant solutions and manifolds.}

The remainder of this paper is organized as follows. In Section II, we introduce the system model, summarize our previous ECS-based analysis of active nematic channel flow, and review experimental and theoretical work on flow reversals in active and passive fluids. In Section III, we use the symmetries of the 2D AN equations to carry out an equivariant bifurcation analysis \an{in a SPU}, tracing primary and secondary local and global bifurcations, and constructing a symmetry-organized catalog of dynamically relevant ECSs. In Section IV, we use this catalog to characterize transient flow reversals in the preturbulent regime, showing how a small number of ECS families and \pg{potential heteroclinic connections} between them organize transitions between oppositely directed, nearly unidirectional states. In Section V, we extend this picture into the active turbulent regime in the two \an{SPUs}, \an{characterize the chaotic sets that arise at high activity, and examine the extent to which persistent two-way reversals are organized by a reduced ECS skeleton, relating its members to the preturbulent branches through their symmetry signatures}.

\section{Model Description and Prior Work} 
\subsection{Governing Equations}
We consider the 2D Beris–Edwards continuum model for AN confined in a 2D periodic channel, a setup that has been used in several previous studies
 \cite{shendruk2017dancing, giomi2012banding, Thampi2014, Wagner2022, Wagner2023}. It models the evolution of the velocity field $\mathbf{u}(\mathbf{x},t) = (u,v)$ and the nematic alignment tensor $\mathbf{Q}(\mathbf{x},t)$. The field $\mathbf{Q}$ is a traceless tensor defined as $Q_{ij} = q(n_i n_j - 0.5\,\delta_{ij})$, where $q$ is the scalar order parameter describing the degree of nematic alignment, and the unit vector $\mathbf{n}$ denotes its direction.
\an{
 We employ the nondimensionalization of \cite{Wagner2023}, using the characteristic
length and time scales:
\[
\ell = \sqrt{\frac{K}{A}}, \qquad
\tau = \frac{\gamma}{A},
\]
where $K$ is the nematic elastic constant, $A$ is the coefficient of the quadratic bulk free-energy term, setting
the strength of the bulk nematic free-energy term, and $\gamma$ is the
nematic rotational viscosity. The corresponding characteristic velocity
scale is
\[
v_0 = \frac{\ell}{\tau}
= \frac{\sqrt{KA}}{\gamma}.
\]
Thus, the spatial coordinates, time, and velocity are nondimensionalized as
$\mathbf{x}=\mathbf{x}_{\mathrm{dim}}/\ell$, 
$t=t_{\mathrm{dim}}/\tau$, and
$\mathbf{u}=\mathbf{u}_{\mathrm{dim}}/v_0$, respectively, while the pressure
is scaled as $p=p_{\mathrm{dim}}\tau/\eta$, where $\eta$ is the shear viscosity. In these nondimensional coordinates, the computational domain is taken to be
a two-dimensional channel of size $[0,\tilde{L}]\times[0,h]$, where
$x\in[0,\tilde{L}]$ is the streamwise (horizontal) direction and
$y\in[0,h]$ is the wall-normal (vertical) direction. With these definitions, the dimensionless governing equations are

\begin{align}
&\mathrm{Re}_{\mathrm{n}}\left(\partial_t + \mathbf{u} \cdot \boldsymbol{\nabla}\right)\mathbf{u} = -\boldsymbol{\nabla} p + 2 (\boldsymbol{\nabla} \cdot \mathbf{E})  - \frac{\mathrm{R}_{\mathrm{a}}}{\mathrm{Er}} \left( \boldsymbol{\nabla} \cdot \mathbf{Q} \right), \nonumber \\
&\left(\partial_t + \mathbf{u} \cdot \boldsymbol{\nabla}\right)\mathbf{Q} + \mathbf{W} \cdot \mathbf{Q} - \mathbf{Q} \cdot \mathbf{W} = \lambda \mathbf{E} +  \mathbf{H}, \label{eqs:main}\\
&\boldsymbol{\nabla} \cdot \mathbf{u} = 0, \nonumber \\
&\mathbf{H} = \mathbf{Q} - b \,\mathrm{Tr}\left(\mathbf{Q}^2\right) \mathbf{Q} + \nabla^2 \mathbf{Q}.
\end{align}

where $\mathbf{E}$ and $\mathbf{W}$ denote the symmetric rate-of-strain tensor and antisymmetric vorticity tensor, respectively:
\begin{align}
E_{ij} &= \frac{1}{2} \left(\partial_i u_j + \partial_j u_i \right); \qquad W_{ij} = \frac{1}{2} \left(\partial_i u_j - \partial_j u_i \right).\nonumber \end{align}

We impose no-slip boundary conditions on the velocity field $\mathbf{u}$ and perpendicular (homeotropic) anchoring on the nematic director $\mathbf{n}$ at the confining walls, and periodic boundary conditions along the streamwise direction:
\begin{itemize}
    \item \textbf{Velocity:} No-slip at \( y = 0, h \), and periodic in \( x \):
    \[
    \mathbf{u}(x, 0, t) = \mathbf{u}(x, h, t) = \mathbf{0}, \quad \mathbf{u}(x + \tilde{L}, y, t) = \mathbf{u}(x, y, t).
    \]
    
    \item \textbf{Nematic tensor:} Homeotropic anchoring with \( q = 1 \) at \( y = 0, h \) and periodic in \( x \):
    \[
    \mathbf{Q}(x, 0, t) = \mathbf{Q}(x, h, t) = \frac{1}{2} \begin{pmatrix} -1 & 0 \\ 0 & 1 \end{pmatrix}, \quad \mathbf{Q}(x + \tilde{L}, y, t) = \mathbf{Q}(x, y, t).
    \]
\end{itemize}

In Eq.~\eqref{eqs:main}, $\mathbf{W}\cdot\mathbf{Q}-\mathbf{Q}\cdot\mathbf{W}$ captures the effect of flow rotation on the nematic director, and $\lambda\mathbf{E}$ is the flow-alignment term. The $\mathbf{H}$ term is the nematic molecular field associated with the free-energy functional and drives the system toward lower free energy, where $b$ is the dimensionless coefficient of the quartic term in the bulk free-energy density. We neglect the passive elastic stress which is subdominant compared to active and viscous stress. Here, $\mathrm{Re}_{\mathrm{n}}$ is the Reynolds number, $\mathrm{R}_{\mathrm{a}}$ the dimensionless activity, and $\mathrm{Er}$ the Ericksen number, defined as:
\[
\mathrm{Re}_{\mathrm{n}} = \frac{\rho v_0 \ell}{\eta},
\qquad
\mathrm{Er} = \frac{\eta}{\gamma},
\qquad
\mathrm{R}_{\mathrm{a}} = \frac{\alpha}{A},
\]
where $\rho$ is the fluid density and $\alpha$ is the active-stress
coefficient.

Following \cite{Wagner2023, Doostmohammadi2017}, we fix $\mathrm{Re}_{\mathrm{n}} = 0.0136$ and $\mathrm{Er} = 1$. As in our previous work \cite{Wagner2023}, we fix $b=5$ and neglect flow alignment to focus on the essential aspects of the system, setting $\lambda=0$. Finally, we set the channel size to $\tilde{L}=80$ and $h=20$, corresponding to a channel
aspect ratio $\tilde{L}/h=4$.}

\subsection{Exact Coherent Structure Framework for AN Channel Flow}
In our previous works \cite{Wagner2022, Wagner2023}, we recast the equations~\eqref{eqs:main} as a dynamical system 
\begin{equation}\label{Dynamical_System}
    \frac{dX}{dt} = F(X;R_a),
\end{equation}
where $X = [\mathbf{u}, \mathbf{Q}]$, the activity number $R_a$ is taken as the bifurcation parameter, and the associated flow map is given by $f^t(X_0) = X_0 + \int_0^t F[X(\tau)]d\tau$. We developed an open-source computational toolbox titled `Exact Coherent Structures in Active Matter' (ECSAct) \citep{ecsact} using the open-source pseudospectral code Dedalus \citep{burns2020dedalus} to compute ECSs and heteroclinic connections. The problem of computing each ECS is posed as a fixed-point equation (FPE), and these high-dimensional FPEs are solved via modified Newton-Raphson algorithms using adaptive `hookstep' method and the matrix-free GMRES method \citep{viswanath2007recurrent}. \an{In the present work, all ECSs are computed using $N_x \times N_y = 32 \times 32$ Fourier and Chebyshev modes in the $x$ and $y$ directions, respectively. The principal ECSs involved in flow-reversals in preturbulent and turbulent regimes are additionally recomputed at $64 \times 64$ resolution to verify that the bifurcating and numerically identified solutions persist under increased spatial resolution. Time integration for both the dynamical simulations and FPE calculations uses the two-stage implicit/explicit Runge--Kutta method `RK222' implemented in Dedalus, with a nominal time step of $dt=0.01$. All computed ECSs satisfy the corresponding FPEs to an $L^2$-norm tolerance of $10^{-10}$. }

In the first work \citep{Wagner2022}, this computational framework was used to identify 46 ECSs (up to symmetry), including the unidirectional flow (UNI), the bidirectional flow (LAN), 11 unstable POs, and 33 RPOs in the preturbulent regime. Several potential heteroclinic connections between various ECSs were also computed, allowing the phase space to be organized as a directed graph with ECSs as nodes. The follow-up work \cite{Wagner2023} confirmed the previously reported general sequence of attractors that are observed while increasing activity: zero flow, unidirectional flow, vortex lattice, and the transition to turbulence. \an{Representative velocity fields of ECSs identified across the preturbulent and turbulent regimes in these two studies are shown in Fig.~\ref{fig:Symmetry_Three}, illustrating the variety of coherent structures and their associated symmetries. The UNI and LAN states serve as organizing structures for these ECSs, with the other solutions appearing to lie on branches related to UNI and LAN through symmetry-breaking bifurcations.} In the preturbulent regime, we identified several `shadowing events', which are defined as finite time intervals over which a trajectory stays close to an ECS, and the trajectory and the ECS have quantitatively similar time evolution. While more than 200 ECS were found in the turbulent regime, no evidence of shadowing was found in that regime.

\subsection{Flow Reversals in Active Nematics and Other Active Fluids}

\begin{figure}[htbp]
    \centering
    \includegraphics[width=\linewidth]{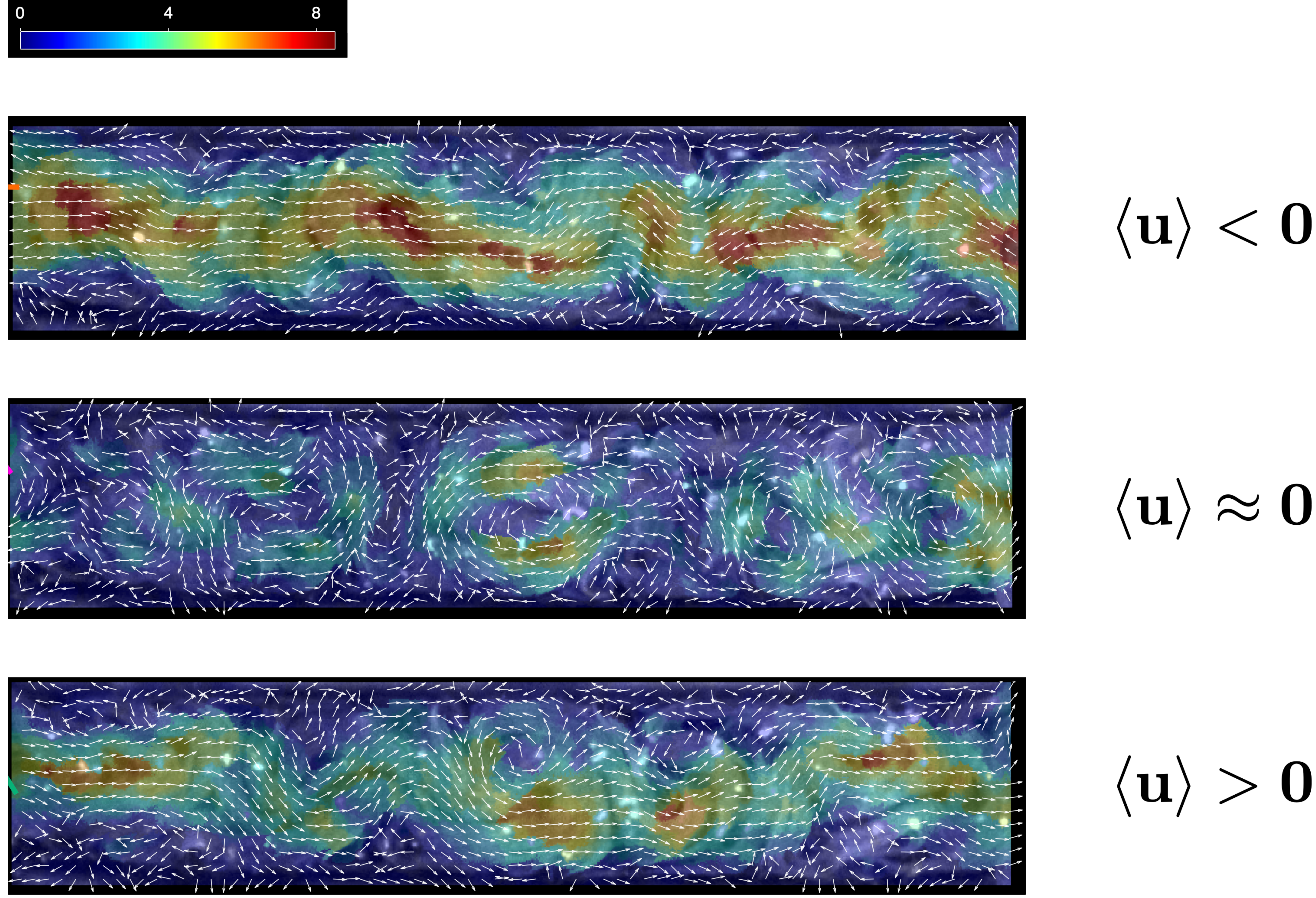}
    \caption{\footnotesize Velocity field snapshots from active nematic experiments of Dor{\'e} \citep{dore2022active} of a reversal event in a channel of size $500 \mu m \times 100 \mu m$. The colormap shows the speed (µm/s). The flow evolves from a left-flowing UNI-like state ($\langle u \rangle <0$) to a right-flowing UNI-like state ($\langle u \rangle >0$) via an intermediate vortex-dominated configuration ($\langle u \rangle \approx 0$). Reproduced with permission of the author.}
    \label{fig:dore}
\end{figure}

The Ph.D. thesis \citep{dore2022active} reported experimental results on AN systems in long, narrow channels whose open ends were connected to an external chaotic AN reservoir. Persistent, nearly unidirectional flows were observed, and the direction of these uniaxial flow states (UNI) was selected at random via spontaneous symmetry breaking of the zero-flow (ZF) state, as expected. More importantly, spontaneous and repeated reversals were also observed between the left- and right-flowing UNI states, and the reversal frequency depended on the channel dimensions. \an{An example of such a reversal observed experimentally is shown in Fig. \ref{fig:dore}.} The author used the channel-averaged streamwise velocity $\langle u \rangle$ as an order parameter to quantify these reversals.  In these experiments, a typical reversal proceeded via an intermediate state: the UNI state with a parabolic velocity profile ($\langle u \rangle \approx \pm u_0$) first collapsed into a vortex-dominated configuration with negligible net flow ($\langle u \rangle \approx 0$), and then evolved into a UNI state whose mean flow was opposite in direction to the original one.

These reversals were attributed to a bend-instability mechanism. Starting from a nematic that is predominantly aligned along the streamwise direction, a bend instability will develop in the transverse direction \citep{duclos2018spontaneous} and reorient the director accordingly. The resulting transversely aligned state is, in turn, unstable to bend perturbations along the streamwise direction, leading to repeated switching between the two streamwise-aligned configurations. To capture this behavior statistically, the author derived and used a single-degree-of-freedom Langevin dynamics model with a double-well potential, fitted to the experimental data. The well minima corresponded to the two UNI states, and noise-driven fluctuations induced transitions between them.

AN flow reversals have also been observed in simulations of an (infinitely long) AN layer sandwiched between viscoelastic walls \citep{mori2023viscoelastic}. For sufficiently soft confinement, the authors found that the zero flow state undergoes a Hopf bifurcation to an oscillating flow state, in which the channel-averaged velocity reverses sign \emph{periodically} in time, whereas stiffer walls produce steady unidirectional flow.

In polar active fluids, recent experimental, computational and analytical work \citep{nishiguchi2025vortex} has argued that flow reversals are a precursor to active turbulence. The authors found that a dense suspension of swimming Bacillus subtilis confined to a cylindrical well exhibit a single vortex steady flow at small radii. This state, which is analogous to the UNI in channel flow, loses stability as the radius is increased beyond a threshold. In the resulting oscillatory state, the vortex \emph{periodically} reverses its direction. Using simulations and weakly nonlinear analysis of a Toner–Tu–Swift–Hohenberg model, it was shown that the onset of reversals is controlled by a global infinite-period bifurcation mediated by a resonant nonlinear interaction of the three lowest modes of the flow. 

\subsection{Possible Generic Dynamical Mechanisms Underlying Flow Reversals }
Beyond active fluids, flow reversals have been extensively observed and studied in various settings \citep{glatzmaier1999role,araujo2005wind,sommeria1986experimental,Yang2025}. A prominent example is that of turbulent Rayleigh-Benard convection (RBC) \citep{ahlers2009heat}, in which large-scale convection (LSC) reversals between the clockwise and anticlockwise vortical states are robustly observed \citep{mishra2011dynamics}. In addition to numerical simulations, such reversals have been studied using both deterministic and stochastic reduced-order dynamical models \cite{bai2016ability,araujo2005wind}.

Regardless of their physical origins, systems that exhibit flow reversals are expected to share certain dynamical features. Typically, a quiescent base state first loses stability through spontaneous symmetry breaking, most naturally via a supercritical pitchfork bifurcation, producing two mirror-symmetric steady states. Exactly periodic reversals then imply the existence of a stable periodic orbit that alternately approaches these symmetry-related states, which may arise through a heteroclinic bifurcation or an infinite-period (global) bifurcation involving the two steady states. In turbulent regimes, where reversals are irregular and no simple periodic attractor exists, several possibilities arise. The turbulent trajectory may shadow a \emph{robust heteroclinic cycle} \citep{krupa1997robust,kreilos2013edge,armbruster1988heteroclinic,holmes1993symmetries} built from the two symmetry-related states, possibly together with an intermediate state, although such cycles typically require strong symmetry constraints. 

A more likely scenario is that turbulent trajectories are shadowing two-way heteroclinic connections between a pair of mirror-image unstable periodic orbits, each of which passes close to one of the steady states. Similar heteroclinic structures, in which each of two unstable periodic orbits passes near a different steady state and the two orbits are linked by two-way heteroclinic connections, have been found in numerous other \emph{low-dimensional} settings \citep{koon2000heteroclinic,gabern2006application}. This reversal scenario could also involve additional `intermediate' states and their invariant manifolds. This mechanism was shown to be the dynamical skeleton responsible for LSC reversals in a 2D forced Navier-Stokes system recently \citep{Suri2024}. In that work, the author computed the two mirror-image asymmetric periodic orbits, and a symmetric `intermediate' orbit. In a reduced phase space projection, turbulent trajectories are shown to shadow one asymmetric periodic orbit, then depart along its unstable manifold towards the intermediate periodic orbit, and finally either fall back to the original side, or cross-over towards the other asymmetric periodic orbit.

The main goal of the current work is to identify such dynamical structures organizing the reversals in the active turbulent regime, and establish their shadowing by typical turbulent trajectories. As a first step, the next section presents a local and global bifurcation analysis in the low-activity, preturbulent regime, clarifying how symmetry dictates the emergence of the most common ECS families in active nematic channel flows.
\an{Because large-scale flows are constrained by the smallest channel dimension, we previously observed threefold or fourfold symmetry of vortex lattice states in our chosen domain \citep{Wagner2023}. Motivated by the idea underlying the minimal-flow-unit approach, namely that reduced periodic domains can isolate dynamically relevant structures, we perform the equivariant analysis in a shorter SPU \cite{Jimenezi1991, Kawahara2011}, specifically $ L = \tilde{L}/3$ . We also briefly discuss $L = \tilde{L}/4$ since a particular bifurcation is more clearly resolved numerically in this domain, while we expect the underlying mechanism persist beyond this particular domain size.}

\section{Local and Global Bifurcations in 2D AN Channel Flow: Role of Symmetry} \label{sec:eqbif}
The systematic way to perform bifurcation analysis of systems with symmetry is by using equivariant bifurcation theory \cite{Golubitsky2002, Chossat2000}. \pg{The basic concepts from group theory and results in equivariant bifurcation theory used in this work are reproduced in the Appendix.} In this section, we begin by listing the symmetries of the AN 2D channel flow system. Then we revisit the equivariant analysis of primary bifurcations of the zero flow (ZF) state \cite{Marzorati2023}. This prior work explains the emergence of UNI and LAN-like solutions in AN channel flows. We significantly extend the analysis of \cite{Marzorati2023} to secondary local and global bifurcations, providing a partial but insightful explanation of the origin of dynamically relevant solutions and a classification based on their symmetry properties. Along the way, we highlight the consequences of symmetry-breaking, symmetry-preserving and symmetry-restoring bifurcations. The Table~\ref{tab:ECS} in the appendix lists the main ECS. Videos of the spatiotemporal evolution of these ECSs are provided in the Supplemental Material.


\subsection{Symmetries in the active nematics equations}

The governing equations (\ref{eqs:main}) with the imposed boundary conditions are equivariant under the continuous symmetries $\tau_{x}(\ell)$ corresponding to $x-$translation, the discrete symmetry $\sigma_x$ corresponding to reflection about the midplane $x-$axis, and the discrete symmetry $\sigma_y$ corresponding to reflection about the $y-$ axis. The corresponding actions on the physical fields are: \begin{align}\label{symmetries}\sigma_x[u,v,Q_{11},Q_{12}](x,y)=[u,-v,Q_{11},-Q_{12}](x,h-y),\nonumber \\
\sigma_y[u,v,Q_{11},Q_{12}](x,y)=[-u,v,Q_{11},-Q_{12}](L-x,y),\nonumber\\
\tau_{x}(\ell)[u,v,Q_{11},Q_{12}](x,y)=[u,v,Q_{11},Q_{12}](x+\ell,y).
\end{align}

Due to the periodic boundary conditions in $x$, the set of translations $\tau_x(\ell)$ is isomorphic to the rotational symmetry group $SO(2)$. The reflection symmetry $\sigma_x$ generates a cyclic group of order 2, which we denote as $\mathbb{Z}_2 (\sigma_x)$. Likewise, the reflection symmetry $\sigma_y$ generates the cyclic group denoted by $\mathbb{Z}_2 (\sigma_y)$. The x-translations and y-reflections do not commute, so they form $O(2) = SO(2) \rtimes \mathbb{Z}_2$ (a semidirect product). The x-reflection $\mathbb{Z}_2 (\sigma_x)$ commutes with this $O(2)$. Finally, the full symmetry group of the system is
\[
G = \mathbb{Z}_2(\sigma_x) \times O(2).
\]

Aside from the work of Marzorati and Turzi~\cite{Marzorati2023}, which we discuss in the next subsection, prior stability and bifurcation studies of active nematic channel flows have not systematically exploited these symmetries. This omission obscures the underlying structure of the infinite-dimensional dynamical system \cite{Cvitanovic2020Chaos}. In this work, we use equivariant bifurcation theory to systematically predict the emergence of new steady and time-dependent solution branches and to determine the symmetries of these solutions. 

Some solutions of Eq. (\ref{eqs:main}) have already been classified by their symmetries previously \cite{Wagner2022}. \pg{The ZF inherits all the symmetries of $G$. Both UNI and LAN retain continuous streamwise translation symmetry, but UNI preserves $\sigma_x$ while breaking $\sigma_y$, and LAN preserves $\sigma_x\sigma_y$ while breaking both individual reflections $\sigma_x$ and $\sigma_y$.} More generally, the plethora of ECSs found in \cite{Wagner2022}
exhibit a rich variety of patterns with observable spatiotemporal symmetries (see Fig. \ref{fig:Symmetry_Three}). The system selects preferable discrete translation symmetries to form vortex solutions \cite{Wagner2022,Wagner2023}, many of them corresponding to oscillatory solutions which maybe traveling waves, periodic or relative periodic orbits. In the rest of this section, we use equivariant bifurcation theory to identify the origins of key ECSs, organize the associated solution branches by their shared symmetries, and assemble a partial isotropy lattice that connects symmetry-distinct branches through sequences of symmetry-breaking bifurcations.

\begin{figure}[htbp]
\centering
\includegraphics[width=\textwidth]{Solutions_symmetries_II.pdf}
\caption{\footnotesize
\an{Snapshots of the velocity fields for selected ECSs organized according to symmetry, with the colormap representing vorticity. (a) The ZF state is $G$-invariant and therefore possesses all the symmetries of the system. The left and right columns then group solutions according to their respective UNI-like and LAN-like symmetry structures. (b) The UNI solution breaks $\sigma_y$, while remaining invariant under $\sigma_x$ and the continuous symmetry $\tau_x(\ell)$. (c) The LAN solution is invariant under $\sigma_x\sigma_y$ and $\tau_x(\ell)$. In (d) and (e), two solutions with broken continuous symmetry in the preturbulent regime at $\mathrm{R_a}=1.5$ are shown. The RPO in (d) is invariant under the group generated by $\sigma_x\tau_x(L/2)$, whereas the RPO in (e) is invariant under the group generated by $\sigma_x\sigma_y$ and $\tau_x(L/3)$. Finally, (f) and (g) show two RPOs in the turbulent regime at $\mathrm{R_a}=4.5$, whose symmetry generators are $\sigma_x\tau_x(L/4)$ and $\sigma_x\sigma_y$, respectively.}
}
\label{fig:Symmetry_Three}
\end{figure}

\subsection{Primary bifurcation in active nematics}
The generalized equivariant branching lemma (EBL) \cite{Chossat2000,Hoyle2006,Golubitsky2002} (see Appendix) is a fundamental tool in equivariant bifurcation analysis, and provides the conditions for the existence of a symmetric fixed point branch emerging from a singular fixed point $X^*$ at a bifurcation value $R_a=R_a^*$. To apply this framework to our high-dimensional system (Eqs. \ref{Dynamical_System}), we restrict the system to the center manifold, where the evolution is given by $\dfrac{dY}{dt}=F_c(Y,\mu)$, where $\mu=R_a-R_a^*$, and the system satisfies $F_c(0,0)=0$, and $D_YF_c(0,0)=0$. The reduced system $F_c$ is $\Gamma-$equivariant, where $\Gamma$ is the isotropy group of $X^*$, i.e., $\Gamma=\{\sigma\in G|\sigma(X^*)=X^*\}$. Note that if $X^*$ has all the symmetries of $G$, then $\Gamma=G$, but otherwise, $\Gamma$ is a proper subgroup of $G$. A key observation is that $\Gamma$ commutes with the Jacobian $L\coloneq D_XF(X^*,R_a^*)$, which implies that $\Gamma$ maps the center subspace back to itself. This fact allows us to study the bifurcation problem via the reduced system $F_c$ by analyzing the action of $\Gamma$ on the center subspace.

Using the EBL, Marzorati and Turzi~\cite{Marzorati2023} performed an equivariant bifurcation analysis of the 2D AN channel system to obtain solutions bifurcating from the zero-flow (ZF) state. We discuss their calculations of this primary bifurcation to illustrate the application of the EBL and set the stage for calculation of secondary bifurcations that follow.

By restricting the analysis to (x-) translation invariant solutions, they only considered equivariance with respect to the Klein four-group $\hat{G}=\mathbb{Z}_2(\sigma_x) \times \mathbb{Z}_2(\sigma_y)$. The ZF solution has the full symmetry, and hence $\Gamma=\hat{G}$. The kernel of the Jacobian at ZF has two dimensions that we associate to $\mathbb{R}^2$ spanned by the two eigenvectors corresponding to zero eigenvalues. In this subspace, the matrix representations of the group $\Gamma=\{I,\sigma_x,\sigma_y,\sigma_x\sigma_y\}$ were computed in the eigenvector basis, and found to be
\[
I = \begin{pmatrix}
1 & 0 \\
0 & 1
\end{pmatrix}, \quad
\sigma_x = \begin{pmatrix}
1 & 0 \\
0 & -1
\end{pmatrix}, \quad
\sigma_y = \begin{pmatrix}
-1 & 0 \\
0 & -1
\end{pmatrix}, \quad
\sigma_y\sigma_x = \begin{pmatrix}
-1 & 0 \\
0 & 1
\end{pmatrix}.
\]

The four proper subgroups of $\Gamma$ are $\{I\}$, $\mathbb{Z}_2(\sigma_x)$, $\mathbb{Z}_2(\sigma_y)$, and $\mathbb{Z}_2(\sigma_x\sigma_y)$. The corresponding fixed point subspaces are 

\begin{align*}
\text{Fix}(\{I\}) &= \mathbb{R}^2 \text{ : 2D fixed point subspace,}  \\
\text{Fix}(\mathbb{Z}_2(\sigma_x)) &= \{ (a, 0) \in \mathbb{R}^2 \mid a \in \mathbb{R} \} \text{ : 1D fixed point subspace} , \\
\text{Fix}(\mathbb{Z}_2(\sigma_y)) &= \{ (0,0) \} \text{ : 0D (trivial) fixed point subspace}, \\
\text{Fix}(\mathbb{Z}_2(\sigma_x\sigma_y)) &= \{ (0, b) \in \mathbb{R}^2 \mid b \in \mathbb{R} \} \text{ : 1D fixed point subspace.}
\end{align*}

\begin{figure}[t]
    \centering
    
    \includegraphics[width=\textwidth]{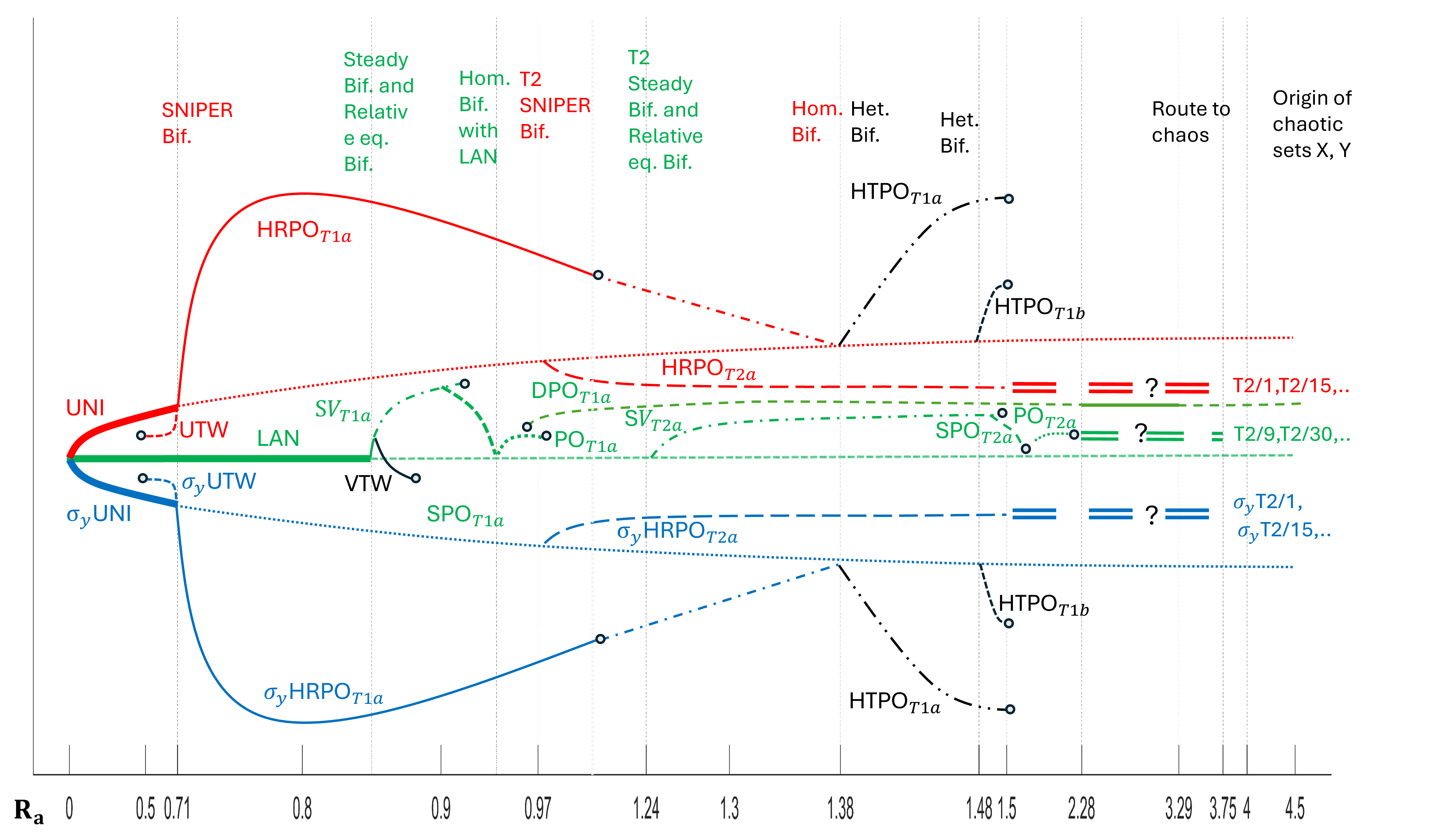}
\caption{\footnotesize \pg{Partial bifurcation diagram in SPU with $L=\tilde{L}/3$. The vertical placement of the branches is loosely guided by $\langle u\rangle$ and is intended primarily to convey the qualitative mean-flow direction: red and blue branches correspond, respectively, to $\langle u\rangle>0$ and $\langle u\rangle<0$ and share $\sigma_x\tau_x(L/2)$ or $\sigma_x\tau_x(L/4)$ symmetry in preturbulent regime, whereas green branches have $\langle u\rangle\approx 0$ and share $\sigma_x\sigma_y$ symmetry in preturbulent regime. Some branches with similar or identical values of $\langle u\rangle$ are vertically displaced for clarity and to avoid overlap. Although the HTPO branches are invariant under $\sigma_x\tau_x(L/2)$, they are plotted twice and in black for illustration, since their orbits are also invariant under $\sigma_y$ and are therefore related to both UNI and $\sigma_y$UNI. The vortex-traveling wave branch VTW is likewise plotted in black to indicate that it has broken $\sigma_x\sigma_y$ symmetry. Solid lines indicate stable branches and all other line styles indicate unstable branches. Circles mark the endpoints of branches for which continuation was not carried to completion. Question marks indicate conjectured connections between preturbulent branches at low activity and turbulent branches at high activity}}

    \label{fig:bifurcation_scheme}
\end{figure}

Hence, there are two axial (i.e. with 1D fixed point subspaces) subgroups $\Sigma$, and by the EBL, we have two pitchfork bifurcations. The first bifurcation leads to UNI-like solutions with reflection symmetry $\Sigma_1= \sigma_x $, and the second to LAN-like solutions with symmetry $\Sigma_2=\sigma_x\sigma_y $. Due to the pitchfork nature of these bifurcations, each new solution branch has a symmetric counterpart obtained by applying $ \sigma_y $. Therefore, a total of four nontrivial solution branches emerge.
\an{
Building on this result, we carry out a secondary bifurcation analysis of the UNI and LAN states. A partial bifurcation diagram is shown in Fig. \ref{fig:bifurcation_scheme}; the various branches appearing in the diagram are introduced and discussed in the subsequent sections.}

\subsection{Secondary bifurcations from UNI}
\an{Since UNI (and LAN) are invariant under the continuous translational symmetry, the eigenvectors of the system linearized about these states are of the form $g(y)e^{2\pi i n_xx/L}$. The UNI solution loses stability when a (`first') pair of complex eigenvalues crosses the imaginary axis, which corresponds to the eigenvectors with $n_x=1$. These eigenvectors break translation symmetry and have isotropy group $\Sigma_{p_1}(U) = \operatorname{gen}\{\sigma_x\tau_x(L/2)\}$, where $\operatorname{gen}\{\cdot\}$ denotes the group generated by the enclosed group operation. This corresponds to the ‘shift-reflect’ symmetry \citep{Cvitanovic2020Chaos}.  Here, we use the notation $p_{1(\text{U})}$ to denote the first complex pair of eigenvalues (and the corresponding eigenspace) of the UNI. }
The imaginary nature of the eigenvalues suggest the possibility of a Hopf bifurcation. This bifurcation is expected to lead to oscillatory solutions that respect the underlying symmetry $\Sigma_{p_{1(\text{U})}}$. Such oscillatory states are typically small perturbations of the UNI configuration, and have been reported in previous studies~\cite{giomi2012banding,Opathalage2019,Ramaswamy2016,shendruk2017dancing,Giomi2011}.
Here, we explore a scenario in which traveling waves emerge from this destabilization of UNI.

\an{In equivariant systems, secondary bifurcations are often more challenging to analyze than primary bifurcations, as the standard equivariant bifurcation theorems typically do not apply directly. This difficulty arises when a solution undergoing bifurcation has less symmetry than the full group $G$ and has a continuous group orbit. The presence of a continuous group orbit can dramatically modify the bifurcation scenario, since the solution undergoing bifurcation is not isolated in state space, and directions tangent to the group orbit are neutrally stable. These neutral directions can couple to the bifurcation dynamics, leading, for example, to drift along the group orbit and to bifurcation scenarios not captured directly by the standard equivariant bifurcation theorems.

For bifurcations from UNI (and LAN), however, the situation is simpler: although these equilibria have smaller symmetry than $G$, their group orbits are discrete and consist of two points each, related by the y-reflection ($\sigma_y$). Hence, there is no continuous group orbit direction that needs to be accounted for in the bifurcation analysis. This allows the bifurcations of UNI (and LAN) to be treated directly within the standard framework of equivariant bifurcation theory.}

\subsubsection{Traveling waves emerge from UNI}
\pg{To predict bifurcating oscillatory solutions with symmetry at a singular fixed point with a pair of purely imaginary eigenvalues, we use the equivariant Hopf theorem \cite{Hoyle2006,Golubitsky2002,Chossat2000} (EHT, see Appendix). As in the EBL case, we consider the reduced-dynamics on this 2D center subspace, which we identify as $\mathbb{R}^2$ spanned by the two eigenvectors. The reduced system $\dot{Y}=F_c(Y,\mu)$ is $\Gamma-$equivariant, where $\Gamma$ is the isotropy subgroup of the fixed point $X^*$. Moreover, in the equivariant Hopf setting, the symmetry group of a periodic solution is viewed as a subgroup of the spatiotemporal symmetry $\Gamma \times SO(2)$, where the additional $SO(2)$ factor represents time translations (or phase shifts) along the periodic orbit. Let the eigenvalues be $\lambda(\mu)=\sigma(\mu)\pm i \rho(\mu)$, with bifurcation occurring at $\mu=0$.
    
To apply EHT to the bifurcation of UNI, we consider the isotropy subgroup of UNI as the base group, denoted by $\Gamma = \mathbb{Z}_2(\sigma_x) \times \mathrm{SO}(2) \subset G$. Here, the $\mathrm{SO}(2)$ factor represents continuous translations in the periodic direction. According to the EHT, we seek a subgroup $\Sigma \subset \Gamma \times \mathrm{SO}(2)=\mathbb{Z}_2(\sigma_x) \times \mathrm{SO}(2)_x\times \mathrm{SO}(2)_t$ that fixes pointwise a two-dimensional subspace, which in this case is the entire $\mathbb{R}^2$, where we have used the subscripts to differentiate between the two $\mathrm{SO}(2)$ groups}.

Next, we obtain the representation of $\Gamma \times \mathrm{SO}(2)$ on $\mathbb{R}^2$ \pg{in a real orthogonal basis} \an{$B=\{v_1,v_2\}$}. On this subspace, we find that reflection $\sigma_x$ acts as

\begin{equation}\label{action_sigmax_HF}
\sigma_x =
\begin{pmatrix}
-1 & 0 \\
0 & -1
\end{pmatrix},
\end{equation}
 and the spatial translations act as the rotation matrices
\begin{equation} \label{action_translation_HF}
R(\phi) =
\begin{pmatrix}
\cos\phi & -\sin\phi \\
\sin\phi & \cos\phi
\end{pmatrix},
\quad \phi = \frac{2\pi \ell}{L}, \quad \ell \in [0, L), \quad \phi \in [0, 2\pi),
\end{equation}

 where the angle $\phi$ parametrizes the spatial rotation. The representation of the temporal group is
 \begin{equation}
R_t(\theta) =
\begin{pmatrix}
\cos\theta & \sin\theta \\
-\sin\theta & \cos\theta
\end{pmatrix},
\quad \theta \in [0, 2\pi),
\end{equation} \an{where $\theta$ parametrizes phase shifts of the periodic solutions.}

\begin{figure}[htbp]
    \centering
    \includegraphics[width=\textwidth]{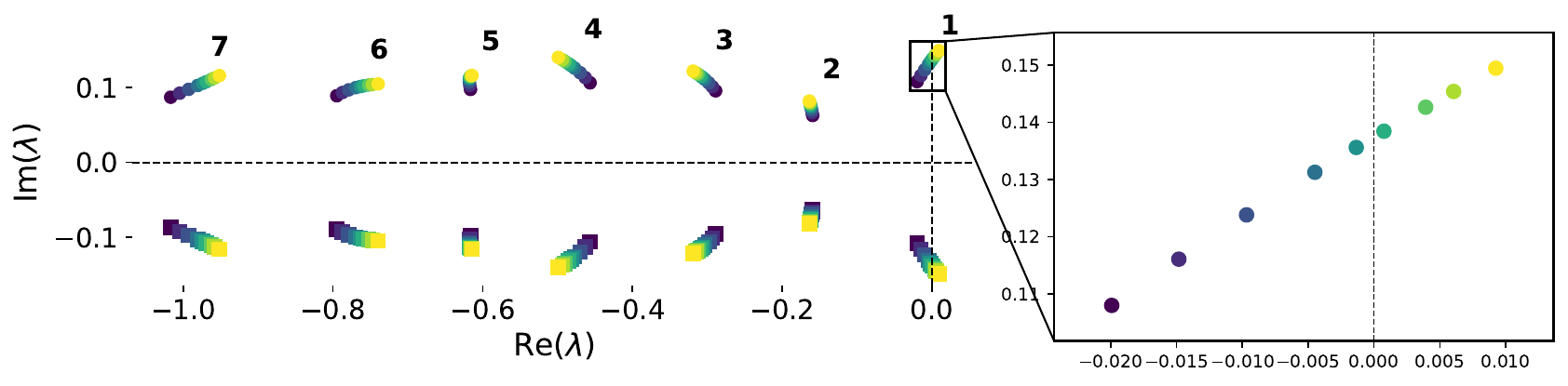}
    \caption{\footnotesize \an{Spectrum near the first Hopf bifurcation from UNI in the SPU with $L=\tilde{L}/3$.The seven leading eigenvalues are shown at selected values of the activity parameter, $\mathrm{R_a}=0.51$, $0.56$, $0.61$, $0.66$, $0.69$, $0.71$, $0.74$, $0.76$, and $0.79$, where the color encodes $\mathrm{R_a}$. The inset shows the leading eigenvalue crossing the imaginary axis at the Hopf bifurcation; the critical value is $R_{a_c}\approx0.702$. The imaginary part predicts the phase velocity $c$ of the emerging traveling wave. For $n_x=1$ and $L=80/3$, $c=\rho(\mathrm{R}{a_c})/(2\pi n_x/L)\approx0.587$.}}

    \label{fig:eigenvalue_spectrum}
\end{figure}

\begin{figure}[t]
    \centering
    \includegraphics[width=\linewidth]{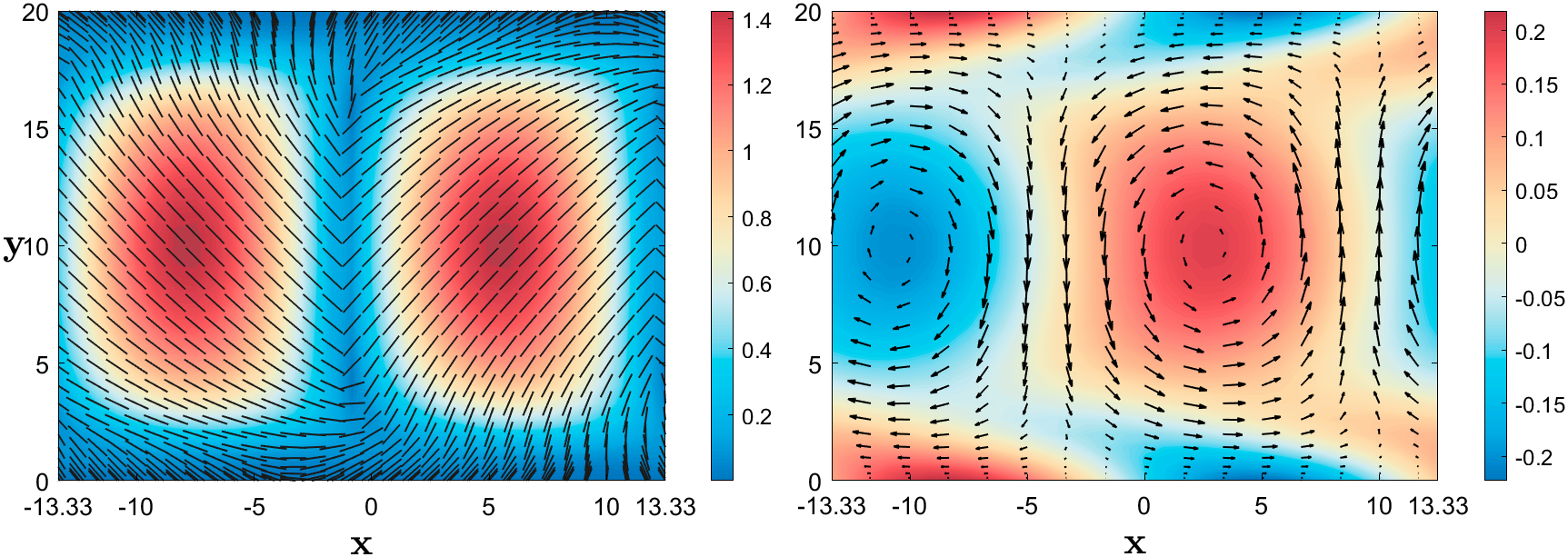}
    \caption{\footnotesize{\an{Nematic director field (left) and velocity field (right) corresponding to the critical basis vector \(v_1\) at first Hopf bifurcation from UNI for the SPU with \(L=\tilde{L}/3\). The second basis vector is obtained from \(v_1\) by the translation \(v_2=\tau_x(L/4)v_1\)}. Both are invariant under the symmetry $\sigma_x\tau_x(L/2).$ }}
    \label{fig:UNI_eigenvector_v1}
\end{figure}

\an{For the SPU with $L=\tilde{L}/3$, the critical eigenvalue associated with the first Hopf bifurcation from UNI occurs at $\mathrm{R_a}\approx 0.702$, see Fig.~\ref{fig:eigenvalue_spectrum}. The nematic director and velocity fields associated with the first basis vector $v_1$ at the bifurcation point are shown in Fig.~\ref{fig:UNI_eigenvector_v1}. The second basis vector, orthogonal to $v_1$, is obtained from $v_1$ by a translation, $v_2=\tau_x(L/4)v_1$. These vectors clearly break translation symmetry and are odd under the reflection $\sigma_x$, i.e., $\sigma_x v_1=-v_1$ and $\sigma_x v_2=-v_2$. Both vectors are invariant under the shift-reflect operation $\sigma_x\tau_x(L/2)$.}

The representation of $\Gamma$ is irreducible but not absolutely irreducible on $\mathbb{R}^2$, i.e., it acts simply on $\mathbb{R}^2$. Indeed, the only subspaces invariant under the action of all group elements given by Eqs.  (\ref{action_sigmax_HF}) and (\ref{action_translation_HF}) are the origin, and the whole space $\mathbb{R}^2$, and these matrices commute with each other. 

The spatiotemporal isotropy subgroup $\Sigma$ that fixes $\mathbb{R}^2$ (pointwise) consists of all group elements whose representation is the identity matrix, which yields:
 \begin{equation}
\Sigma =\{I,\sigma_xR(\pi), \sigma_xR_t(\pi), R(\theta)R_t(\theta)\}.
\end{equation}

Recall that $R(\pi)$ is the $L/2$ spatial shift, and hence the purely spatial discrete symmetry $\sigma_x R(\pi)$ is nothing but the shift-reflect symmetry $\Sigma_{p1(U)}$. 
As rotation in time $R_t(\pi)$ has the same representation as the spatial $L/2$ shift $R(\pi)$, the discrete spatiotemporal symmetry $\sigma_xR_t(\pi)$ is also the shift-reflect symmetry. Finally, the group elements of the form $R(\theta) R_t(\theta)$ describe continuous spatiotemporal symmetries in which each shift (or rotation) in space is compensated by a corresponding rotation in time.

Then, the EHT says that a unique branch of small-amplitude \textbf{time-periodic} solutions bifurcates from the UNI state, with spatiotemporal symmetry group $\Sigma$. Concretely, $R(\theta) R_t(\theta)$ symmetry implies the solutions are (small-amplitude) traveling (or equivalently, rotating) waves, while the other two symmetries require that these traveling waves possess the purely spatial shift-reflect symmetry $\sigma_x\tau_x(L/2)$.

\begin{figure}[h!]
    \centering

    \begin{minipage}[t]{0.45\textwidth}
        \centering
        \includegraphics[width=\textwidth]{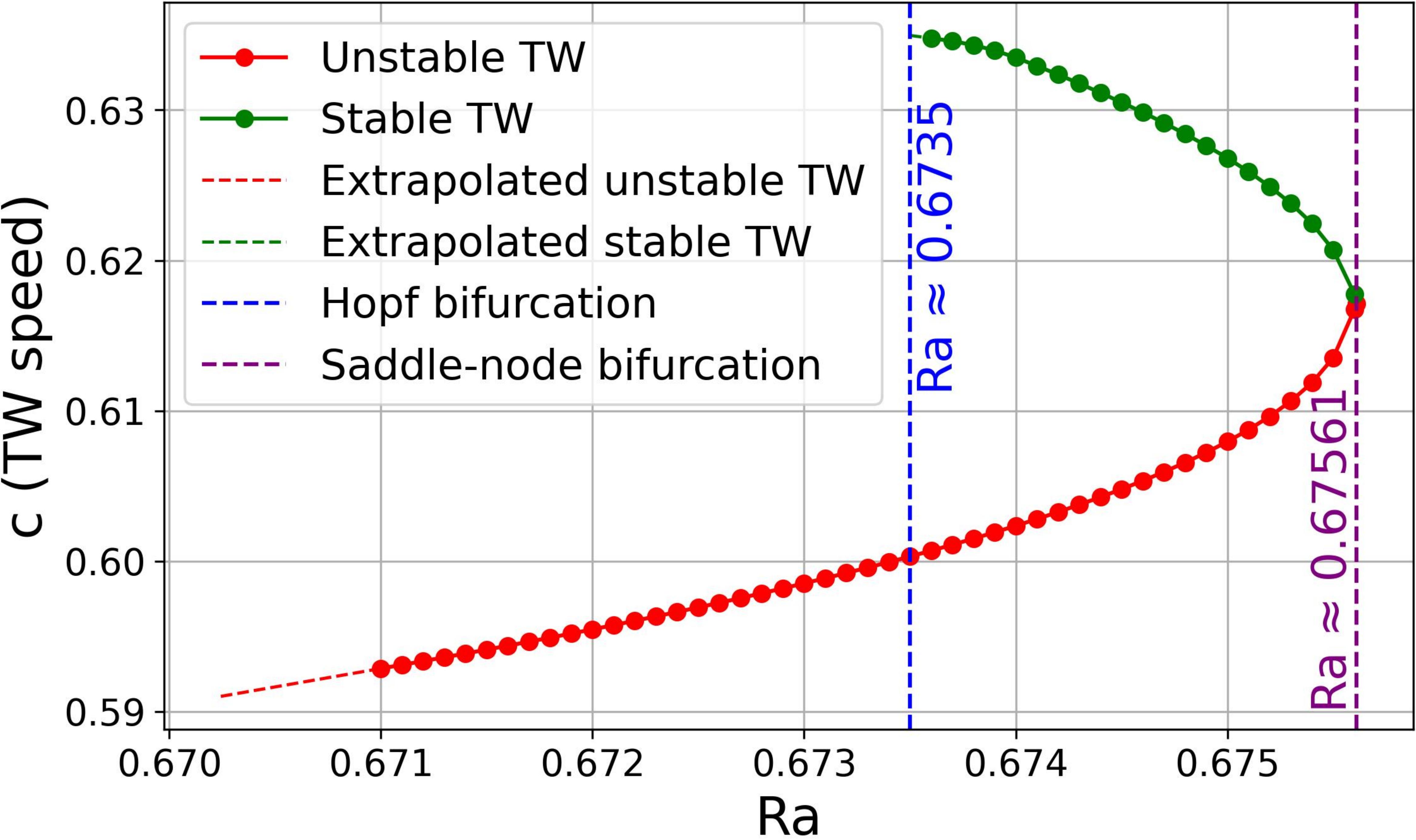}
        
        \vspace{0.1cm}
        {\footnotesize (a) Bifurcation diagram of TW speed versus $\mathrm{R}_{\mathrm{a}}$.}
    \end{minipage}
    \hfill
    \begin{minipage}[t]{0.50\textwidth}
        \centering
        
        \includegraphics[width=\textwidth]{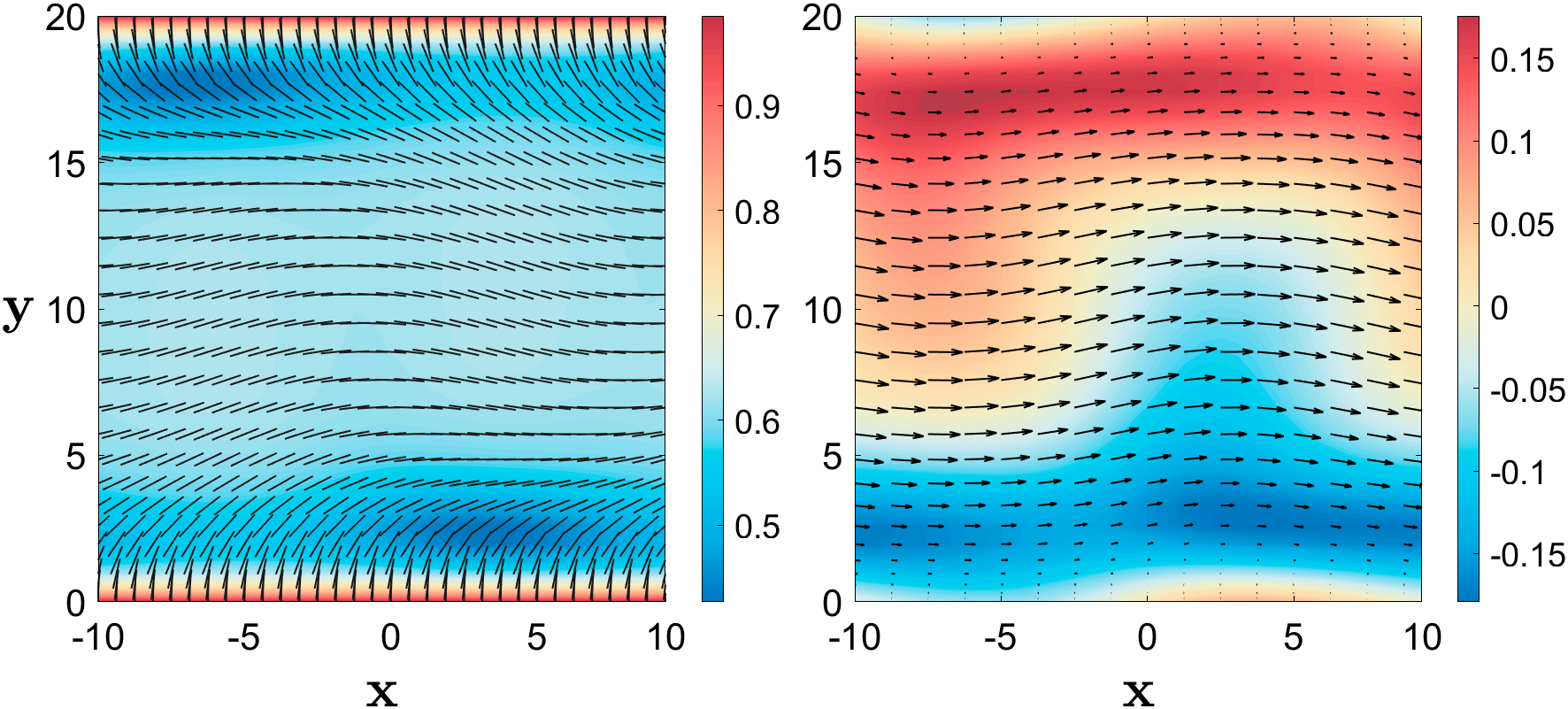}
        
        \vspace{0.1cm}
      
        {\footnotesize (b) Stable TW at $\mathrm{R}_{\mathrm{a}}=0.6755$\an{, invariant under the shift-reflect symmetry $\sigma_ x\tau_x(L/2) $.}}
    \end{minipage}

    \vspace{0.6cm}

    \begin{minipage}[t]{0.45\textwidth}
        \centering
        \includegraphics[width=\textwidth]{SNIPER.pdf}
        
        \vspace{0.1cm}
        {\footnotesize (c) Projection onto the reduced phase space showing the TWs and nearby potential heteroclinic connection.}
    \end{minipage}
    \hfill
    \begin{minipage}[t]{0.5\textwidth}
        \centering
        \includegraphics[width=\textwidth]{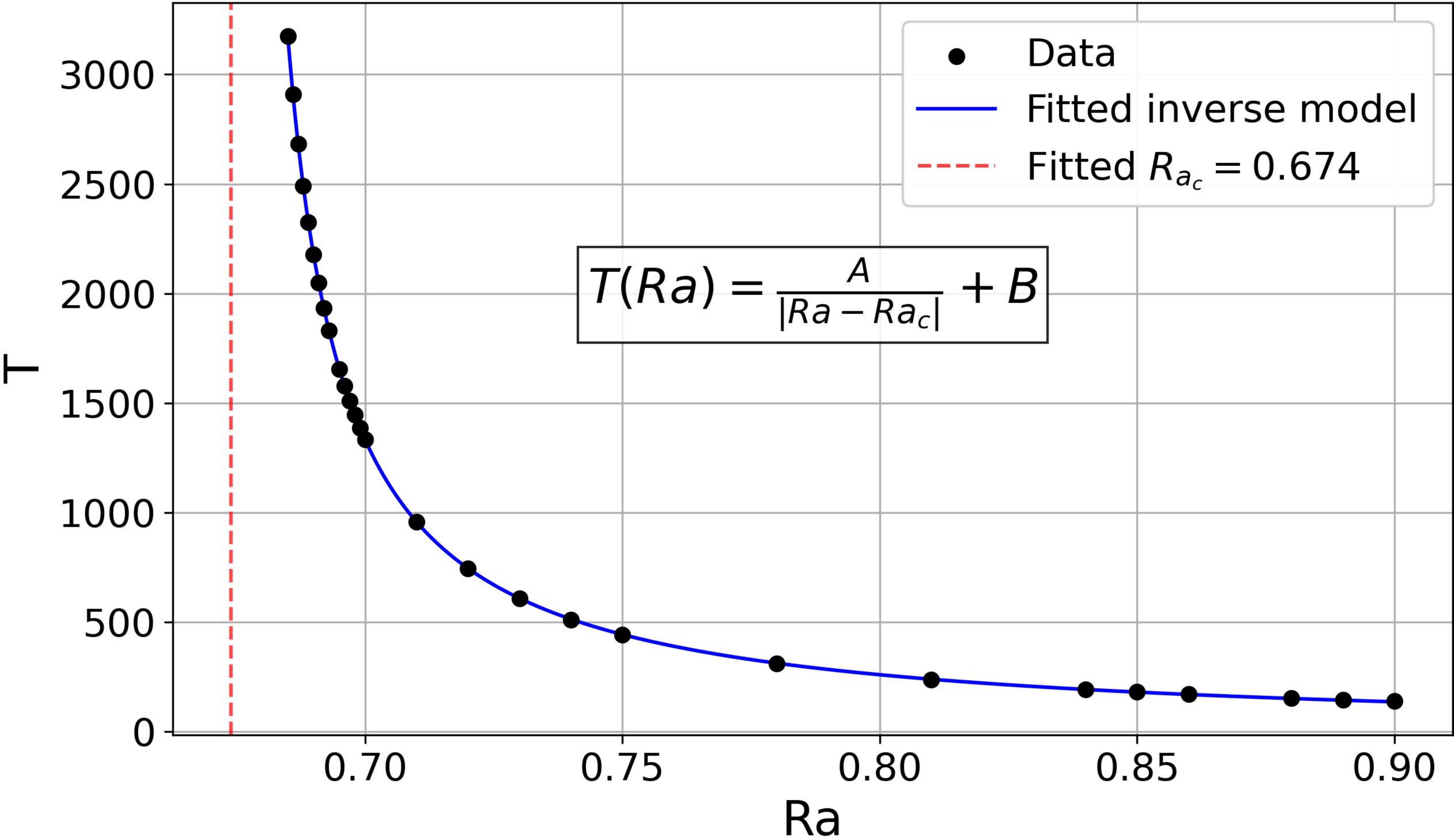}
        
        \vspace{0.1cm}
        {\footnotesize (d) Inverse-scaling-law fit for the period of the emerging RPO.}
    \end{minipage}

    \caption{\footnotesize
    Saddle-node infinite-period (SNIPER) bifurcation of traveling waves following the Hopf bifurcation of the UNI equilibrium in the SPU with $L=\tilde{L}/4$.
    (a) Bifurcation diagram showing the TW speed as a function of $\mathrm{R}_{\mathrm{a}}$.
    (b) Snapshot of the nematic director and velocity fields of the stable TW.
    (c) Projection onto the reduced phase space $(\langle Q_{11}\rangle,\langle v^2\rangle)$ showing the stable and unstable TWs and the nearby heteroclinic connection preceding the SNIPER bifurcation.
    (d) Inverse scaling of the period of the emerging RPO with the distance from the bifurcation point, supporting the degenerate SNIPER scenario.
    }
    \label{fig:SNIPER_main}
\end{figure}

\subsubsection{Saddle-node bifurcation of traveling waves}
\an{The predicted traveling-wave branch from the above-mentioned Hopf bifurcation in the SPU with $L=\tilde{L}/3$ proved elusive to detect numerically, as we discuss later. Meanwhile, we briefly turn to the SPU with $L=\tilde{L}/4$, where the emergence of the traveling-wave branch can be numerically resolved particularly clearly. The numerically computed stable TW is shown in Fig.~\ref{fig:SNIPER_main} in the context of a global bifurcation scenario, which we develop in the next subsection. Fig.~\ref{fig:SNIPER_main}(a) shows the stable TW branch, whose nematic and velocity fields at $\mathrm{R}{\mathrm{a}}\approx0.67558$ are shown in panel (b). The stable branch exists over the range $\mathrm{R}{\mathrm{a}}\approx0.6735$ to $\mathrm{R}_{\mathrm{a}}\approx0.67561$, where it collides with an unstable traveling wave (UTW), and disappears in a saddle-node bifurcation. This is a saddle-node bifurcation of traveling-wave solutions, which are \textit{continuous group orbits}. As remarked previously, we need to use an extension of the standard equivariant bifurcation theory that accounts for the effects of the continuous symmetries~\cite{krupa1990bifurcations} of the traveling waves in order to confirm this bifurcation.}

\begin{figure}[htbp]
    \centering
    \includegraphics[width=0.5\textwidth]{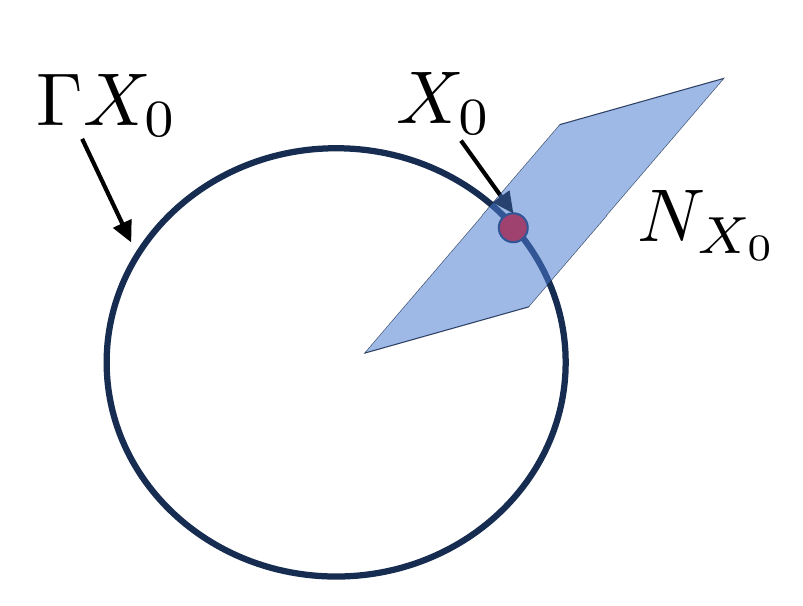}
    \caption{\footnotesize \pg{Local decomposition near a group orbit $\Gamma X_0$. The normal slice $N_{X_0}$, transverse to the group orbit at $X_0$, isolates the dynamics normal to the orbit. The dynamics restricted to the normal slice are $\Sigma_{X_0}$ equivariant and EBL applies directly.}}
    \label{fig:grouporbit}
\end{figure}

\pg{For a \( \Gamma \)-equivariant dynamical system, suppose the group orbit $\Gamma X_0$ of a point $X_0$ (with isotropy group $\Sigma_{X_0}$) has dimension at least one so that the orbit forms a continuous manifold rather than a discrete collection of symmetry-related states. Two common examples are: (a) the continuous family generated by translating a non-translation invariant equilibrium and (b) the group orbit associated with a traveling wave. In case (a), the vector field vanishes at every point on the group orbit, whereas in case (b), the flow is tangent to the group orbit. Thus, in both cases, $\Gamma X_0$ is flow invariant. The TW and UTW solutions are examples of case (b) and are referred to as relative equilibria.

The bifurcation analysis from a group orbit involves decomposing the vector field \( F \) into the sum of a normal and a tangential component: \( F = F_N + F_T \), near a point on the group orbit ~\cite{krupa1990bifurcations}. This decomposition separates the neutral directions associated with the continuous symmetry from the directions in which an actual bifurcation can occur. In the first step, we pick any point $X_0$ on the group orbit, and consider the dynamics restricted to the normal slice $N_{X_0}$ at that point, given by $\dot{X}=H(X,\mu)$, where, $H=F_N|_{N_{X_0}}$, see Fig. \ref{fig:grouporbit}. A classical equivariant bifurcation analysis is then performed on $H$ using the EBL/EHT, with the $H$ dynamics being $\Sigma_{X_0}$ equivariant. For some of the bifurcations discussed below, we will also use the Proposition \ref{prop:trivial_action_bif} provided in Appendix, which addresses the EBL/EHT cases where \( \Sigma_{x_0} \) acts trivially on a 1D or 2D center subspace 
\cite{krupa1990bifurcations}.

Once the bifurcation in the normal slice has been determined, the corresponding solution must be reconstructed in the full phase space by adding back the dynamics along the group orbit. To do so, we use the dimension-counting result stated in Theorem \ref{thm:relative_equilibrium} in the Appendix. The theorem determines the number of independent frequencies of the reconstructed solution, which, together with its symmetry properties, identifies the nature of the bifurcating state in the full phase space.}
For the purported bifurcation, we approximate the relevant eigenspace by computing the subspace corresponding to the eigenvalue with smallest non-zero real part for linearization around the UTW, close to the bifurcation point. As expected, this space is one-dimensional, associated with a small real eigenvalue, and is invariant under the symmetry \( \sigma_x \tau_x(L/2) \). This implies that the action of the isotropy subgroup \( \Sigma_{x_0} \) on the eigenspace is trivial, indicating that no symmetry is broken at the bifurcation. \pg{By Proposition~\ref{prop:trivial_action_bif}, we conclude that the bifurcation is indeed a saddle-node bifurcation of relative equilibria. This result is consistent with our numerical finding that the two branches of traveling waves coalesce at the bifurcation point.}

 \pg{Under the bifurcation scenario of Proposition \ref{prop:trivial_action_bif}, the resulting equilibria ($X$) inherits the symmetry of $X_0$. Hence in such situations, the Theorem \ref{thm:relative_equilibrium} dimension count can be replaced by $k = \operatorname{rank}\left( N(\Sigma_{X_0}) / \Sigma_{X_0} \right)$. In the present case $\Gamma=\mathbb{Z}_2(\sigma_x) \times \mathrm{SO}(2)$ (inherited from UNI), and the isotropy group is the shift-reflect $\Sigma_{X_0}=\mathbb{Z}_2(\sigma_x\tau_x(L/2))$, hence we get $k=1$. This is consistent with our prior observation that the bifurcating solutions are traveling waves, where time evolution coincides with translations $\tau_x(l)X$ ($0\leq l \leq L$).}

\subsubsection{Saddle-node Infinite-period bifurcation and the origin of homoclinic-like relative periodic orbits} The saddle-node bifurcation of TWs discussed above is part of a global bifurcation scenario involving a saddle-node infinite-period bifurcation (SNIPER), as illustrated in Fig.~\ref{fig:SNIPER_main}. In the interval \( 0.6737 \leq \mathrm{R}_{\mathrm{a}} \leq 0.67561 \), UNI, TW, and UTW coexist, with evidence of heteroclinic connections between them. To investigate whether transversal or tangential intersections among their manifolds are possible, we focus on the activity value \( \mathrm{R}_{\mathrm{a}} = 0.675 \). In an $N-$ dimensional space, an unstable manifold of dimension $n_u$ and a stable manifold of dimension $n_s$ generically intersect transversely in  a $d=n_u+n_s-N$ dimensional submanifold \citep{guillemin2025differential}.

UNI and UTW each have two-dimensional unstable manifolds, which allows for transversal intersections with the stable manifold of TW. However, a transversal connection between the unstable manifold of UTW and the stable manifold of UNI is not possible. Perturbing UNI along its unstable manifold leads to TW, while perturbing UTW in an unstable direction also drives the system toward TW. Interestingly, we also found a potential heteroclinic connection from UTW to UNI very close to \( \mathrm{R}_{\mathrm{a}} = 0.675 \), see \an{Fig.~\ref{fig:SNIPER_main}(c).}

As the activity increases, the two traveling waves get closer and disappear in the saddle-node bifurcation at \( \mathrm{R}_{\mathrm{a}} \approx 0.6756 \). Beyond this point, a family of relative periodic orbits (RPOs) with isotropy group \( \Sigma_{p_1(U)} \) emerges, passing very close to UNI. \an{A representative RPO whose shape closely resembles the vanished heteroclinic network, is also shown in Fig.~\ref{fig:SNIPER_main}(c). The periods of these RPOs follow the inverse scaling law of a degenerate SNIPER bifurcation \citep{strogatz2024nonlinear,kuehn2009scaling}, as shown in Fig.~\ref{fig:SNIPER_main}(d).}

\begin{figure}[htbp]
    \centering
    \subfloat[\footnotesize
Reduced phase space projection of four members of the homoclinic-like RPO family HRPO$_{T1a}$ at $\mathrm{R_a} = 0.72$, $0.90$, $1.02$, and $1.24$, shown in purple, blue, black, and green solid curves, respectively. Their $\sigma_y$ counterparts are shown as dashed curves. For HRPO$_{T1a}$ at $\mathrm{R_a}=1.24$, we show velocity snapshots at two locations along the orbit: one corresponding to a vortex-lattice flow state and the other to a nearly unidirectional flow state. \an{UNI and $\sigma_y$UNI are indicated by red and blue asterisks, respectively, near the SNIPER bifurcation point, $\mathrm{R_a}\approx 0.71$.}]{%
        \includegraphics[width=.97\textwidth]{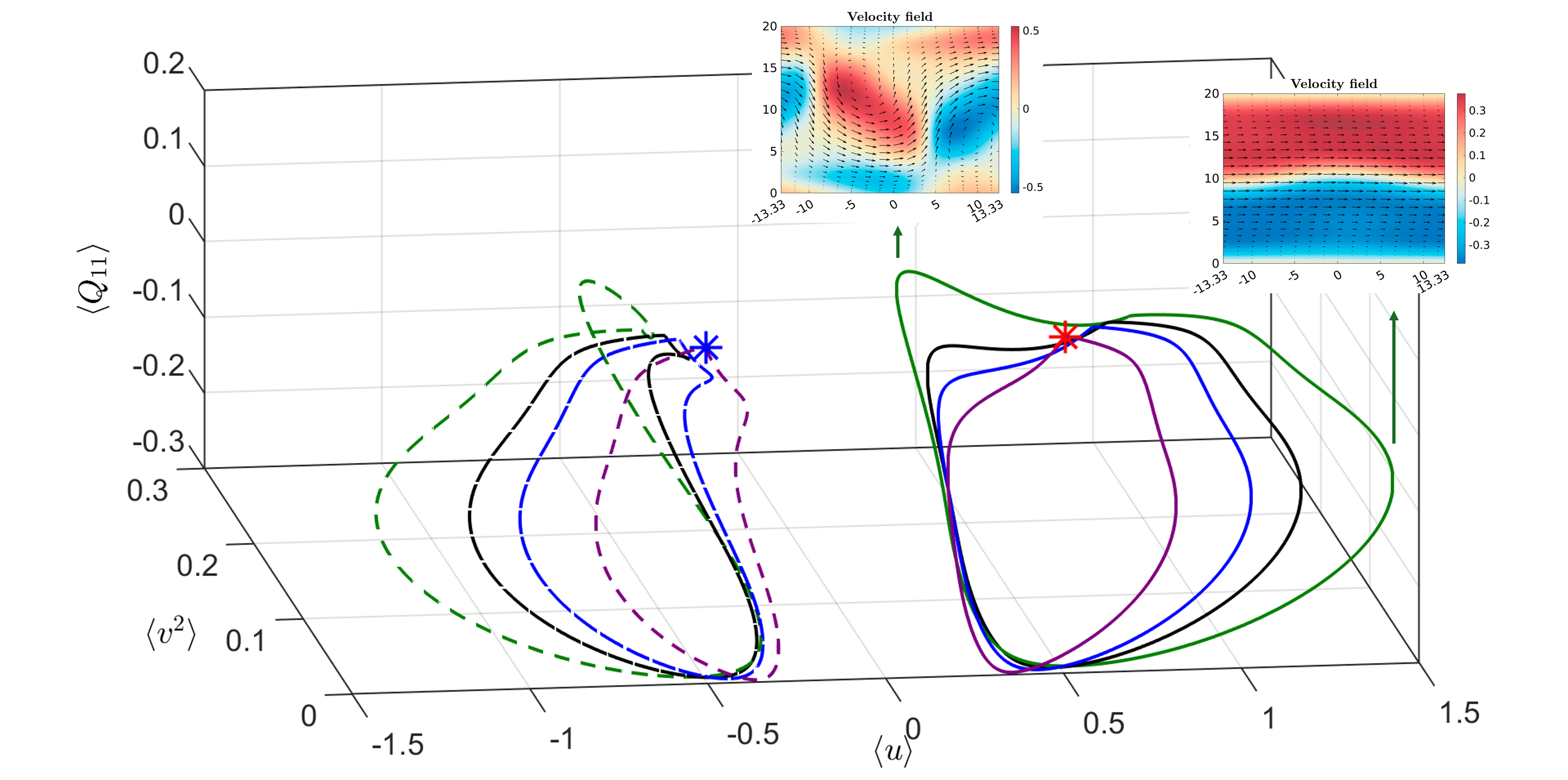}%
        \label{fig:Homoclinic_like_RPO}%
    }\\[0.5em]
    \subfloat[\footnotesize Inverse law scaling fit between the period and the activity. Here, $A \approx 40.291$, $B \approx -15.671$, and $\mathrm{R_{a_c}} \approx 0.702$.]{%
        \includegraphics[width=.6\textwidth]{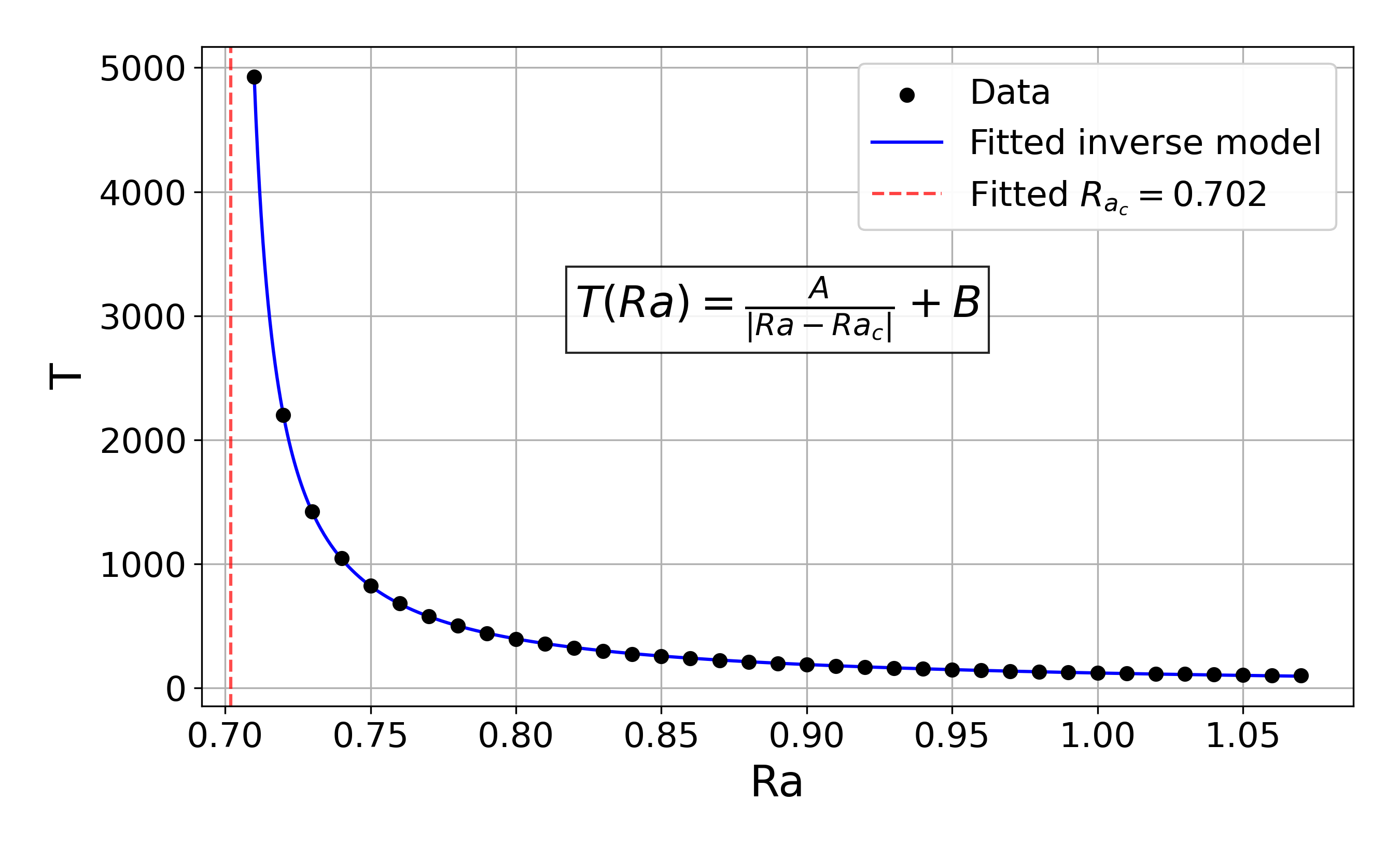}%
        \label{fig:SNIPER_LAW}%
    }

    \caption{\footnotesize Confirmation that homoclinic-like RPOs (of the HRPO$_{T1a}$ family) originate from a SNIPER bifurcation in the \an{SPU} A  ($L = \tilde{L}/3$). \an{Note that the estimated SNIPER critical value \(\mathrm{R}_{\mathrm{a_c}}\approx0.702\) is essentially identical to the Hopf bifurcation value of UNI, making the stable TW difficult to detect in practice.}}
    \label{fig:SNIPER_Lover_3}
\end{figure}

\an{We now resume our analysis of SPU with $L=\tilde{L}/3$, where the SNIPER bifurcation is not as clearly resolved as in the domain with $L=\tilde{L}/4$. The unstable traveling wave UTW, which was continued down to approximately $\mathrm{R}_{\mathrm{a}} \approx 0.55$, disappears at the higher-activity end just beyond the Hopf bifurcation point of UNI. Hence, the TW branch emerging from UNI is expected to exist over a very narrow range of activity before colliding with UTW in the saddle-node bifurcation. We suspect that this narrow range is responsible for our inability to compute the stable TW to the desired accuracy. In support of the SNIPER scenario, we confirm the existence of the expected RPOs, several of which are shown in Fig.~\ref{fig:Homoclinic_like_RPO}. Near \( \mathrm{R}_{\mathrm{a}} \approx 0.71 \), the RPO branch also exhibits the same degenerate SNIPER scaling as before, as shown in Fig.~\ref{fig:SNIPER_LAW}. This fit predicts $\mathrm{R}_{\mathrm{a_c}}\approx0.702$, very close to Hopf bifurcation point, hence the difficulty of resolving the TW branch. Taken together, the evidence strongly suggests that a SNIPER bifurcation occurs in the SPU with \( L = \tilde{L}/3 \), despite the difficulty in resolving the short TW branch. The observation of the same scenario for different domain sizes suggests that the SNIPER mechanism for the origin of the RPO branches is robust and is not a consequence of a particular domain size.
}

Due to their proximity to UNI in phase space, we refer to these RPOs as \emph{homoclinic-like RPOs} characterized by recurrent processes of defect creation and annihilation. We refer to the RPO family in Fig. \ref{fig:SNIPER_Lover_3} as HRPO$_{T1a}$, where $Tk$ in the subscript implies the $L/k$ translation symmetry, and the following letter distinguishes different families $(a,b,\dots)$. Since the dynamics leading to flow reversals are closely associated with the exact coherent structures (ECSs) that emerge in the channel with \( L = \tilde{L}/3 \), we henceforth restrict our analysis to this domain (and subdomain \( L = \tilde{L}/6 \)) for the remainder of the paper. The \an{SPU} with \( L = \tilde{L}/3 \) is henceforth referred to as \an{SPU} A.




\subsubsection{From Homoclinic-like RPOs to Heteroclinic-like POs }
\begin{figure}[htbp]
    \centering

    \subfloat[\footnotesize Two homoclinic-like RPO members of the HRPO$_{T1a}$ family at $\mathrm{R}_{\mathrm{a}}= 1.35 \text{ and } 1.37$ shown as solid red and blue curves, respectively, with their $\sigma_y$ symmetry-related counterparts shown as dashed curves. And two heteroclinic-like PO members of the HTPO$_{T1a}$ family at  $\mathrm{R}_{\mathrm{a}}= 1.39 \text{ and } 1.41$ shown as black and green curves, respectively. The orbits of these POs are invariant under $\sigma_y$. The black arrow shows the direction of increasing  activity ($\mathrm{R}_{\mathrm{a}}$). \an{UNI and $\sigma_y$UNI are indicated by red and blue asterisks, respectively, for $\mathrm{R}_{\mathrm{a}}= 1.38$.}]{%
        \includegraphics[width=\textwidth]{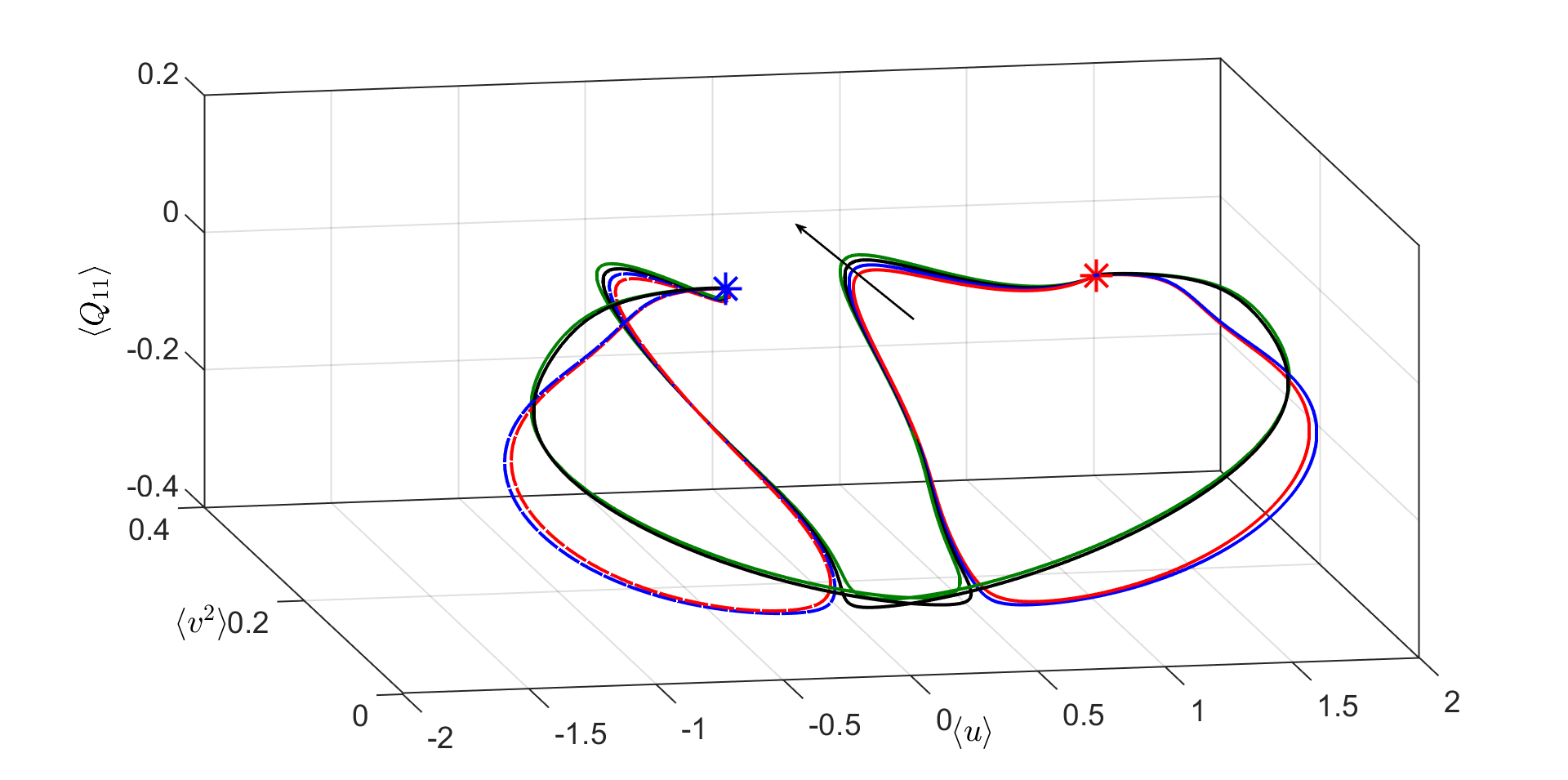}%
        \label{fig:HO_HT_transition}%
    }\\[1em]

    \subfloat[\footnotesize Logarithmic scaling of period near homoclinic bifurcation in which the HRPOs disappear. Here, $A\approx 13.03$, $B\approx29.63537$, and $\mathrm{R}_{\mathrm{a_c}}\approx 1.38087$.]{%
        \includegraphics[width=0.48\textwidth]{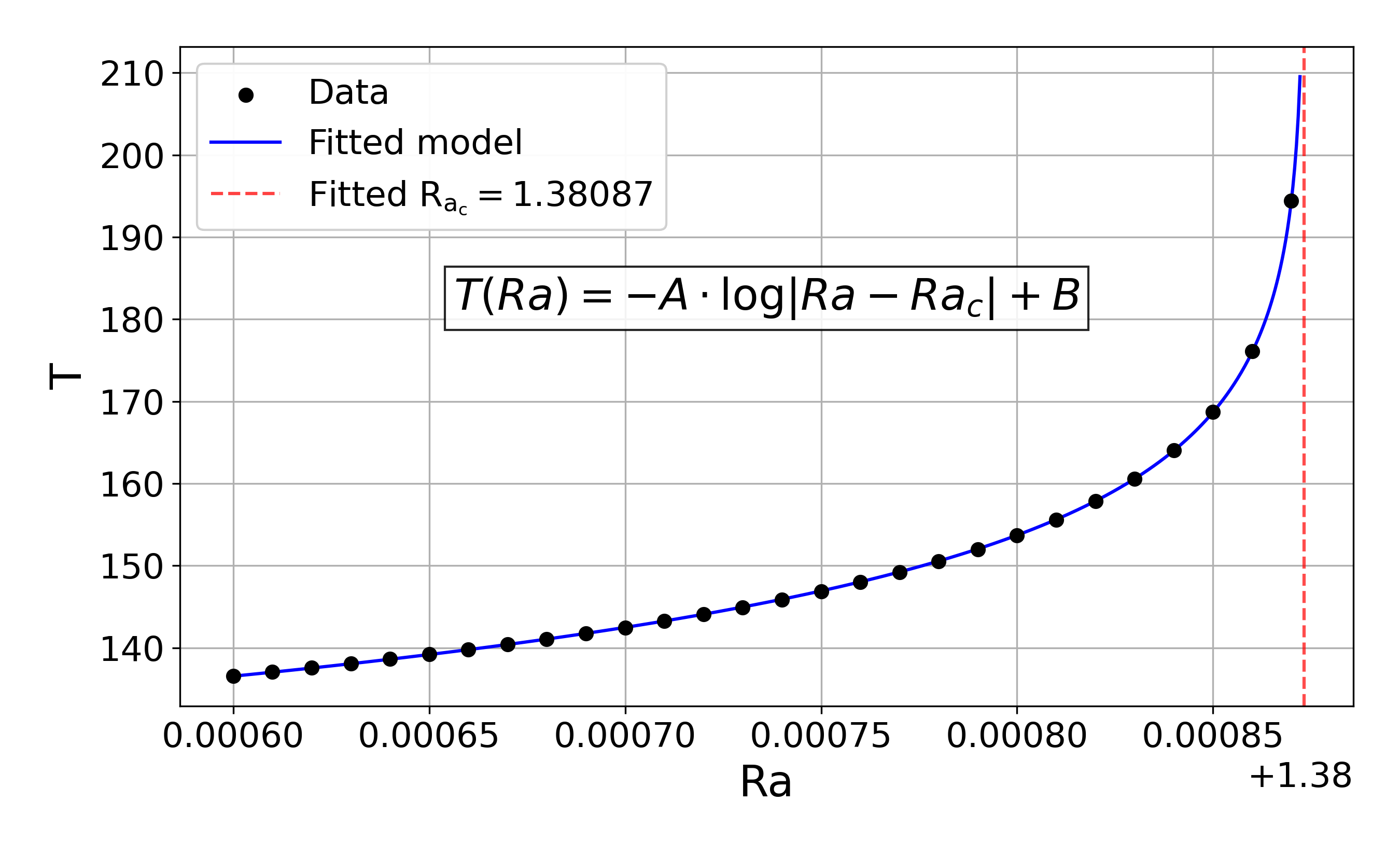}%
        \label{fig:log_homo_RPO}%
    }\hfill
    \subfloat[\footnotesize Logarithmic scaling of period near heteroclinic bifurcation after which the HTPOs emerge. Here, $A\approx 19.56307$, $B\approx 86.35135$, and $\mathrm{R}_{\mathrm{a_c}}\approx 1.38087$.]{%
        \includegraphics[width=0.48\textwidth]{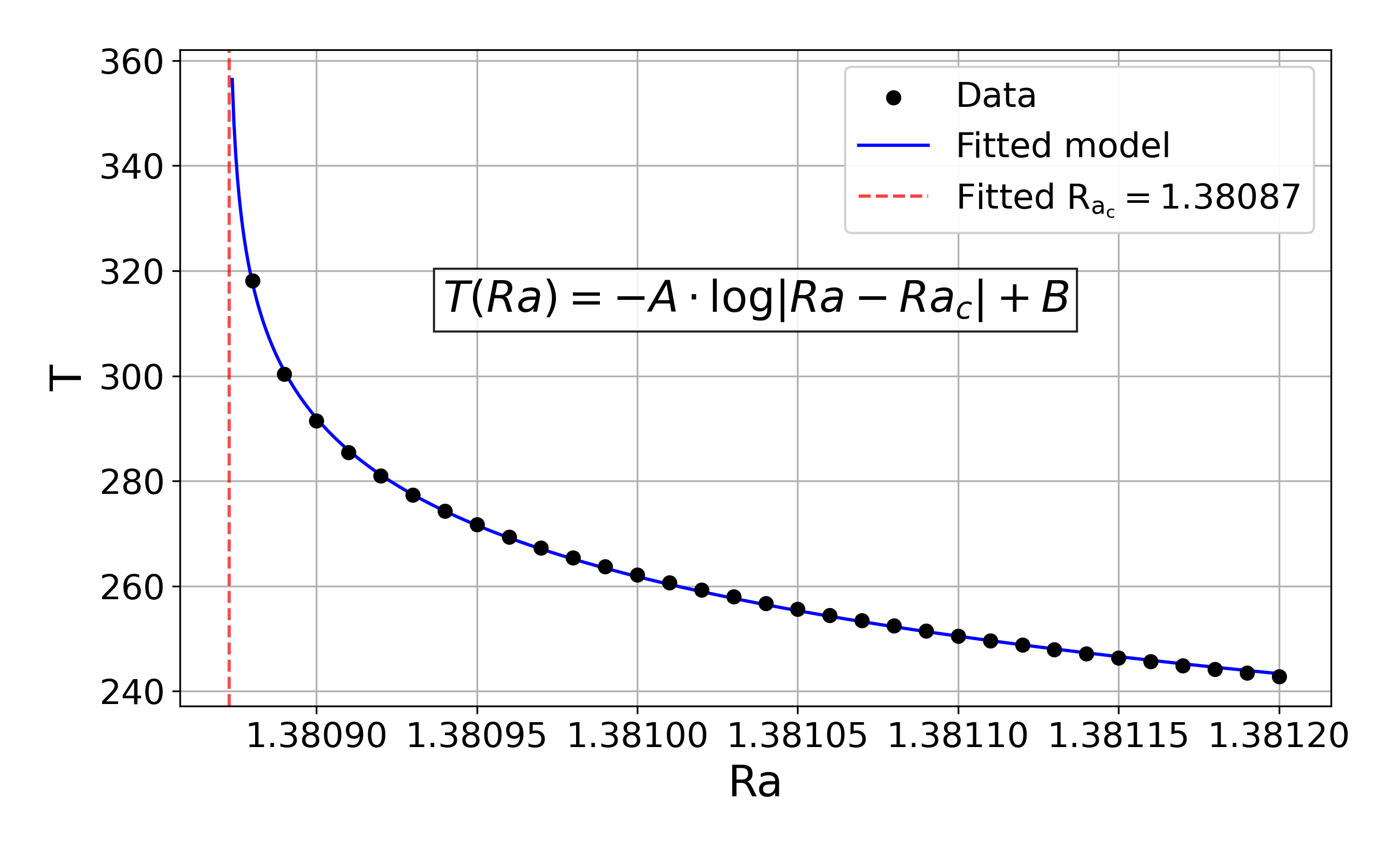}%
        \label{fig:log_homo_PO}%
    }

    \caption{\footnotesize Two consecutive global bifurcations, both with logarithmic scaling consistent with homoclinic/heteroclinic bifurcations. The first (homoclinic) bifurcation results in the vanishing of homoclinic-like RPOs. The second (heteroclinic) bifurcation leads to emergence of heteroclinic-like POs. In both cases, the RPOs and POs collide with the UNI (and $\sigma_y$ UNI) solution at the respective bifurcation points. Note that the estimated bifurcation values  $\mathrm{R}_{\mathrm{a_c}}$ \an{are approximately the same.}}
    \label{fig:Homoclinic_RPO_PO_transition}
\end{figure}

\an{As activity increases beyond the SNIPER bifurcation point, \( \mathrm{R}_{\mathrm{a}} \gtrsim 0.71 \) (for $L = \tilde{L}/3$)}, the periods of homoclinic-like RPOs initially decrease (as predicted by the inverse law), but eventually the RPOs start to become larger again, until the RPO branch collides with \texttt{UNI} and disappears through a homoclinic bifurcation at approximately \( \mathrm{R}_{\mathrm{a}} \approx 1.381 \). \an{Two such RPOs close to the bifurcation are shown in Fig.~\ref{fig:HO_HT_transition}, while Fig. \ref{fig:log_homo_RPO} confirms that near the bifurcation, the RPO periods along this branch exhibit the logarithmic scaling characteristic of a homoclinic bifurcation.

Immediately beyond this point, a branch of heteroclinic-like periodic orbits emerges via a heteroclinic bifurcation involving \texttt{UNI} and $\sigma_y\texttt{UNI}$. Together with the disappearing RPOs, two of the emerging POs are shown in Fig.~\ref{fig:HO_HT_transition}. The periods again follow a logarithmic scaling, as shown in Fig.~\ref{fig:log_homo_PO}. This new family of periodic orbits is denoted as HTPO$_{T1a}$, where we use the same naming convention as before. Their origin is via a \emph{symmetry-increasing} bifurcation \citep{dellnitz1995admissible,chossat1988symmetry} in which the two RPOs, related by the $\sigma_y$ symmetry, disappear to create a PO whose orbit is $\sigma_y$-symmetric. These ECSs are additionally invariant under the symmetry \( \sigma_x \tau_x(L/2) \) inherited from the HRPOs. These solutions are important because they connect the $\langle u \rangle > 0 $ and \( \langle u \rangle < 0 \) regions}. 

\begin{figure}[htbp]
    \centering
    \subfloat[\footnotesize
A member of the second branch of heteroclinic-like periodic orbits, HTPO$_{T1b}$, which is $\sigma_y$-symmetric, shown close to the second heteroclinic bifurcation at $Ra \approx 1.5$, together with UNI and $\sigma_y$UNI, indicated by red and blue asterisks, respectively. Also shown are three velocity snapshots along the orbit: two exhibiting opposite unidirectional flows and one with a vortical flow state.]{%
        \includegraphics[width=\textwidth]{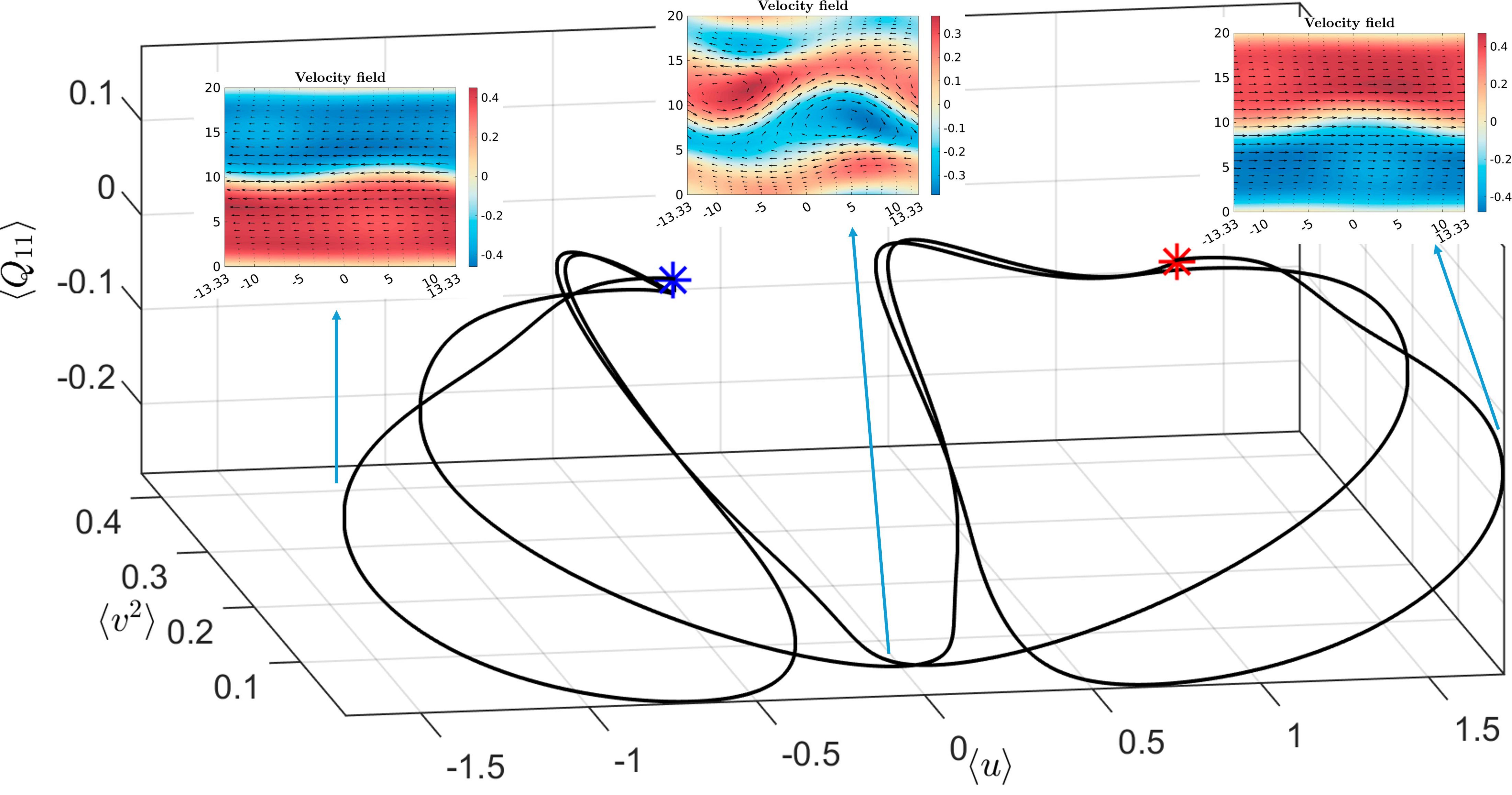}%
        \label{fig:HTPO_T1b}%
    }\\[0.5em]
    \subfloat[\footnotesize \an{Logarithmic scaling of period near the second heteroclinic bifurcation.} Here, $A\approx 25.13308$, $B\approx  97.805602$, and $\mathrm{R}_{\mathrm{a_c}}\approx 1.485646$.]{%
        \includegraphics[width=.6\textwidth]{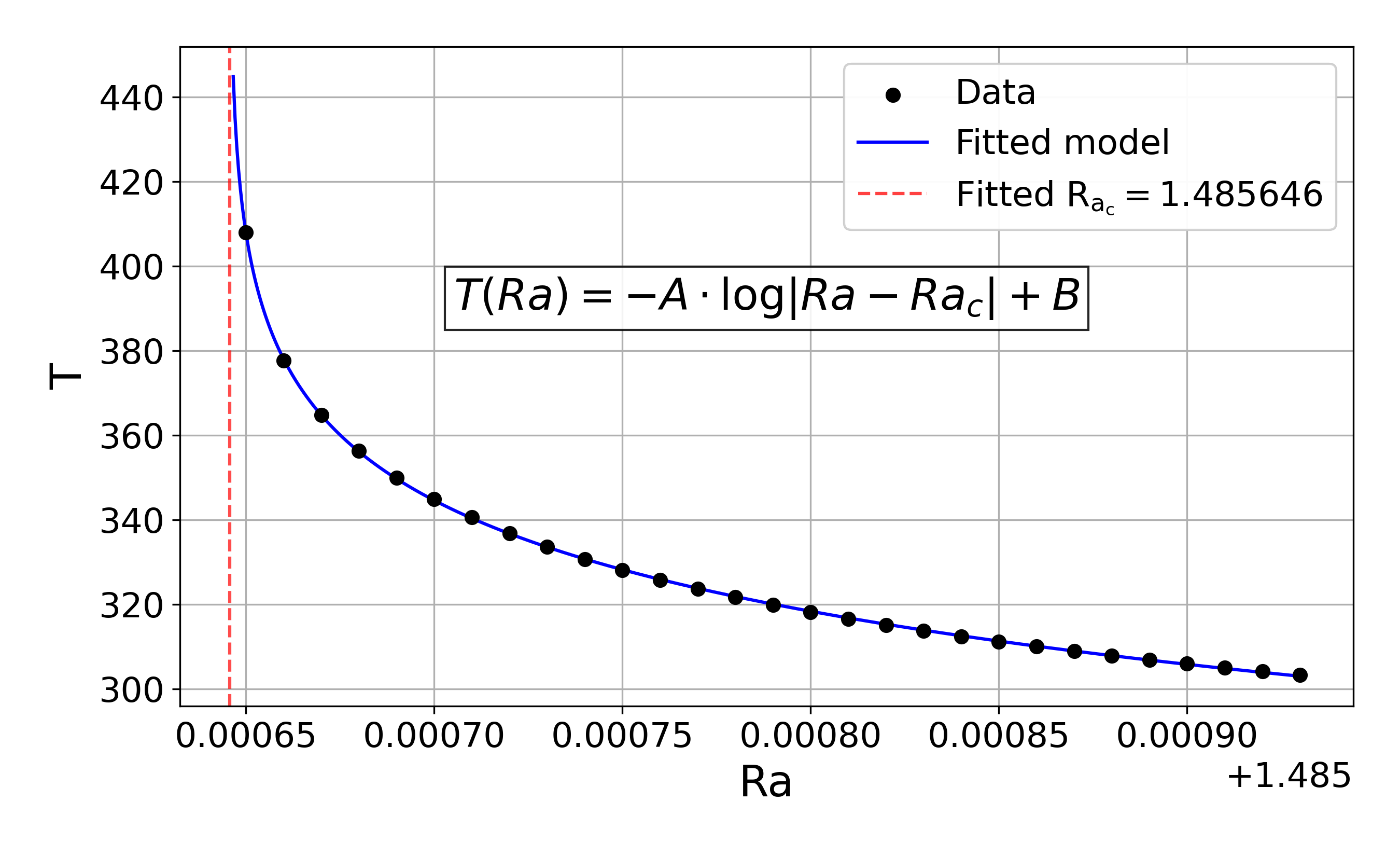}%
        \label{fig:log_HTPO2}%
    }

    \caption{\footnotesize A second branch of heteroclinic-like periodic orbits, HTPO$_{T1b}$, and associated logarithmic scaling near the second heteroclinic bifurcation point.}
    \label{fig:HTPO2_pair}
\end{figure}

\an{A second heteroclinic bifurcation, again involving \texttt{UNI} and \( \sigma_y \texttt{UNI} \), occurs at approximately \( \mathrm{R}_{\mathrm{a}} \approx 1.486 \), giving rise to a second branch of heteroclinic-like POs, HTPO$_{T1b}$. A member of this branch is shown in black in Fig.~\ref{fig:HTPO_T1b} together with the logarithmic fit of the corresponding periods in Fig.~\ref{fig:log_HTPO2}. The solution cycles between nearly unidirectional flows in opposite directions, passing through a vortex-like flow, as illustrated by the representative channel plots in the Fig.~\ref{fig:HTPO_T1b}. We have observed that the two branches, HTPO$_{T1a}$ and HTPO$_{T1b}$, coexist over a range of activity numbers. In particular, they coexist at $\mathrm{R}_{\mathrm{a}}\approx 1.5$.}




\subsubsection{Secondary Bifurcations Associated with Higher Unstable Eigenvalues of UNI}

Our analysis so far has been limited to bifurcations from the primary unstable eigenspace of UNI. As the activity number is increased to \(\mathrm{R}_{\mathrm{a}} = 4.5\), a total of seven pairs of complex eigenvalues eventually cross the imaginary axis at various intermediate activity values. The corresponding secondary bifurcations give rise to additional ECSs, each embedded within a distinct symmetry subspace. While a full analysis is out of scope, it is crucial to identify how the symmetries of these other eigenspaces shape the bifurcation structure and the global dynamics deep into the turbulent regime.

\an{Each crossing corresponds to a Hopf bifurcation, and the symmetries of each unstable 2D eigenspace are listed as isotropy groups in Table~\ref{tab:isotropy_groups_UNI} in the Appendix. These symmetries constrain the nature of solutions that can emerge from the corresponding bifurcations. We performed a detailed continuation and bifurcation analysis for the first eigenvalue pair, as described above. For the remaining eigenvalue pairs, we do not pursue the same exhaustive analysis of the resulting branches, since the bifurcation sequences observed for the first pair suggest that similar mechanisms recur for the other symmetry subspaces. Instead, we focus on identifying the role of additional symmetries present in these branches.}

\begin{figure}[htbp]
    \centering
    \includegraphics[width=1.0\textwidth]{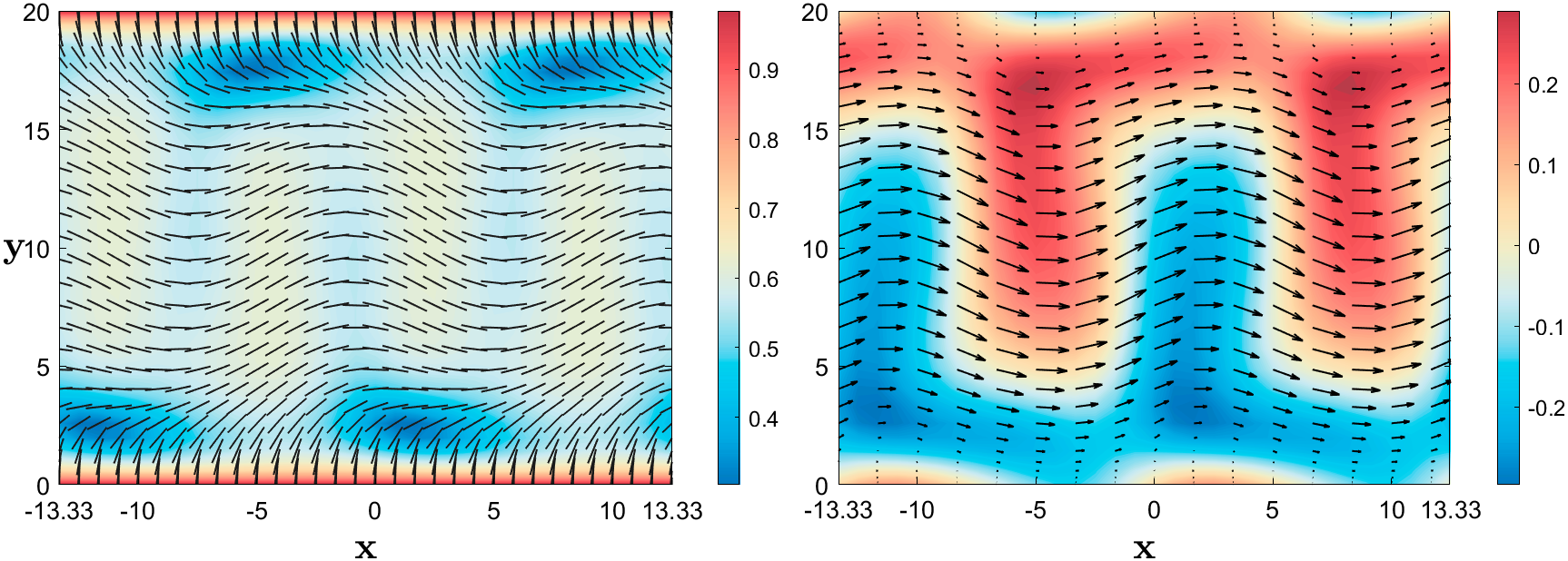}
    \caption{\footnotesize A snapshot of the nematic director (left) and velocity fields (right) of a traveling wave solution with isotropy group \(\Sigma_{p_{2(\text{U})}}\), computed at \(\mathrm{R}_{\mathrm{a}} = 1.0\). This solution originates from UNI via an equivariant Hopf bifurcation at \(\mathrm{R}_{\mathrm{a}} \approx 0.97\), associated with the second complex-conjugate pair of eigenvalues crossing the imaginary axis.}
    \label{fig:TW_UNI_T2}
\end{figure}

The second unstable eigenvalue pair leads to a bifurcation scenario closely resembling that of the first pair, resulting in a branch of traveling waves with isotropy group \(\Sigma_{p_{2(\text{U})}}\). \an{A representative solution from this branch is shown in Fig.~\ref{fig:TW_UNI_T2}, where its field structure resembles two copies of the $TW$ from the first pair stacked horizontally. This branch also} undergoes a saddle-node bifurcation and subsequently a SNIPER bifurcation. The result is a branch of homoclinic-like RPOs that we call HRPO$_{T2a}$. However, in contrast to the first eigenvalue pair, no subsequent heteroclinic connections between UNI and \(\sigma_y\)UNI were observed \an{as the activity was increased}.

The branches arising from bifurcations associated with the higher eigenvalues populate the phase space with solutions that inherit the underlying symmetries of corresponding critical eigenspaces. In particular, the \pg{second and fourth} unstable eigenspaces have the symmetry \(\tau_x(L/2)\), and therefore give rise to solutions confined to the invariant subspace defined by this transformation. \an{This invariant subspace is equivalent to the phase space of an SPU with half the domain length, i.e., a subdomain of length $\tilde{L}/6$, where, as we show in the following sections, \an{ECSs with the $\tau_x(L/2)$ symmetry provide a dynamical framework for the turbulent reversal dynamics.}}

\subsection{Secondary bifurcations from LAN}
In this subsection, we carry out an equivariant bifurcation analysis of the secondary instabilities emerging from the LAN state in \an{SPU} A. Our main aim is to clarify the origin of dynamically relevant vortex–lattice–like ECSs. Oscillatory vortex states have been reported in several experimental and numerical studies~\cite{Ramaswamy2016,giomi2012banding,shendruk2017dancing,Wagner2022,Wagner2023}; in particular, Wagner \textit{et al.} found vortex patterns whose characteristic length scale adjusts to the channel aspect ratio. A subset of these, termed `dancing disclinations'~\citep{shendruk2017dancing,tan2019topological,klein2025spontaneous,Mitchell2024}, exhibit $+1/2$ defects that form a space–time braid, and appear as POs and RPOs~\cite{Wagner2022,Wagner2023}. Here we show that the relevant vortex–lattice solutions are not connected to the previously identified secondary bifurcations of the UNI equilibrium, and instead most likely arise from secondary bifurcations of the LAN equilibrium.

\subsubsection{One-vortex equilibria}
\begin{figure}[t]
    \centering
    \includegraphics[width=\linewidth]{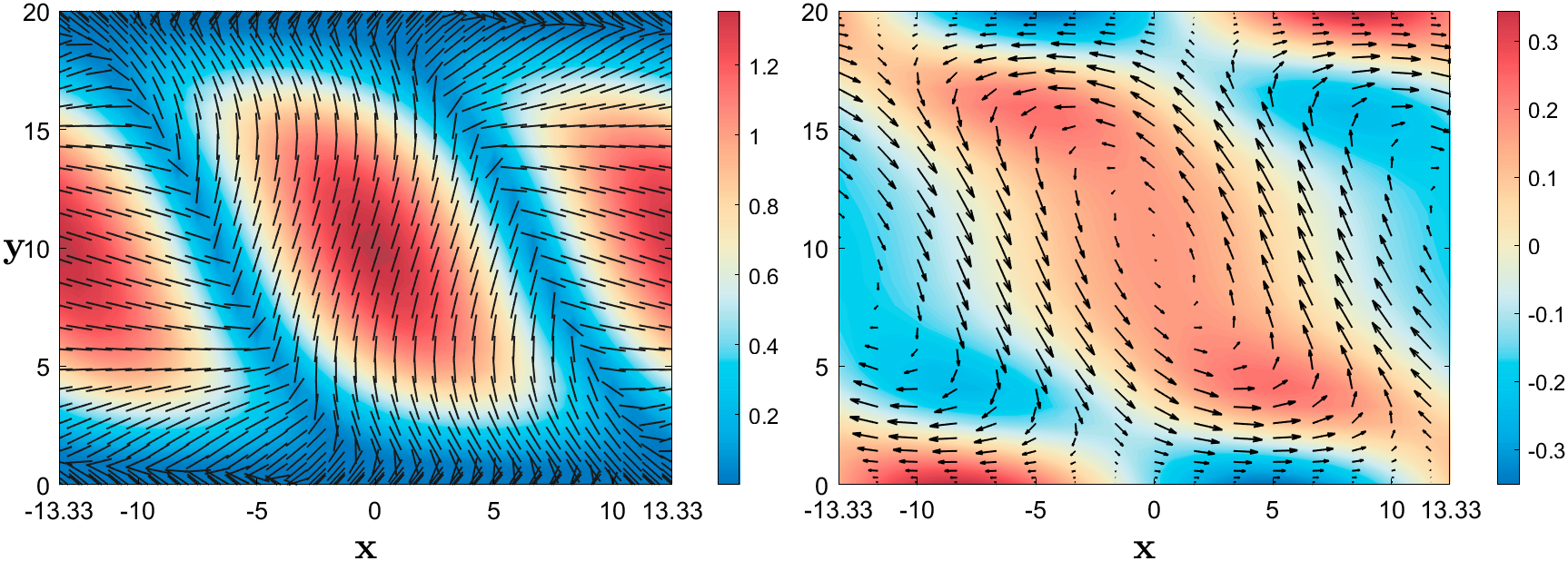}
    \caption{\footnotesize{Nematic director field (left) and velocity field (right) of critical \(\sigma_x\sigma_y\)-invariant eigenvector \(v_1\) at first pitchfork bifurcation from LAN. The companion orthogonal eigenvector is obtained from \(v_1\) by the translation \(v_2=\tau_x(L/4)v_1\).}}
    \label{fig:LAN_eigenvector_v1}
\end{figure}

We consider the isotropy subgroup of LAN, $\Gamma_{LAN} = SO(2) \rtimes \mathbb{Z}_2 \cong O(2) $, as the base group  for the equivariant analysis of bifurcations from LAN. The LAN solution first loses stability when a pair of real eigenvalues simultaneously crosses the imaginary axis at  $\mathrm{R}_{\mathrm{a}}\approx 0.85$, \an{corresponding to eigenvectors with $n_x=1$}. 
Since the eigenvalues are real, this is a steady bifurcation, and the two-dimensional center subspace is taken as $\mathbb{R}^2$. In this center subspace, we can choose real orthogonal eigenvectors $v_1$ and $v_2$ such that the reflection $\sigma_x\sigma_y$ is represented by


\begin{equation}\label{action_sigmax_sigma_y_ST}
\sigma_x\sigma_y =
\begin{pmatrix}
1 & 0 \\
0 & -1
\end{pmatrix}.
\end{equation}
\an{Fig.~\ref{fig:LAN_eigenvector_v1} shows the nematic director and velocity fields of the $\sigma_x\sigma_y$-invariant eigenvector $v_1$. The second eigenvector $v_2$ is obtained by translating $v_1$ by $L/4$, $v_2=\tau_x(L/4)v_1$. In the representation given by Eq.~\ref{action_sigmax_sigma_y_ST} on $\mathbb{R}^2$, $\sigma_x\sigma_y$ acts as a reflection about the $x$-axis. Its fixed-point subspace is therefore the $x$-axis, corresponding to the one-dimensional subspace spanned by $v_1$.}

The translations act as the rotation matrices,
\begin{equation} \label{action_translation_ST}
R(\phi) =
\begin{pmatrix}
\cos\phi & -\sin\phi \\
\sin\phi & \cos\phi
\end{pmatrix},
\quad \phi = \frac{2\pi \ell}{L}, \quad \ell \in [0, L), \quad \phi \in [0, 2\pi).
\end{equation}

Every subgroup generated by the conjugate reflection $\Sigma_{\phi}=R(\phi)\sigma_x\sigma_yR(-\phi)$ is axial, and has the line (passing through origin) with slope $\tan \phi$ as its fixed point subspace. The EBL predicts an infinite number of (pitchfork) branches of solutions bifurcating from LAN, one for each $\phi$, and possessing the corresponding $\Sigma_{\phi}$ symmetry. 

\begin{figure}[htbp]
    \centering
    \subfloat[\footnotesize The nematic director and velocity field of a one-vortex \textbf{equilibrium} solution bifurcated from the LAN at $\mathrm{R}_{\mathrm{a}}=0.85$. Its set of spatial translations constitutes a \textbf{relative equilibrium}.]{%
        \includegraphics[width=0.30\textwidth]{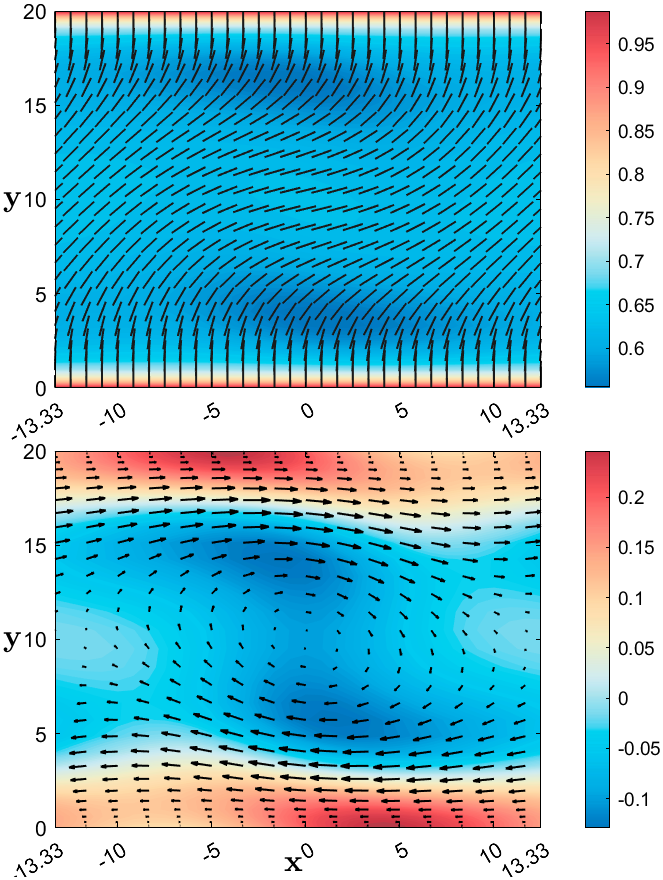}%
        \label{fig:One_vortex}%
    }
    \hspace{0.02\textwidth}
    \subfloat[\footnotesize Two snapshots from a time simulation at \( \mathrm{R}_{\mathrm{a}} = 0.851 \) of the stable \textbf{one-vortex wave} traveling leftward \an{at a non-dimensional speed of $0.27$}, separated by a time interval of \( 50\tau \). This wave branch emerges via a steady bifurcation (of the corresponding normal vector field) from the one-vortex relative equilibrium.]{%
        \includegraphics[width=0.60\textwidth]{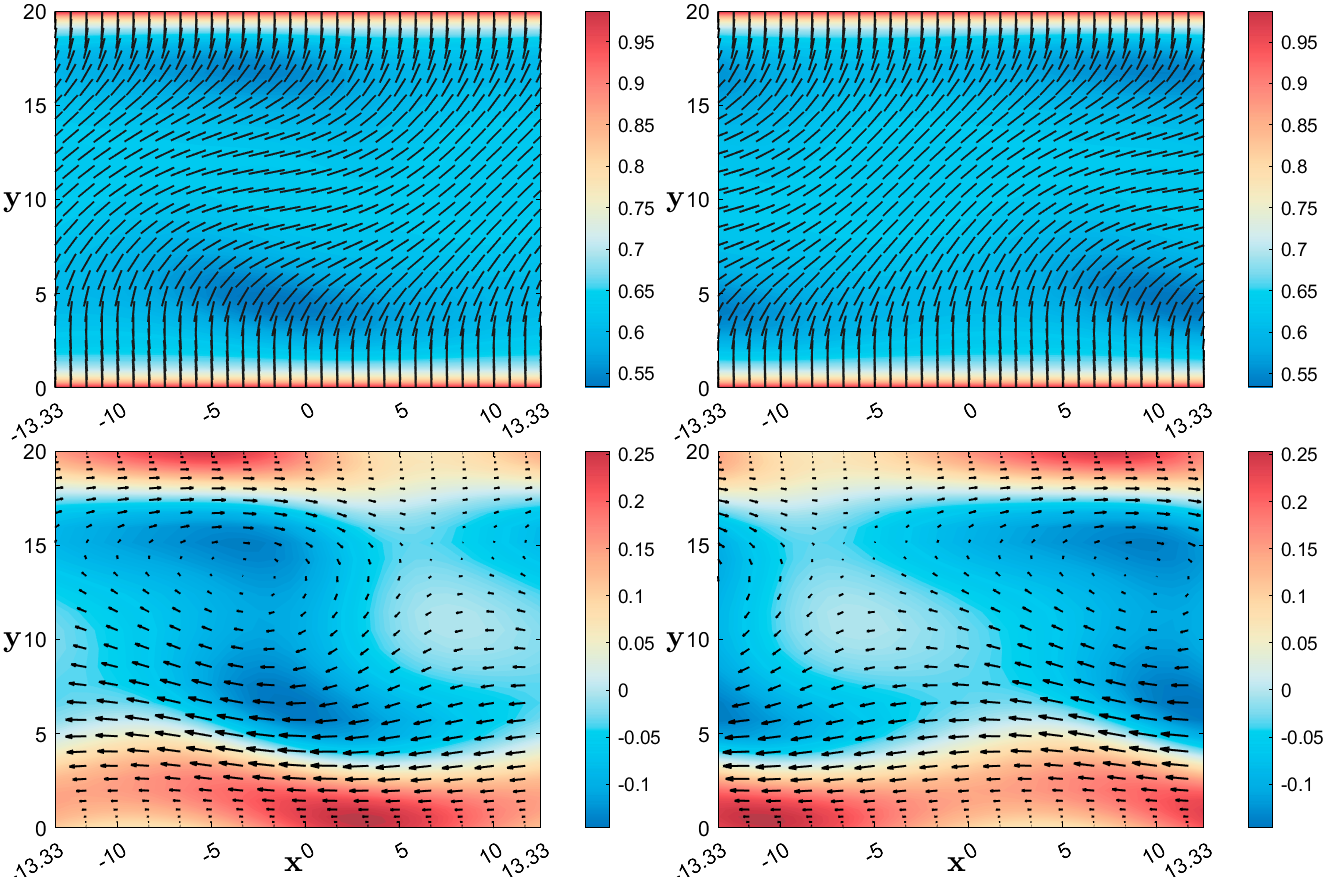}%
        \label{fig:one_vortex_TW}%
    }
    \caption{\footnotesize One-vortex solutions: relative equilibria and traveling wave originating from the LAN solution in \an{SPU} A.}
    \label{fig:one_vortex_solutions}
\end{figure}

Indeed, numerically we found a branch of steady one-vortex equilibrium solution (see Fig. \ref{fig:One_vortex}), invariant under the reflection $\Sigma_0=\sigma_x\sigma_y$, emerging from the LAN equilibrium. The other branches of solutions with $\Sigma_{\phi>0}$ symmetry are simply translations of this one-vortex solution. These solutions form a group orbit of equilibria, i.e., a relative equilibrium. 

\subsubsection{One-vortex traveling waves via symmetry-breaking bifurcation}
The steady vortex solutions are stable for a small range of activity, \(0.85 \lessapprox \mathrm{R}_{\mathrm{a}} \lessapprox 0.851\). Like in the case of TW discussed earlier, the bifurcation occurring at \(\mathrm{R}_{\mathrm{a}} \approx 0.851\) from the group orbit of the one vortex solution requires analyzing the vector field normal to the group orbit.

We denote any one-vortex equilibrium by \( X_0 \), and consider the kernel of Jacobian of the normal vector field, \( d_XH(X_0, 0) \), which is one-dimensional and associated with a real eigenvalue in this case. \an{The isotropy group acts nontrivially on this critical eigenspace, with the action of \( \sigma_x \sigma_y \) given by $\sigma_x\sigma_y(v)=-v$, where $v\in\ker d_XH(X_0,0)$ is a basis vector of the one-dimensional kernel} \cite{Golubitsky2002, krupa1990bifurcations}. To apply the Equivariant Branching Lemma, we identify the axial group \(\Sigma = \{I\}\), and predict a branch of solution with \emph{no symmetry}. To find the group orbit of this solution, we apply Theorem ~\ref{thm:relative_equilibrium} with \(\Sigma = \{I\}\) and $\Gamma=O(2)$, the base group for LAN. This yields \( N(\{I\}) / \{I\} \cong O(2) \), which implies that the trajectory on the group orbit of this bifurcating solution is rank 1, and hence a \emph{traveling (rotating) wave}, consistent with the general analysis for \( O(2) \)-equivariant systems \cite{krupa1990bifurcations}.


\begin{figure}[htbp]
    \centering
    \subfloat[\footnotesize The SPO$_{T1a}$ family of periodic solutions bifurcating from the one-vortex steady state via a symmetry-preserving Hopf bifurcation, with the arrow indicating the parent steady solution. The colormap indicates activity. These periodic orbits grow in size as the activity is increased, and eventually collide with LAN and disappear via a homoclinic bifurcation.]{%
        \includegraphics[width=\textwidth,height=0.3\textheight,keepaspectratio]{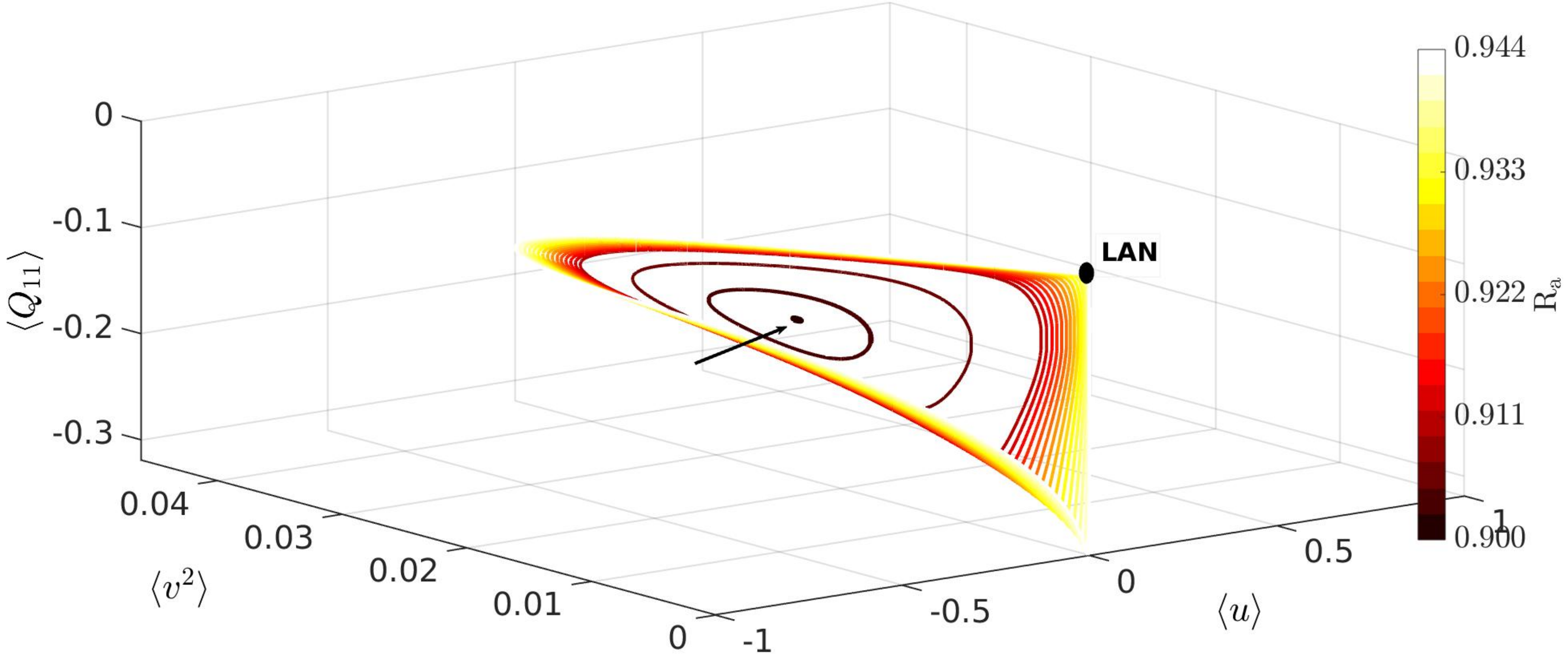}%
        \label{fig:Vortex_Homo_LAN}%
    }\\[0.5em]
    \subfloat[\footnotesize Logarithmic fit of the period near the homoclinic bifurcation, consistent with theory. Here, $A\approx 60.9477$, $B\approx -133.8315$, and$\mathrm{R}_{\mathrm{a_c}}\approx 0.9447$. ]{%
        \includegraphics[width=\textwidth,height=0.3\textheight,keepaspectratio]{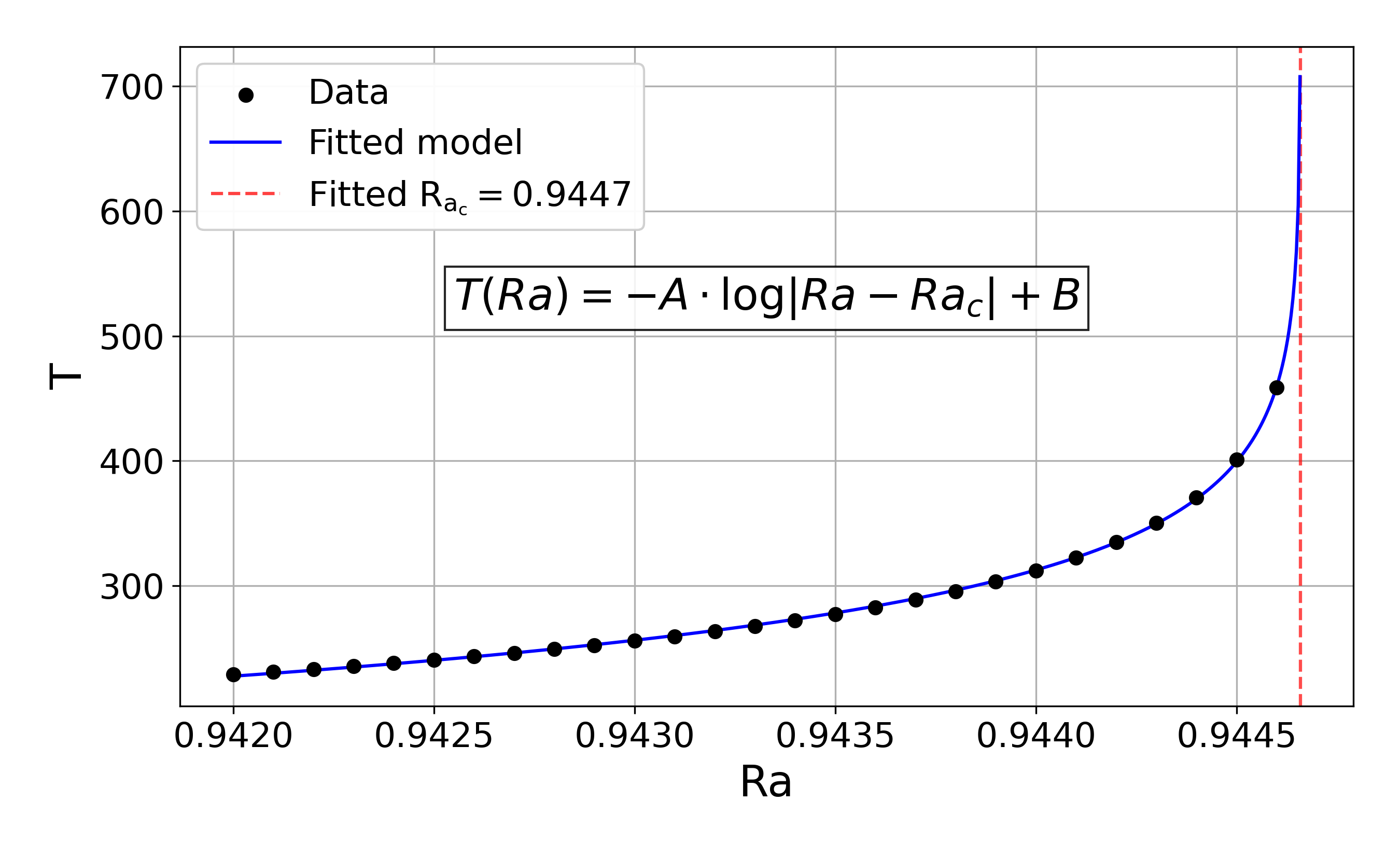}%
        \label{fig:Log_Law_One_Vortex}%
    }
    \caption{\footnotesize One-vortex standing wave POs bifurcating from the steady state and terminating at a homoclinic bifurcation with LAN.}
    \label{fig:Combined_One_Vortex}
\end{figure}

Numerically, we found \an{the predicted} traveling wave solution, \an{denoted as VTW}, at \( \mathrm{Ra} \approx 0.851 \), which we call \emph{one-vortex traveling  wave} (see Fig. \ref{fig:one_vortex_TW}). As predicted, this traveling wave has no symmetry. This solution closely resembles the traveling vortices observed in contractile active polar fluids \cite{Ramaswamy2016}. Note that unlike the UNI case, this TW does \emph{not} arise from a Hopf bifurcation, and this bifurcation is sometimes referred to as a `drift-pitchfork' bifurcation \citep{kness1992symmetry,brauns2024nonreciprocal}. We also note that the ZF$\rightarrow$LAN$\rightarrow$ One-vortex equilibrium $\rightarrow$ one-vortex TW sequence is closely related to the bifurcation scenario observed for one-dimensional excitable media \citep{kness1992symmetry}.

\an{To the best of our knowledge, stable and coherent vortex traveling-wave solutions, such as the one-vortex traveling wave identified here, have not previously been reported in active nematics.}

\subsubsection{One-vortex standing waves via symmetry-preserving bifurcation}
The one-vortex TWs are solutions with no spatial symmetry. However, the previously reported dancing disclination solutions which are vortex dominated POs usually have the $\sigma_x\sigma_y$ symmetry \an{(or a conjugate of it)} \cite{Wagner2023}. To understand their origin, we look for symmetry-preserving bifurcations of the one-vortex equilibrium. We find that this equilibrium undergoes a second bifurcation at $\mathrm{R}_{\mathrm{a}} \approx 0.9$, when a complex pair of eigenvalues crosses the imaginary axis. The symmetry group $\Gamma = \mathbb{Z}_2(\sigma_x \sigma_y)$ acts trivially on the center subspace of the vector field $H$ normal to the group orbit. 

Proposition~\ref{prop:trivial_action_bif} yields that $H$ undergoes a Hopf bifurcation to a unique time-periodic solution. To obtain the group orbit of the bifurcating solution, \pg{we apply Theorem \ref{thm:relative_equilibrium} with $\Sigma_{X_0}=\mathbb{Z}_2(\sigma_x \sigma_y)$ being the isotropy subgroup of the one-vortex solution (since this is inherited by the bifurcating solution due to no symmetry being broken as a result of the action being trivial), and $\Gamma=\Gamma_{LAN}$, to get $k=rank \left(\dfrac{\mathbb{Z}_2\times\mathbb{Z}_2}{\mathbb{Z}_2}\right)+1=rank(\mathbb{Z}_2)+1=1$.} This rules out RPOs (which have dimension 2), and implies that the bifurcating solutions of the full vector field are either standing waves (POs) or traveling waves (POs). An isotropy subgroup calculation similar to that performed for Hopf bifurcation from UNI shows that these solutions are in fact standing waves. \pg{Numerically, we found that these periodic solutions, denoted by SPO$_{T1a}$, cycle between a defect-free vortex lattice and laning-like profiles, and remain confined to the subspace defined by $\langle u \rangle = 0$} .
\begin{figure}[ht]
\centering
\begin{tikzpicture}[node distance=0.5cm, every node/.style={inner sep=0}]

\node (dpo1) {\includegraphics[width=0.48\textwidth]{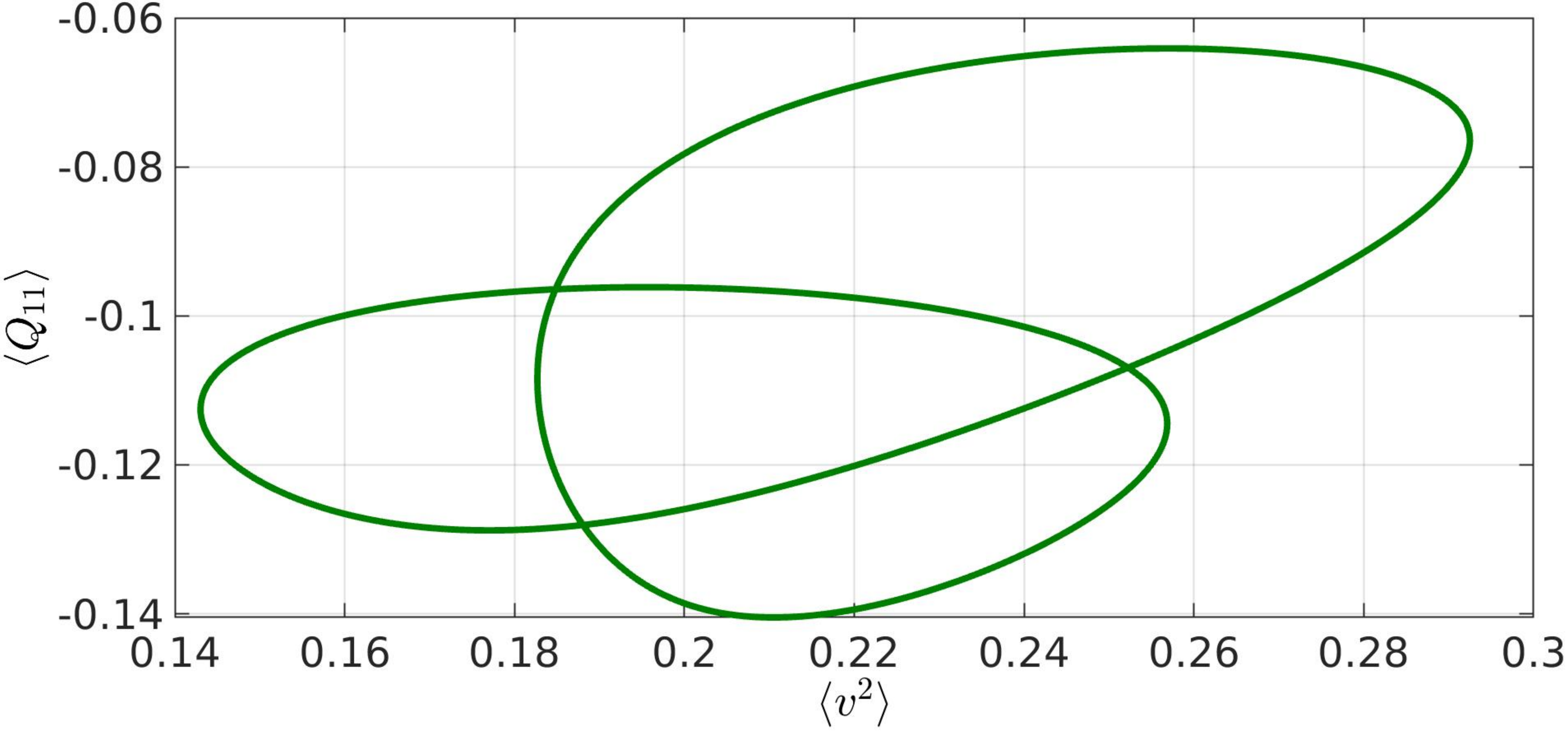}};

\node[right=0.5cm of dpo1] (dpo2) {\includegraphics[width=0.48\textwidth]{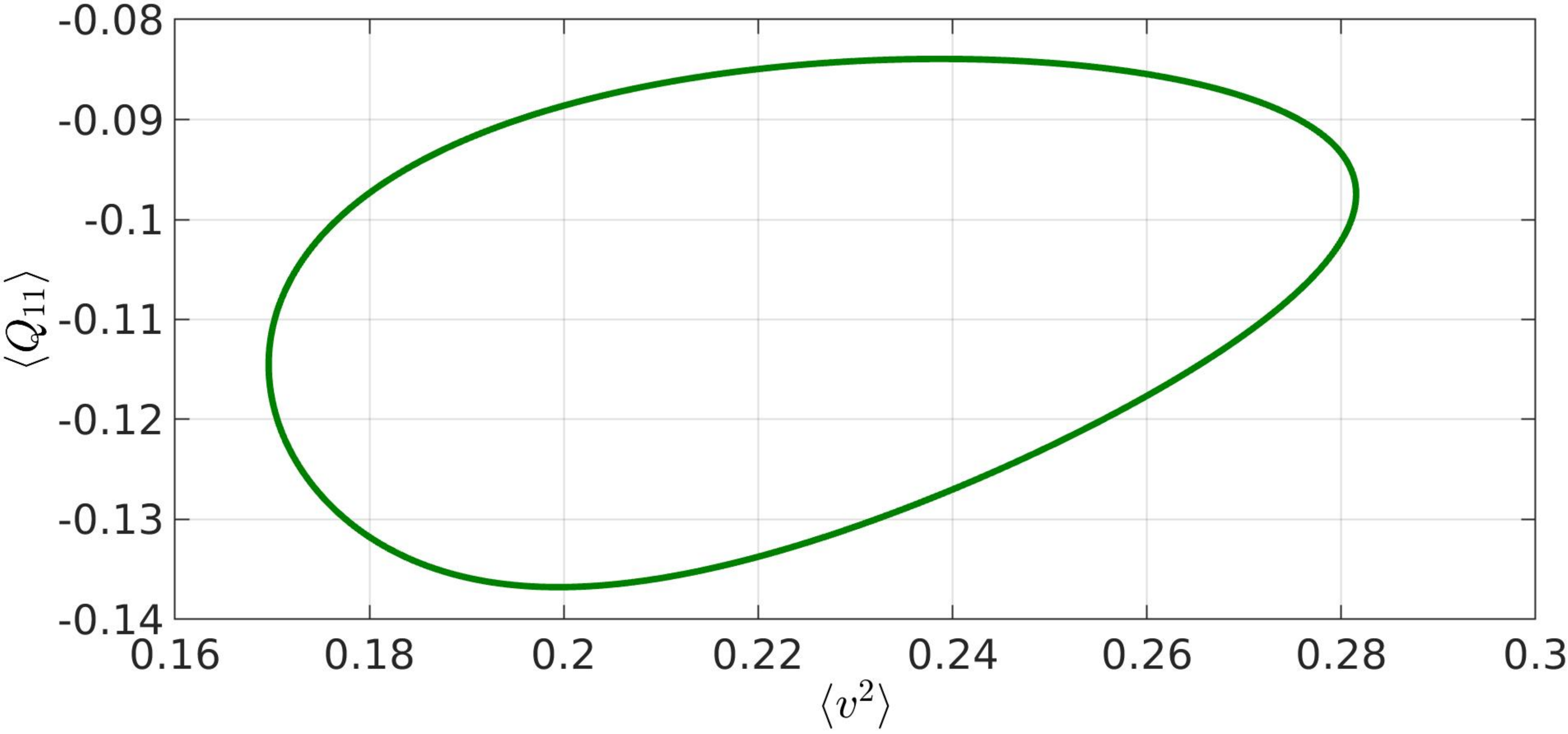}};

\node[below=0.2cm of dpo1] {(a) DPO$_{T1a}$, Ra = 1.50};
\node[below=0.2cm of dpo2] {(b) DPO$_{T1b}$, Ra = 1.50};

\end{tikzpicture}
\caption{\footnotesize Two dancing-disclination-like POs (DPOs) that co-exist at $\mathrm{R_a} = 1.5$, projected onto the reduced phase space ($\langle Q_{11}\rangle$,$\langle v^2\rangle$). Both solutions lie in the $\langle u \rangle = 0$ subspace (i.e., no net streamwise flow) due to the $\sigma_x\sigma_y$ symmetry.}
\label{fig:dpoT1_visuals}
\end{figure}

\subsubsection{Homoclinic bifurcation of the one-vortex POs and the emergence of dancing disclination solutions}
\an{As the activity is increased from $\mathrm{R}_{\mathrm{a}} \approx 0.9$, the periods of  SPO$_{T1a}$ solutions grow till this family eventually collides with LAN and disappears via a homoclinic bifurcation at $\mathrm{R}_{\mathrm{a}} \approx 0.945$, as shown in Fig.~\ref{fig:Vortex_Homo_LAN}. The logarithmic divergence of the period near the bifurcation provides further evidence for the homoclinic nature of this collision (see Fig.~\ref{fig:Log_Law_One_Vortex}).

On the other side of the homoclinic bifurcation, for $\mathrm{R}_{\mathrm{a}}\gtrsim 0.9447$, we find a branch of POs with symmetry conjugate to $\sigma_x\sigma_y$, that cycles between a defect-laden vortex lattice and LAN-like patterns, which we denote as PO$_{T1a}$. The appearance of these cycling solutions suggests the possibility of a second homoclinic connection involving LAN on this side of the parameter range, or, alternatively, an origin through a local bifurcation from the SPO$_{T1a}$ branch before this family disappears. However, the available data are not yet sufficient to establish the occurrence of either bifurcation. At $\mathrm{R}_{\mathrm{a}}\approx0.96$, we find a branch of unstable dancing disclination type POs with $\sigma_x\sigma_y$ symmetry, and denote this family by DPO$_{T1a}$. Attempts to continue this branch toward lower activity were unsuccessful.  

Similarly, we identify another branch of dancing-disclination dynamics DPO$_{T1b}$, which can be continued down to activity as low as $\mathrm{R}_{\mathrm{a}}=0.74$. Both DPO$_{T1a}$ and DPO$_{T1b}$ lie in the invariant subspace associated with a symmetry conjugate to $\sigma_x\sigma_y$ and are therefore constrained to the $\langle u\rangle=0$ subspace, as shown in Fig.~\ref{fig:dpoT1_visuals} for $\mathrm{R}_{\mathrm{a}}=1.5$, where both solutions coexist. Their shared symmetry suggests that these branches may be connected, directly or indirectly, to branches bifurcating from LAN. A detailed analysis of the origin of these branches and their possible connections to the LAN-related homoclinic structures is left for future work.}

\subsubsection{Secondary Bifurcations Associated with Higher Unstable Eigenvalues of LAN}
So far, our analysis has been limited to bifurcations associated with the first unstable eigenspace of LAN. As the activity number is varied from \(\mathrm{R}_{\mathrm{a}} \approx 0.85\) to \(\mathrm{R}_{\mathrm{a}} \approx 4.5\), a total of eight pairs of real eigenvalues eventually cross the imaginary axis. The isotropy subgroups of these eight unstable 2D eigenspaces are listed in Table \ref{tab:isotropy_groups_LAN} in the Appendix. While the study of all bifurcations is outside the scope of this work, we briefly discuss the bifurcations from the second unstable eigenspace, since these solutions play an important role in the turbulent regime we analyze later in this paper. 

\begin{figure}[htbp]
    \centering
    \subfloat[\footnotesize]{%
        \includegraphics[width=0.35\textwidth]{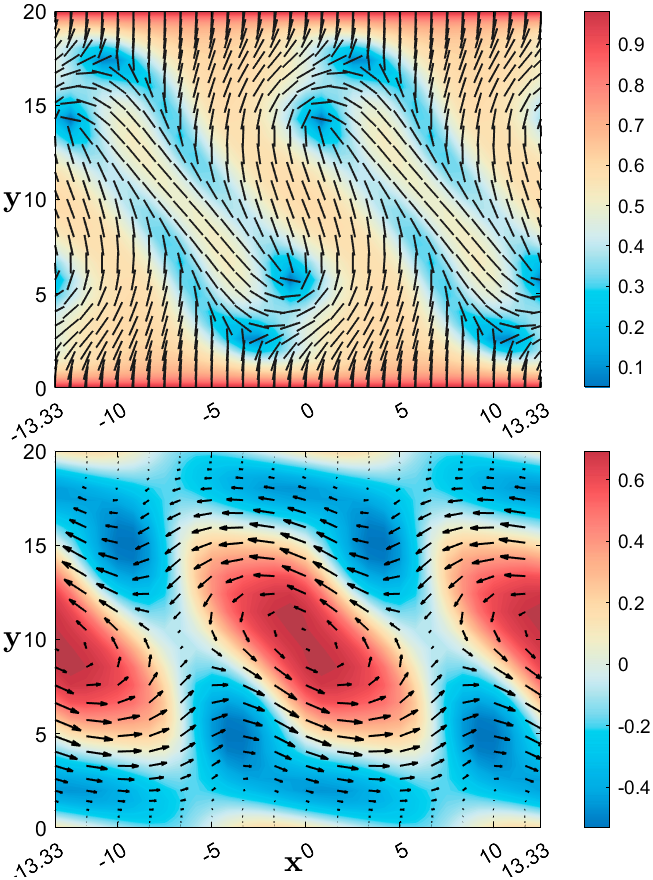}%
        \label{fig:plot_channel}%
    }\hfill
    \subfloat[\footnotesize]{%
        \includegraphics[width=0.60\textwidth]{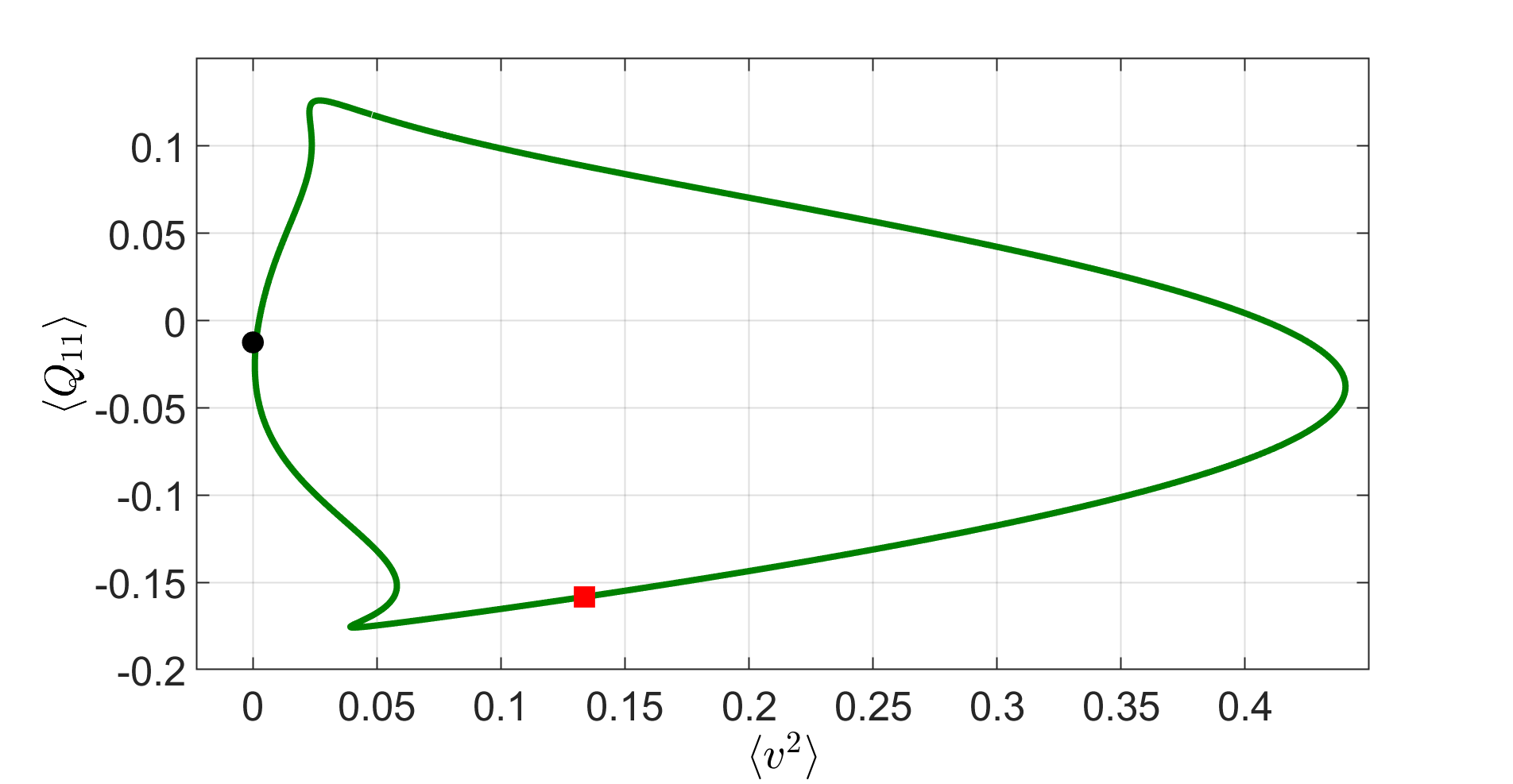}%
        \label{fig:proj_phase_space}%
    }
    
    \caption{\footnotesize \an{Visualization of PO$_{T2a}$ at $\mathrm{R_a}=2.25$. (a) Snapshot of the nematic director and velocity fields, which have the double-reflection $\sigma_x\sigma_y$ and two-fold translation $\tau_x(L/2)$ symmetries. (b) Projection onto the reduced phase space $(\langle Q{11}\rangle,\langle v^2\rangle)$. The red square indicates the position of the state corresponding to the snapshot in (a), while the small black circle marks the position of LAN. The proximity of LAN to the periodic orbit makes the two appear to touch in the projection. Like LAN, this periodic orbit lies in the $\langle u\rangle=0$ subspace (i.e., it has no net streamwise flow) as a consequence of the $\sigma_x\sigma_y$ symmetry.}}
    \label{fig:dpo_visuals}
\end{figure}

This second pair of unstable eigenvectors possesses the \(\tau_x(L/2)\) symmetry. The sequence of bifurcations is analogous to those of the first pair, and the bifurcating solutions possess the extra $\tau_x(L/2)$ symmetry. The steady solution, \an{which originates at $\mathrm{R_a}=1.25$, is a two-vortex equilibrium. A standing-wave PO family originates from this relative equilibrium at $\mathrm{R_a}\approx 1.49$, denoted as SPO$_{T2a}$. We also confirmed the existence of another branch of periodic orbits, denoted as PO$_{T2a}$, with symmetry conjugate to  $\sigma_x \sigma_y$ and translation $\tau_x(L/2)$, that cycles between a LAN-like and defect-laden two-vortex lattice, as illustrated by the representative example shown in Fig.~\ref{fig:dpo_visuals} at $\mathrm{R_a}=2.25$. This branch is found for $\mathrm{R_a}\gtrsim1.8$ and may originate from either a local or global bifurcation involving the $\mathrm{SPO}_{T2a}$ family or LAN, similar to the hypothetical mechanisms discussed for the first pair.} We anticipate that solutions associated with higher modes could \an{also} bifurcate from LAN through similar mechanisms.

    
%


\section{Transient Flow Reversals in Preturbulent Flow}\label{sec:preturb}
In section \ref{sec:eqbif}, we uncovered the origin of several key ECSs that emerge from, or are related to, bifurcations of the UNI and LAN equilibria. We now discuss how these ECSs and their invariant manifolds facilitate robust, transient flow reversals in the preturbulent regime, where reversals are characterized by switching between near-unidirectional states located in the regions $\langle u \rangle < 0$ (left) and $\langle u \rangle > 0$ (right), respectively. By transient, we mean that each trajectory switches direction a finite number of times before settling at a regular attractor (i.e. a stable ECS) or a chaotic attractor. By robust, we mean that the reversal mechanism persists for an open interval of activity values.

\subsection{Transient Reversal Dynamics at Low Activity: Organizing ECSs and Heteroclinic Connections}

We focus on trajectories that begin near UNI or its symmetry counterpart $\sigma_y$UNI. When these ECSs are stable, all nearly uniaxial initial conditions decay to them. When these ECSs lose stability, the homoclinic-like RPOs (HRPO$_{T1a}$ and $\sigma_y$HRPO$_{T1a}$) become the attractors instead, up to $\mathrm{R}_{\mathrm{a}} \approx 1.11$. Beyond this activity, and before the heteroclinic bifurcation that occurs at $\mathrm{R}_{\mathrm{a}} \approx 1.38$, the two RPOs are unstable. The unstable manifolds of these RPOs and the heteroclinic connections between these ECSs and the DPOs shape the flow between nearly uniaxial states.  

\begin{figure}[htbp]
    \centering
    \subfloat[\footnotesize Mechanism 1: Flow reversal with no DPO mediator. The black dashed trajectory, initialized near UNI along its unstable manifold, successively follows the \pg{potential} heteroclinic connections UNI $\rightarrow$ HRPO$_{T1a}$ $\rightarrow$ $\sigma_y$HRPO$_{T1a}$ and eventually settles into the localized chaotic attractor.]{%
        \includegraphics[width=\textwidth]{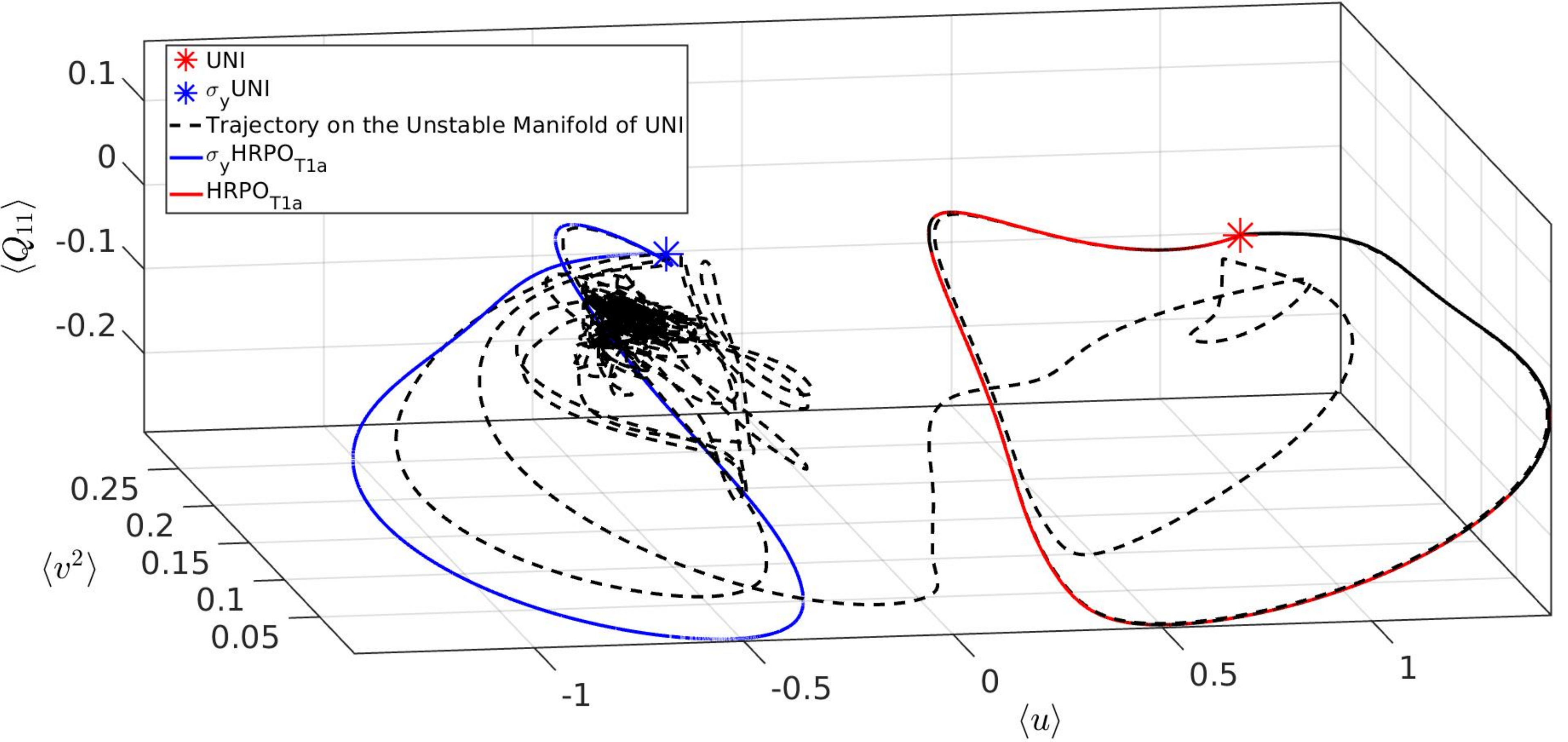}%
        \label{fig:mechanism1}%
    }\\[0.5em]
    \subfloat[\footnotesize Mechanism 2: Flow reversal using the DPO$_{T1a}$ as mediator. The black dashed trajectory follows the \pg{potential} heteroclinic connections UNI $\rightarrow$ HRPO$_{T1a}$ $\rightarrow$ DPO$_{T1a}$ and settles into the localized chaotic attractor.]{%
        \includegraphics[width=\textwidth]{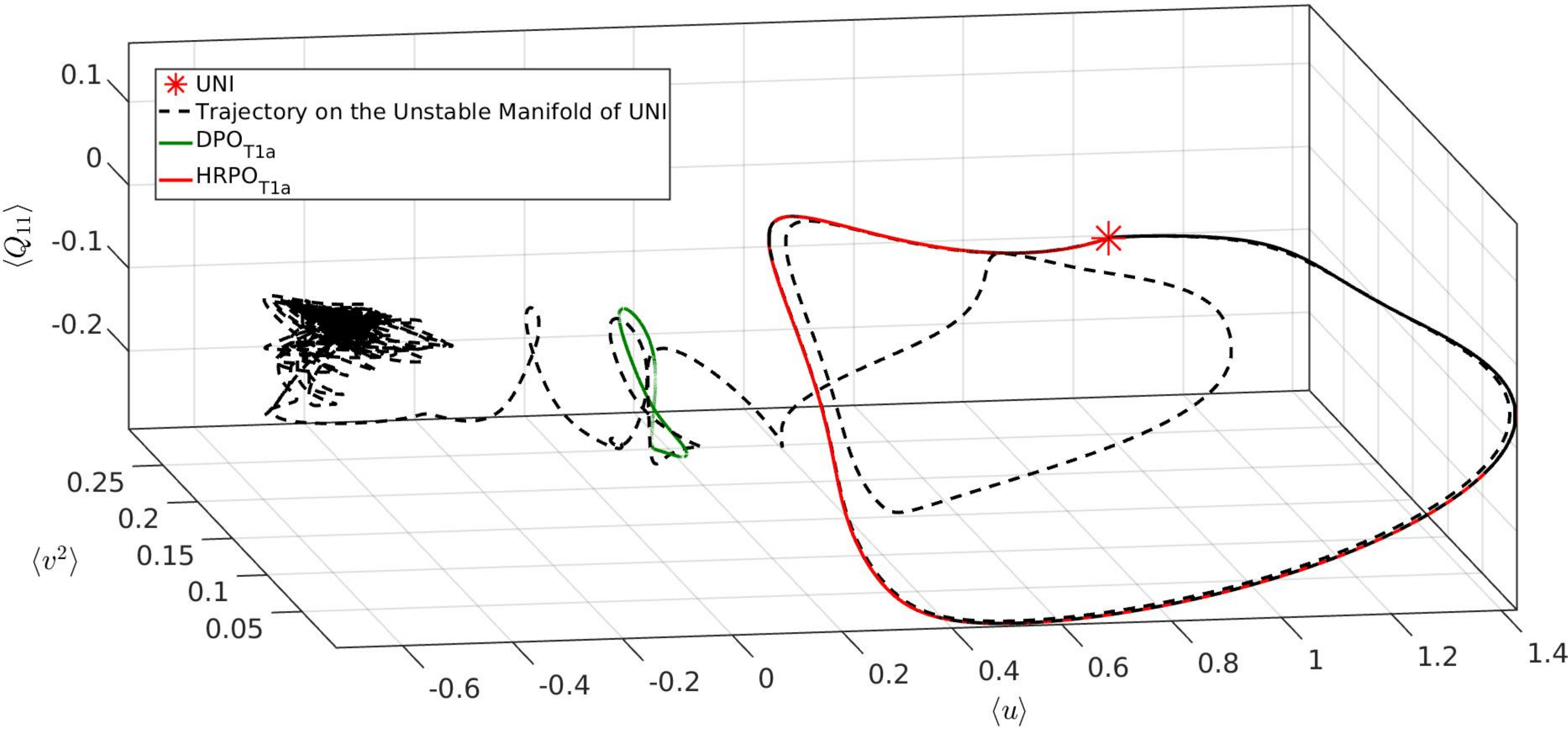}%
        \label{fig:mechanism2}%
    }
    \caption{\footnotesize Two main mechanisms for flow reversals observed at $\mathrm{R}_{\mathrm{a}} = 1.24$ by perturbing UNI.}
    \label{fig:flow_reversal_mechanisms}
\end{figure}

At $\mathrm{R}_{\mathrm{a}} = 1.24$ for instance, these ECSs co-exist with a localized chaotic attractor (and its $\sigma_y$ copy) that has been reported earlier \cite{Wagner2022}. Perturbing UNI along its four unstable directions reveals the two main mechanisms for flow reversals facilitated by potential heteroclinic connections. The first mechanism, which requires no mediators to cross from right to left, is shown in Fig. \ref{fig:mechanism1}. The trajectory shadows the \pg{potential} heteroclinic connections UNI $\rightarrow$ HRPO$_{T1a}$ $\rightarrow$ $\sigma_y$HRPO$_{T1a}$, and eventually settles into the localized chaotic attractor on the left. The second mechanism, shown in Fig. \ref{fig:mechanism2}, shadows the \pg{potential} heteroclinic connections HRPO$_{T1a}\rightarrow$ DPO$_{T1a}$ to end up in the same localized chaotic attractor.  The rest of trajectories originating near UNI (resp. $\sigma_y$UNI ) do not exhibit reversals and settle onto the chaotic attractor on the right (resp. left).


Increasing the activity further reveals additional reversal mechanisms. At $\mathrm{R}_{\mathrm{a}} = 1.5$, for instance, we observe that eight fundamental ECSs organize the reversal dynamics between left- and right-flowing states. These include the two heteroclinic-like HTPO$_{T1a}$ and HTPO$_{T1b}$, the homoclinic-like HRPO$_{T2a}$, the DPO$_{1a}$ and DPO$_{1b}$ associated with LAN, and three additional RPOs. These new RPOs, which we call DRPO$_{T1a,b,c}$, likely emerge from bifurcations of the DPO branches. These are the drifting versions of dancing disclination solutions, as first reported in~\cite{Wagner2022}. Interestingly, at this activity number, no localized chaotic attractor was found. Instead, the three DRPOs occupy approximately the same region of phase space that was previously occupied by the chaotic attractor, and DRPO$_{T1c}$ is the only attractor in this region.

\begin{figure}[htbp]
    \centering
    \subfloat[\footnotesize
Mechanism 3 via HTPO$_{T1b}$. The Black dashed trajectory, initiated on the unstable manifold of UNI, first shadows HTPO$_{T1b}$, and then follows the \pg{potential} heteroclinic connections HTPO$_{T1b}\!\rightarrow$DPO$_{T1a}\!\rightarrow$DRPO$_{T1a}\!\rightarrow$DRPO$_{T1b}\!\rightarrow$DRPO$_{T1c}$. The various ECS visible in the figure are HTPO$_{T1b}$ (red), DPO$_{T1a}$ (green) and DRPO$_{T1c}$ (magenta). Both DRPO$_{T1a}$ and DRPO$_{T1b}$ are hidden beneath the final segment near DRPO$_{T1c}$. Also shown are representative snapshots of UNI-like and one-vortex-like flow states during the reversal in this preturbulent regime.
]{%
        \includegraphics[width=\textwidth]{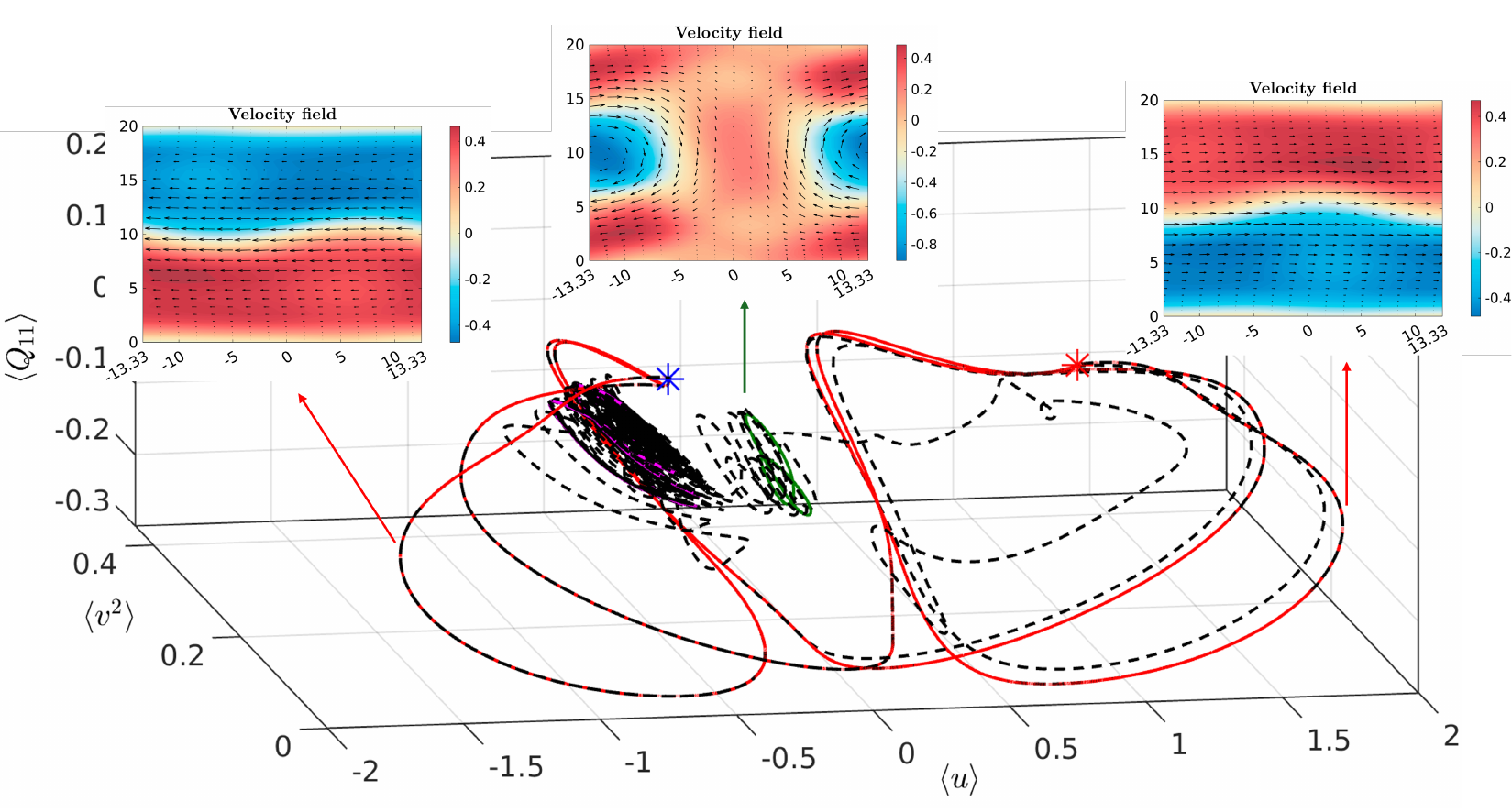}%
        \label{fig:mechanism3}%
    }\\[0.5em]
    \subfloat[\footnotesize Mechanism 1 using HRPO$_{T2a}$. The  black dashed trajectory follows the \pg{potential} heteroclinic sequence UNI $\rightarrow$ HRPO$_{T2a}$ $\rightarrow$ DRPO$_{T1a}\rightarrow$ DRPO$_{T1b}$ $\rightarrow$ DRPO$_{T1c}$.]{%
        \includegraphics[width=\textwidth]{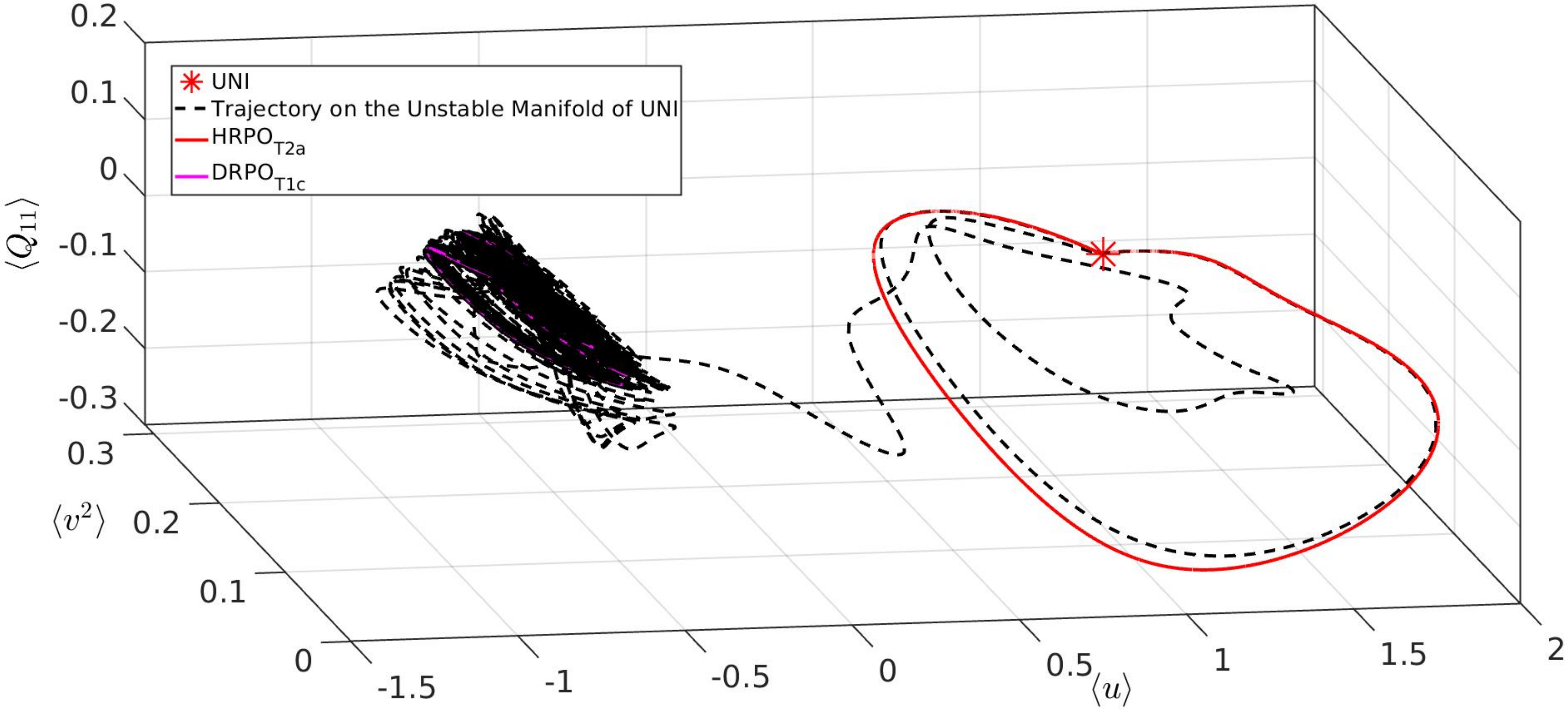}%
        \label{fig:mechanism_HTPO_T2}%
    }

    \caption{\footnotesize Main mechanisms of flow reversals at $\mathrm{R}_{\mathrm{a}} = 1.5$ observed by perturbing UNI along its four dimensional unstable subspace. (a) A mechanism that uses the \textit{HTPO$_{T1b}$} with DPO$_{T1a}$ as the mediator. (b) A mechanism using HRPO$_{T2a}$ without \textit{DPOs} as  mediators.}
    \label{fig:flow_reversal_mechanisms_R15}
\end{figure}

By perturbing UNI along its two pairs of unstable directions at $\mathrm{R}_{\mathrm{a}} = 1.5$, we obtain trajectories that initially shadow either HTPO$_{T1b}$ or HRPO$_{T2a}$. In the first case, the initial transient consists of repeated reversals between the two nearly uniaxial flow states, due to the heteroclinic nature of HTPO$_{T1b}$. Eventually, the trajectory diverges from HTPO$_{T1b}$ and either moves directly to the region populated by the three DRPOs, or passes near the DPO$_{T1a}$ before reaching a symmetry instance of the DRPO family as shown in Fig. \ref{fig:mechanism3}. \pg{DPO$_{T1a}$ has a single unstable direction at this activity, so its stable manifold is codimension-one and can act as an edge separating the two RPO attractors. Perturbations along opposite branches of the unstable direction evolve toward the respective attractors. The reversal mechanism involving DPO$_{T1a}$ here is consistent with heteroclinic connection in which the unstable manifold of the HTPO$_{1b}$ intersects the stable manifold of the DPO$_{T1a}$. A trajectory approaching DPO$_{T1a}$ close to this connection is subsequently carried by the DPO$_{T1a}$'s unstable manifold toward the $DRPO_{T1c}$ attractor on the opposite side. This is similar to the switching mechanism via a leaky heteroclinic tangle described by Suri \citep{Suri2024}.}

In the second case, after transiently shadowing HRPO$_{T2a}$, a trajectory can undergo a reversal event by a \pg{potential} heteroclinic connection to the DRPO$_{T1a}$ copy on the opposite side, or \pg{via a  potential heteroclinic chain involving DPO$_{T1a}$}. All trajectories at $\mathrm{R}_{\mathrm{a}} = 1.5$ eventually end up in the attractor DRPO$_{T1c}$ or its $\sigma_y$ copy.

In summary, for low activity numbers, three main types of reversal mechanisms are identified: the first involves potential heteroclinic connections originating from HRPOs or HTPOs to the relevant attractor; the second mechanism is similar but with DPOs serving as mediators; and the third arises when trajectories shadow either of the two HTPOs themselves. When both the HTPOs exist, we observe that HTPO$_{T1b}$ is more dominant of the two in influencing trajectories originating from perturbations of UNI. Animations of representative reversal trajectories and their shadowed ECS solutions are provided in the Supplemental Material.

\subsection{Transient Flow Reversals at Higher Activity and the Path Toward Turbulence}

\begin{figure}[htbp]
    \centering
    \subfloat[\footnotesize {The unstable DPO$_{T1a}$ at $\mathrm{R}_{\mathrm{a}} = 3.5$ is shown in green. An attracting periodic orbit that emerges from DPO$_{T1a}$ at $\mathrm{R}_{\mathrm{a}} \approx 3.29$ via a period doubling bifurcation is shown in black. Notice that this new PO not in the plane $\langle u \rangle = 0$, i.e., it breaks the $\sigma_x\sigma_y$ symmetry. Also note the extent of the $\langle u\rangle$ axis is much smaller than in earlier plots showing the HRPOs and HTPOs earlier.}]{%
        \includegraphics[width=0.47\textwidth]{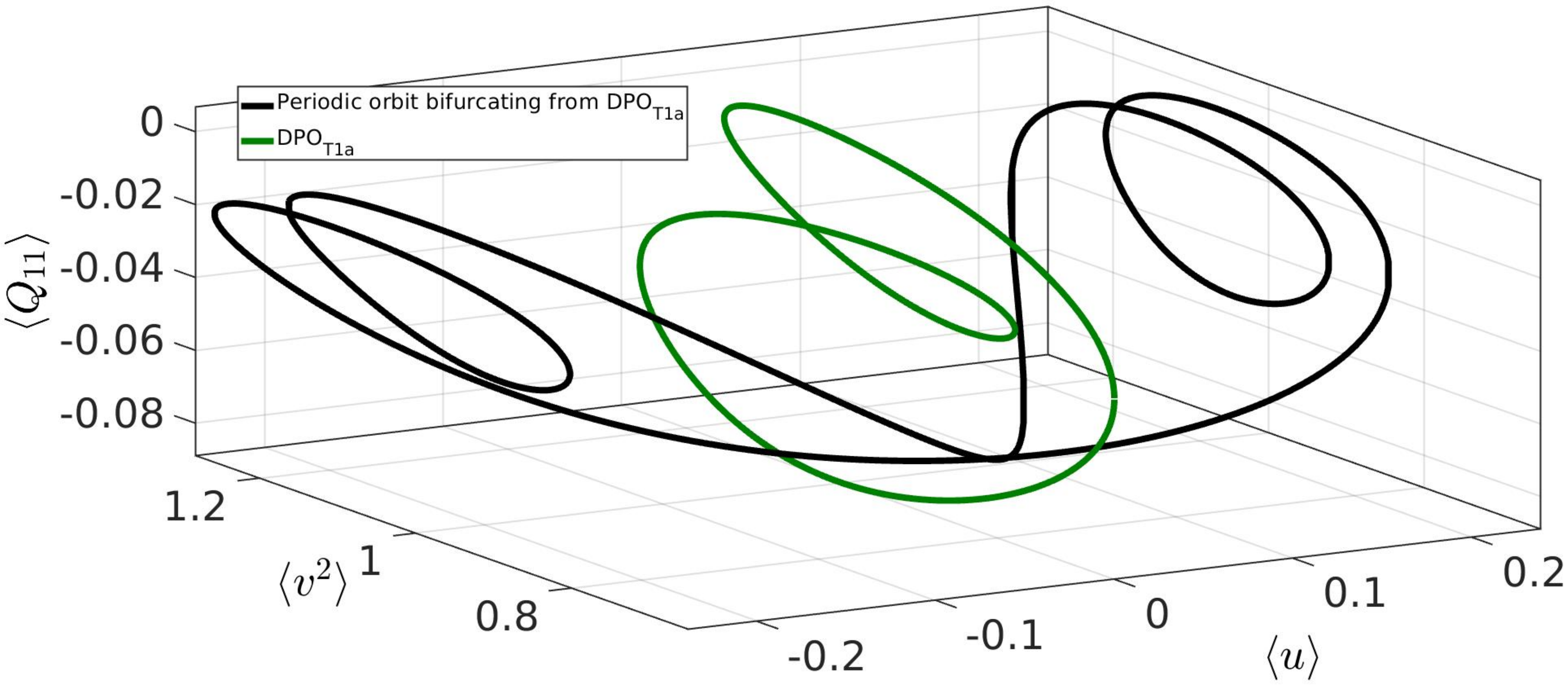}%
        \label{fig:PO_from_DPO}%
    }\hfill
    \subfloat[\footnotesize A trajectory attracted to a region populated by ECSs at $\mathrm{R}_{\mathrm{a}} = 4.0$, serving as a precursor to the chaotic set $\mathcal{Y}$ at $\mathrm{R}_{\mathrm{a}} = 4.5$.]{%
        \includegraphics[width=0.5\textwidth]{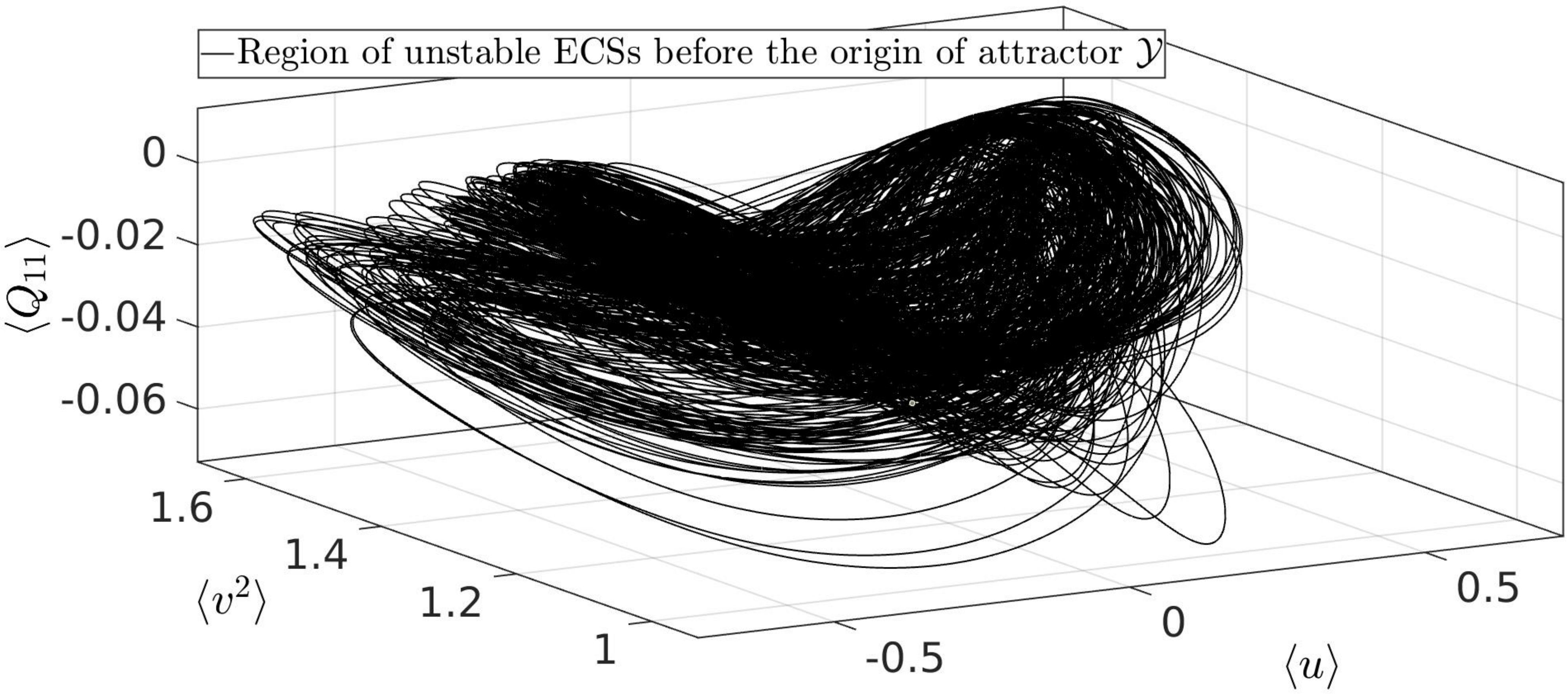}%
        \label{fig:prelude_to_Y}%
    }
    \caption{\footnotesize Phase space visualizations illustrating the evolution from a periodic orbit to the onset of the chaotic set $\mathcal{Y}$.}
    \label{fig:Period-doubling}
\end{figure}

\an{Rather than attempting an exhaustive continuation and bifurcation analysis of all ECS branches into the turbulent regime, we focus on selected branches and bifurcations that reveal how the symmetry-organized phase-space structure evolves with increasing activity.} We have seen earlier that solutions with symmetries conjugated to $\sigma_x\sigma_y$, and branches bifurcating from them, provide mechanisms for reversal. By following the DPO$_{T1a}$ branch into the higher activity region, we observe that at $\mathrm{R}_{\mathrm{a}} \approx 2.38$, this branch becomes stable. Then, at $\mathrm{R}_{\mathrm{a}} \approx 3.29$, it undergoes a period-doubling bifurcation that breaks the $\sigma_x\sigma_y$ symmetry, giving rise to a stable preperiodic orbit \an{shown in Fig.~\ref{fig:PO_from_DPO}}. The resulting solution represents, by itself, a stable mechanism for periodic flow reversals about $\langle u\rangle = 0$. As the activity number reaches $\mathrm{R}_{\mathrm{a}} = 4.0$, new ECSs emerge from a sequence of bifurcations related not only to DPO$_{T1a}$ but also to nearby solutions, which increasingly populate this region of phase space in the form of a localized chaotic attractor shown in Fig. ~\ref{fig:prelude_to_Y}. \an{This sequence preludes the birth of the chaotic set $\mathcal{Y}$ at $\mathrm{R}_{\mathrm{a}} = 4.5$, shown in Fig.~\ref{fig:attractor_summary}.} Within $\mathcal{Y}$, solutions related to DPO$_{T1a}$ \an{may} continue to play an important role in mediating reversals.


\begin{figure}[htbp]
    \centering
    \includegraphics[width=1.0\textwidth]{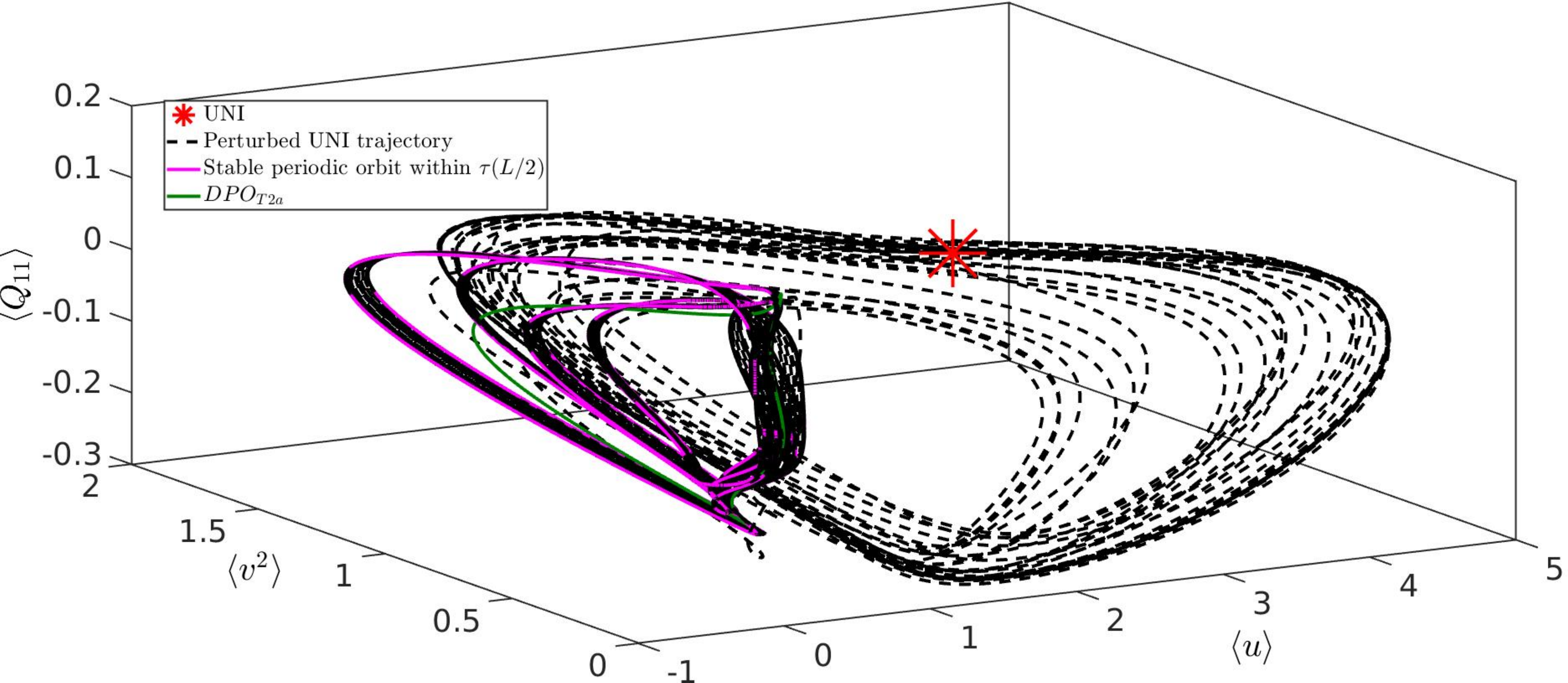}
    \caption{\footnotesize At $\mathrm{R}_{\mathrm{a}} = 3.75$, the black dashed trajectory perturbed from UNI settles into a periodic orbit (magenta) that is stable within the $\tau_x(L/2)$ invariant subspace. This periodic orbit seems to have originated from secondary bifurcations of PO$_{T2a}$. The periodic orbit PO$_{T2a}$ is shown in green. }
    \label{fig:PO_attracting_trajectory}
\end{figure}

\an{The other set of ECSs that remain relevant at higher activity are those with $\tau_x(L/2)$ symmetry, emerging from bifurcations of the UNI and LAN branches. The perturbations of UNI along the \pg{second and fourth} pairs of unstable eigenvectors at $\mathrm{R}_{\mathrm{a}} = 3.75$ end up at a periodic orbit, stable in $\tau_x(L/2)$ invariant subspace, see Fig.~\ref{fig:PO_attracting_trajectory}. This PO appears to be borne from PO$_{T2a}$ through a sequence of bifurcations and intermediate solution branches that break the $\sigma_x\sigma_y$ symmetry. This and other nearby POs undergo bifurcations similar to those that gave rise to $\mathcal{Y}$, and at $\mathrm{R}_{\mathrm{a}} \approx 4.0$, a new chaotic set emerges. We denote this set by $\mathcal{X}$, see Fig.~\ref{fig:attractor_summary}. At this stage, trajectories initialized from UNI and perturbed within the four-dimensional subspace composed of the 2D \pg{second and fourth} unstable eigenspaces can cross into the other region of phase space, near $\sigma_y$UNI, and switch recurrently between these two states. Both chaotic sets $\mathcal{X}$ and $\mathcal{Y}$ exhibit reversals between regions where the flow velocity has a mean value $\langle u \rangle$ that persistently oscillates about positive and negative values, respectively.}


\section{Flow Reversal Mechanisms in Turbulence: Two-Way Persistent Reversals}






\begin{figure}[htbp]
  \centering

  \subfloat[\footnotesize Chaotic set $\mathcal{X}$]{%
      \includegraphics[width=0.48\linewidth]{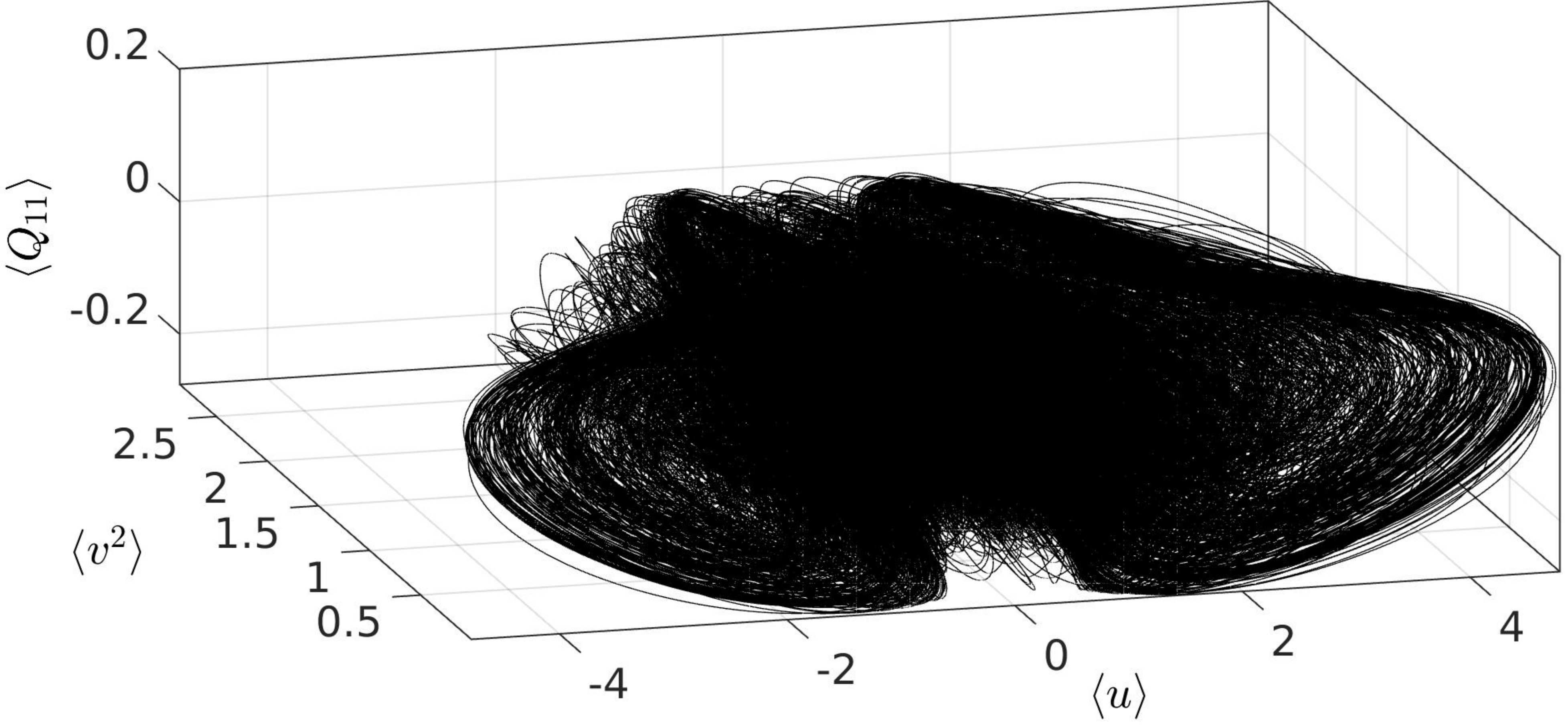}%
  }\hfill
  \subfloat[\footnotesize Chaotic set $\mathcal{Y}$]{%
      \includegraphics[width=0.48\linewidth]{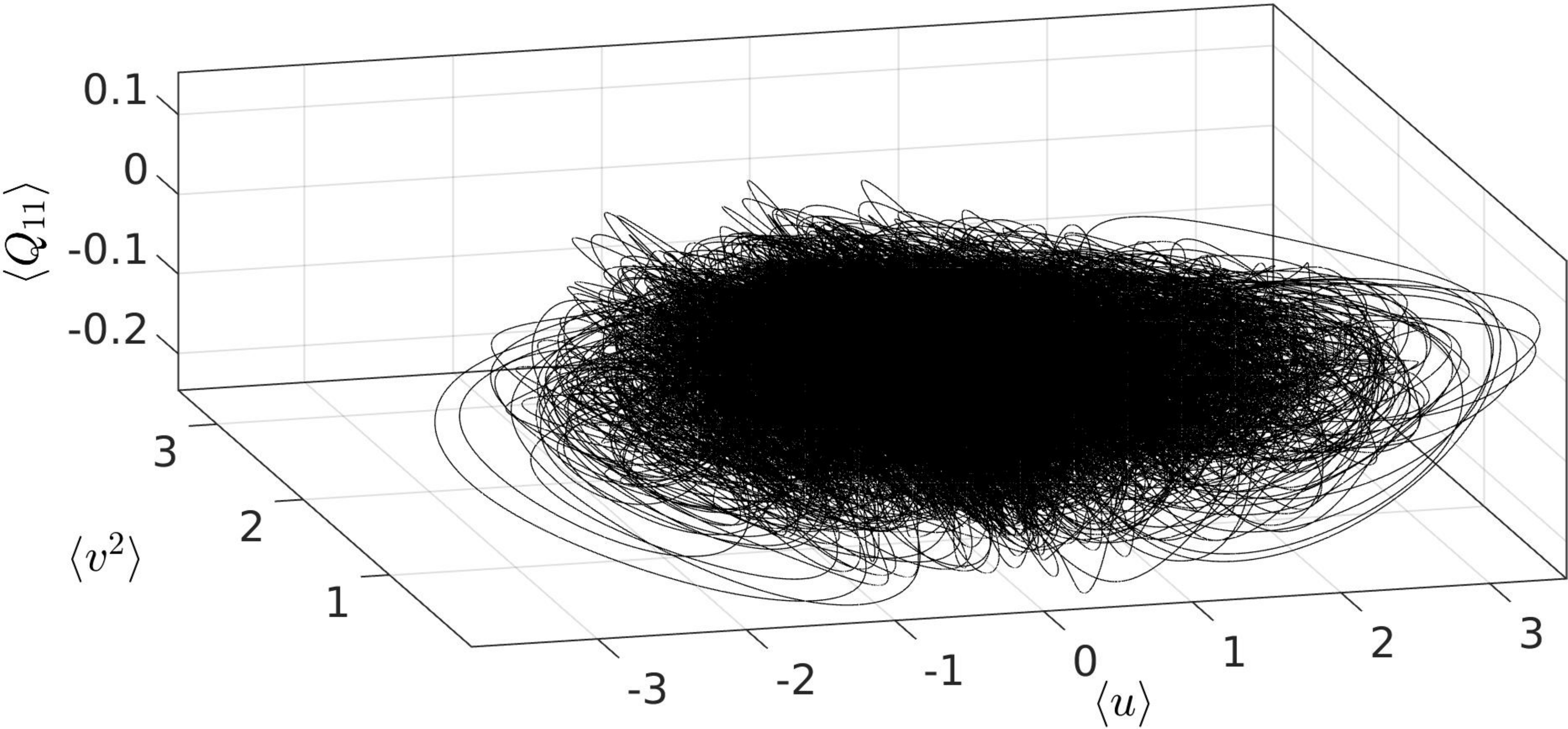}%
  }\\[1em]
  \subfloat[\footnotesize \an{Chaotic} set $\mathcal{Z}$]{%
      \includegraphics[width=0.48\linewidth]{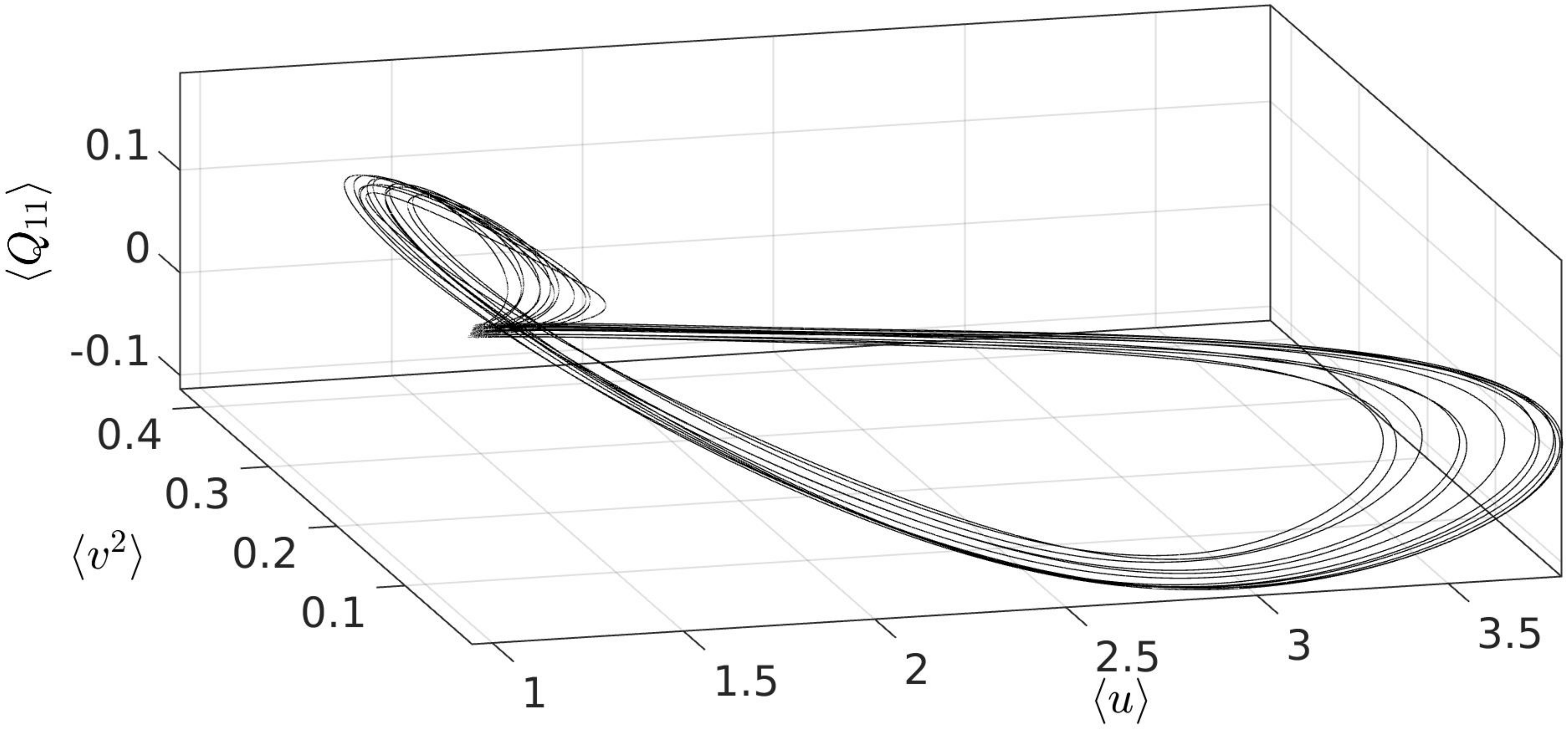}%
      \label{fig:gamma}%
  }

  \caption{\footnotesize All perturbations of UNI along its unstable manifold in the \an{SPU} A end up in one of the three sets $\mathcal{X}$, $\mathcal{Y}$ and $\mathcal{Z}$. The top row shows the chaotic sets $\mathcal{X}$ and $\mathcal{Y}$. The bottom row shows the \an{chaotic} set $\mathcal{Z}$.}
  \label{fig:attractor_summary}
\end{figure}

The results presented for the preturbulent regime, where we have shown that the dynamical trajectories are shaped by the ECSs and their invariant manifolds, motivate us to investigate the nature of the reversal mechanisms within the turbulent regime. In the original full-channel configuration (length $\tilde L$), however, Wagner \textit{et al.}~\cite{Wagner2023} explored the turbulent regime at \(\mathrm{R}_{\mathrm{a}} = 4.5\), reporting no evidence of shadowing among the 210 distinct ECSs and their discrete-symmetry counterparts, leaving the role of ECSs in the dynamics still unclear. In contrast, our use of smaller domains as \an{SPU}s allows us to identify shadowing events in a simplified system with fewer solutions and reduced complexity. These detected shadowing events highlight the pathways through which chaotic trajectories undergo reversals. In this section, we consider two \an{SPU}s: the \an{SPU} A which has been analyzed throughout the previous sections with length $L_A=\tilde{L}/3$, and its halved version \an{SPU} B with $L_B=\tilde{L}/6$. We explore the following questions in the following:
\begin{itemize}
    \item Q1: Where do typical trajectories end up at $\mathrm{R}_a=4.5$ in the two \an{SPU}s ?
    \item Q2: Do the trajectories that end up in one of the two chaotic sets $\mathcal{X}$ or $\mathcal{Y}$ shadow the ECSs embedded in those sets at $R_a=4.5$?
    \item Q3: What are the symmetries and possible origins of ECSs embedded in the chaotic sets \an{$\mathcal{X}$ and $\mathcal{Y}$}?
\end{itemize}
The UNI in the \an{SPU} A at $\mathrm{R}_{\mathrm{a}} = 4.5$ has seven complex pairs of unstable eigenvalues. Perturbations of the UNI within the direct sum of eigenspaces corresponding to the \pg{first, third and fifth} pairs evolve towards the chaotic set $\mathcal{Y}$, while perturbations within the direct sum of eigenspaces of the \pg{second and fourth} pairs evolve towards the chaotic set $\mathcal{X}$. \an{In the remaining two eigenspaces, perturbations lead to a relatively weakly chaotic set which is confined to the region $\langle u\rangle>0$, see Fig. \ref{fig:attractor_summary}. By symmetry, there is another chaotic set $\sigma_y\mathcal{Z}$ confined to the region $\langle u\rangle<0$ that is reached by corresponding perturbations of $\sigma_y$UNI. Since neither set exhibits flow reversals at $\mathrm{R}_{\mathrm{a}} = 4.5$, they were not considered in this study. 

We find that an arbitrary perturbation (i.e., with no symmetry) always ends up in $\mathcal{Y}$, indicating that in \an{SPU} A, $\mathcal{Y}$ is the only almost-global attractor while the other two sets $\mathcal{X}$ and $\mathcal{Z}$ are saddles. Motivated by the apparent two-fold translation (i.e., $\tau_x(L_A/2)$) symmetry of $\mathcal{X}$ itself, we verify that in fact, all initial conditions with $\tau_x(L/2)$ symmetry (that are not $\tau_x(L_A/4)$ symmetric) end up in $\mathcal{X}$ in \an{SPU} A. Hence, we also analyze dynamics in the smaller \an{SPU} B, where $\mathcal{Y}$ is eliminated due to lack of two-fold translation symmetry, and we find that the halved-version of $\mathcal{X}$ is the only almost-global attractor that attracts arbitrary initial conditions almost surely. \an{SPU B contains the half-domain representations of the SPU A ECSs with two-fold translational symmetry. The Lyapunov time of $\mathcal{X}$ is $\tau_{\mathcal{X}}\approx 7.17\tau$ in SPU B, while the more complex chaotic set $\mathcal{Y}$ has a Lyapunov time $\tau_{\mathcal{Y}}\approx 4.34\tau$ in SPU A. Here, the Lyapunov time is defined as $1/\lambda_1$, where $\lambda_1$ is the largest Lyapunov exponent of the corresponding chaotic set at $\mathrm{R}_{\mathrm{a}} = 4.5$.} The weakly-chaotic set $\mathcal{Z}$}, while still present in \an{SPU} B, again only attracts a measure zero set of trajectories that have the $\tau_x(L_A/4)$ translation symmetry. Table \ref{tab:chaos} summarizes this discussion and answers Q1 posed above. Now we turn to answering Q2 and Q3.

\subsection{Establishing shadowing within the chaotic sets $\mathcal{X}$ and $\mathcal{Y}$}

\an{Hereafter, \(t\) denotes dimensional time}. To establish shadowing, we perform long-time simulations (\( t = 30000\tau \)) in \an{SPU} A within each of the two chaotic sets and focus on the final segment, \( [29000\tau, 30000\tau] \), where transient effects have long decayed. Both chaotic sets exhibit persistent flow reversals between two nonzero mean flow states, as evidenced by their reduced phase space projections and the time series of representative trajectories over this interval. We analyze shadowing events involving a dominating set of 75 ECSs in \an{SPU} A.  Within this set, we identify four symmetry-based families: 40 ECSs with no additional translation symmetry (T1), 32 ECSs invariant under \(\tau_x(L_A/2)\) (T2), one ECS invariant under \(\tau_x(L_A/3)\) (T3), and two ECSs invariant under \(\tau_x(L_A/4)\) (T4). Each family includes the corresponding symmetry-related counterparts, and we investigate how these families contribute to organizing the dynamics within the chaotic attractors.

For the plots in this section, we color the ECS based on the spatial average of the streamwise velocity $\langle u \rangle$ as follows:
\begin{itemize}
    \item ECSs confined to $\langle u \rangle > 0$ (red),
    \item ECSs confined to $\langle u \rangle < 0$ (blue),
    \item ECSs crossing the subspace $\langle u \rangle = 0$ (green).
\end{itemize}

To establish shadowing and identify shadowed ECSs, we follow a two step approach \citep{Krygier2021}. In the first step, for the final $1000\tau$ window within these attractors, we analyze the distances between a typical turbulent trajectory and every ECS (including translation-symmetry-related copies) using a continuous-symmetry-reduced distance \citep{Wagner2023},

\an{\begin{equation}\label{eq:distance}
    d(X_1(t),X_E) = \min_{\substack{0 \leq \ell \leq L \\ 0 \leq s \leq T}} 
    \left\| X_1(t) - \tau_x(\ell) X_E(s) \right\|_2
\end{equation}}
\an{where \(X_1(t)=[u,v,Q_{11},Q_{12}](t)\) is the turbulent trajectory, \(X_E(s)=[u,v,Q_{11},Q_{12}]_E(s)\) is the trajectory of the corresponding ECS, \(s\in[0,T]\) is the time along the ECS trajectory, \(T\) is the period of the ECS, and \(||\cdot||_2\) denotes the unweighted \(L^2\) norm.} At each time step, we identify the ECS with the minimum distance to the turbulent trajectory. The turbulent trajectory should exhibit dynamics similar to those of the ECS during the shadowing event ~\cite{Crowley2022, Crowley2023}. In the second step, we confirm shadowing by comparing the dynamical evolutions with the trajectory for the ECSs selected as closest to the trajectory, as we discuss next.


\begin{table}[h]
    \centering
    \begin{tabular}{l c c c c}
        \hline
        & Length & $\mathcal{X}$ & $\mathcal{Y}$ & $\mathcal{Z}$ \\ \hline
        \an{SPU} A & $L_A=\tilde{L}/3$ 
              & Chaotic saddle 
              & Chaotic attractor 
              & \an{Chaotic} saddle \\
              &                     
              & $\tau_x(L_A/2)$ symmetric
              & No symmetry 
              & $\tau_x(L_A/4)$ symmetric \\[2pt]
        \an{SPU} B & $L_B=\tilde{L}/6$ 
              & Chaotic attractor 
              & Doesn't exist 
              & \an{Chaotic} saddle \\
              &                     
              & No symmetry 
              & -- 
              & $\tau_x(L_B/2)$ symmetric \\ \hline
    \end{tabular}
    \caption{Properties of the three invariant sets in the two \an{SPU}s. For \an{SPU} B, the entries refer to the corresponding sets truncated to the reduced domain. Here, symmetry refers to pointwise symmetry.  }
    \label{tab:chaos}
\end{table}

\subsection{Chaotic Set $\mathcal{X}$}
\subsubsection{Dominant ECSs Within $\mathcal{X}$}

\begin{figure}[htbp]
  \centering
  \subfloat[\footnotesize Time series of \( \langle u \rangle \) along the turbulent trajectory $X(t)$ over the interval $29000\tau \le t \le 30000\tau$. Various shadowing events are also marked, colored by the region occupied by the corresponding shadowed ECS. The encounters with ECS that lie completely in the half-space (\( \langle u \rangle > 0 \)) are in red, with ECS that lie completely in the half-space (\( \langle u \rangle < 0 \)) are in blue, and with ECS that lie on both sides of the hyperplane (\( \langle u \rangle = 0 \)) are shown in green.]{%
    \includegraphics[width=.85\textwidth]{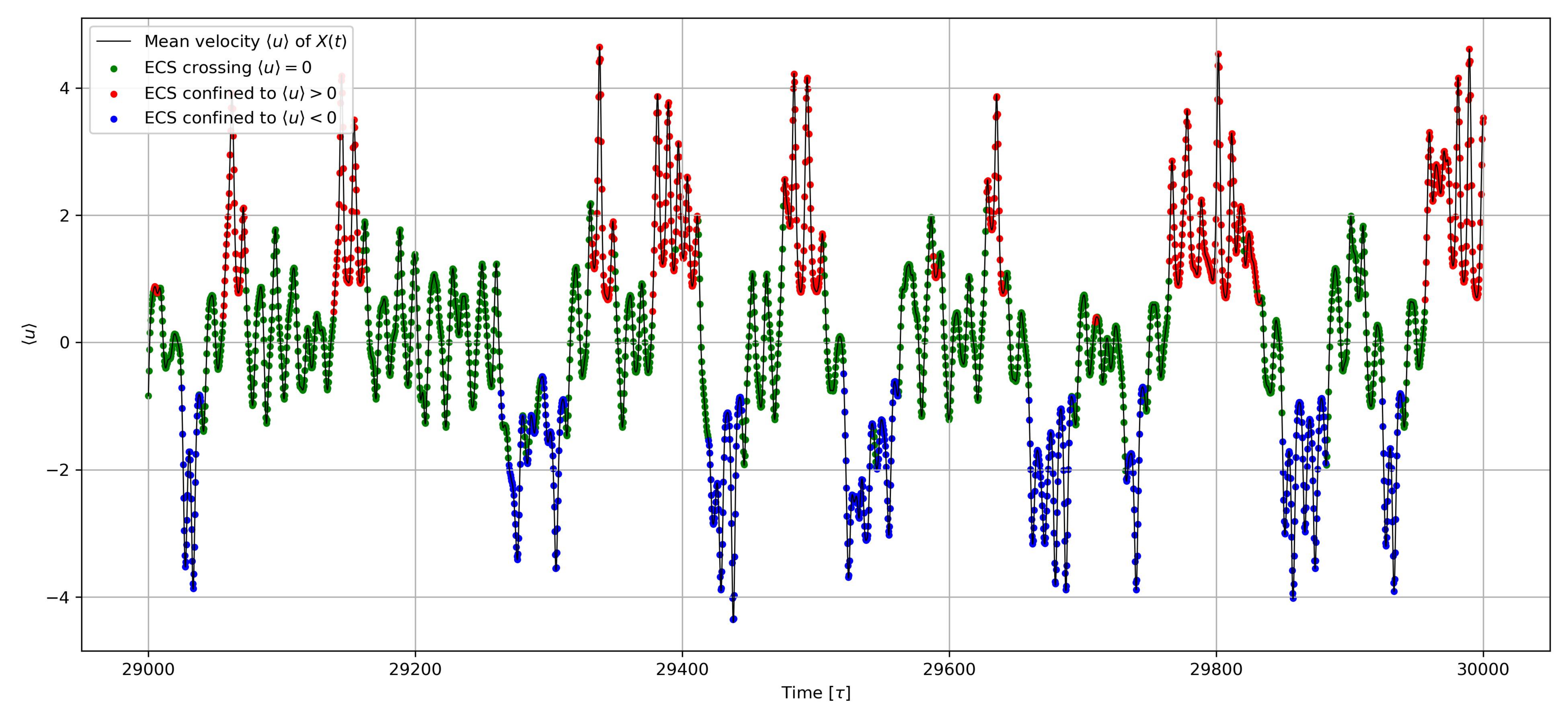}%
    \label{fig:time_series_x}%
  }\\[1em]
  \subfloat[\footnotesize Reduced phase space projection showing the segment turbulent trajectory $X(t)$ in the interval $29000\tau \le t \le 30000\tau$ (black) and the six main ECSs (and their symmetry copies). The color scheme for the ECSs is the same as in (a). The orange circle indicates the beginning of the segment ($t=29000\tau$) and the magenta square indicates the end of the segment ($t=30000\tau$).]{%
    \includegraphics[width=.85\textwidth]{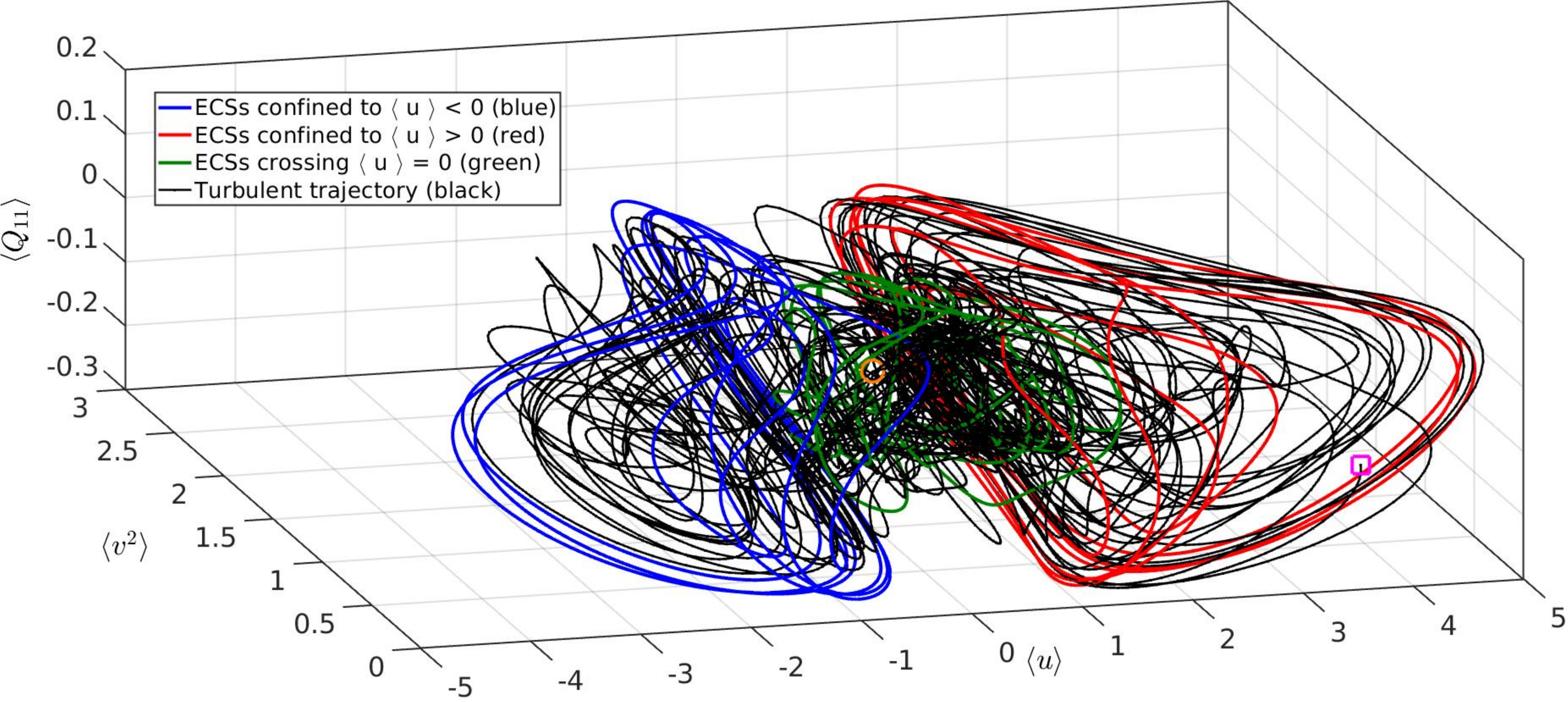}%
    \label{fig:trajectory_ecs}%
  }\\[1em]
  \subfloat[\footnotesize Reduced phase space portrait showing the six main ECSs (and their symmetry copies) shadowed by the turbulent trajectory, along with their symmetry-related copies under $\sigma_y$.]{%
    \includegraphics[width=.85\textwidth]{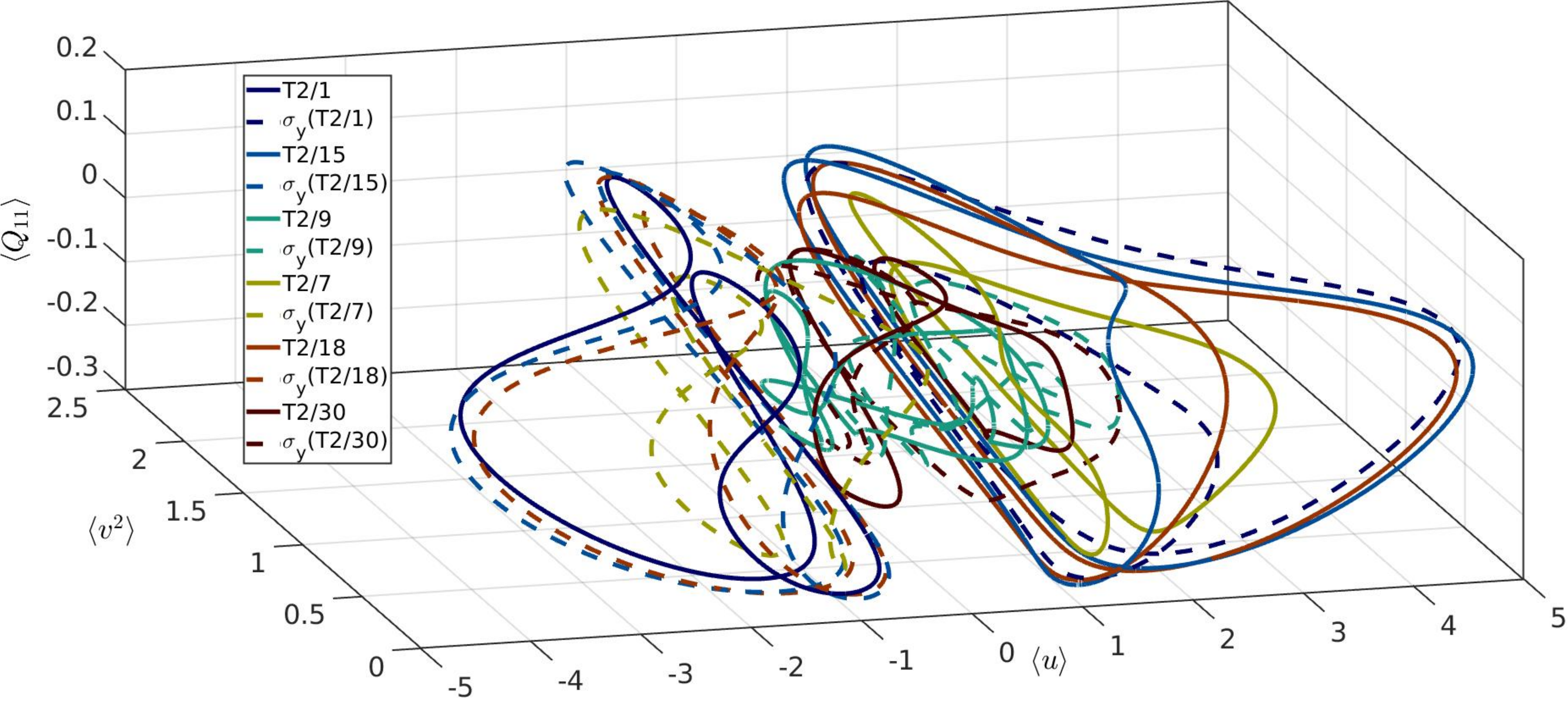}%
    \label{fig:main_ecs}%
  }\
  ;

  \caption{\footnotesize Flow reversals and the relationship between the turbulent trajectory and the shadowed ECSs.}
  \label{fig:ecs_analysis}
\end{figure}

We consider a typical trajectory \( X(t) \) lying in $\mathcal{X}$ in \an{SPU} A over the interval $[29000\tau, 30000\tau]$. This segment of the trajectory exhibits eleven flow reversals. The mean streamwise velocity \( \langle u \rangle \) fluctuates around zero, with extended excursions near \( \langle u \rangle = -2 \) (net negative flow) and \( \langle u \rangle = 2 \) (net positive flow). To investigate the underlying structure, we record the closest ECS to the trajectory every \( 0.4\tau \), yielding 2501 comparison points. These ECSs are color-coded as discussed above. As shown in Fig.~\ref{fig:ecs_analysis}, when the trajectory transitions between red and blue regions, it typically passes near a green ECS.

Across the 2,501 points, 40 of the 75 ECSs (along with their symmetry copies) appear as the closest state at least once. To isolate the most dynamically significant structures, we focus on the ten most frequently encountered ECSs. Together, these ten ECSs account for approximately 62\% of all occurrences (see Table~\ref{tab:ecs_top10} in Appendix). As anticipated, each of these ten ECSs are invariant under $\tau_x(L_A/2)$ symmetry. 
\begin{figure}[htbp]
    \centering
    \includegraphics[width=1.0\textwidth]{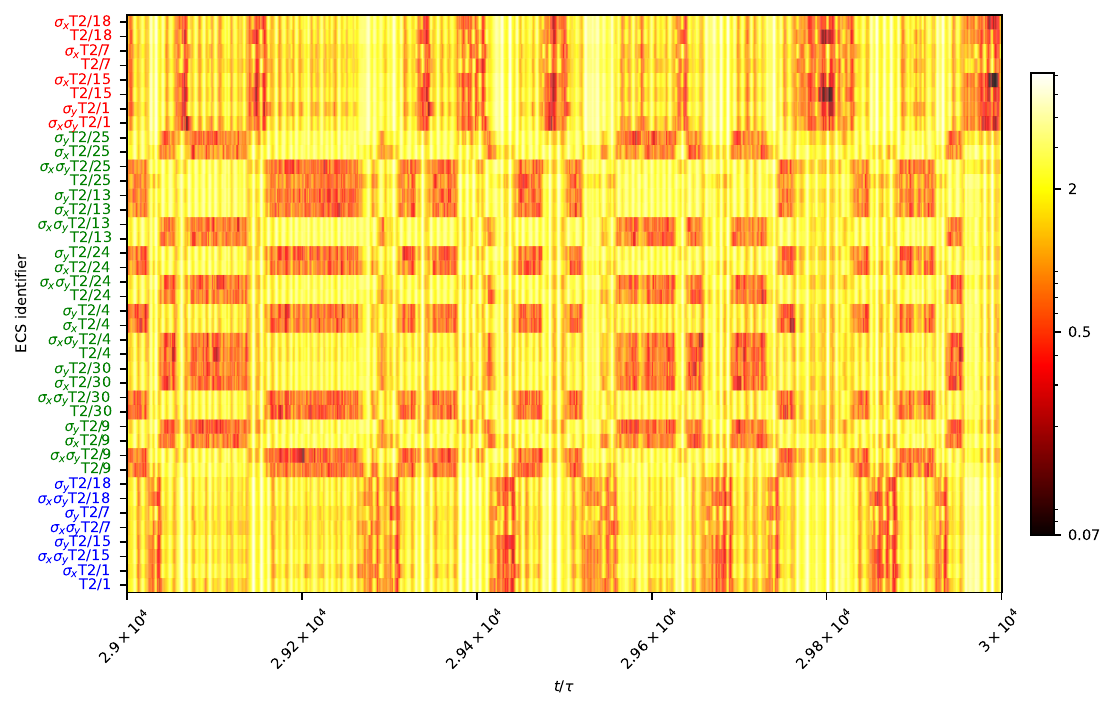}
    \caption{\footnotesize Symmetry-reduced distance plot showing the closest encounters between the chaotic trajectory $X(t)$ and the ten most relevant ECSs and their symmetry copies. The ECS list on the left follows the same color scheme as in Fig. \ref{fig:ecs_analysis}(a). The plot shows a clear pattern of repeated transitions (i.e., reversals) between the Group 1 `left flowing' (blue) and 'right flowing' (red) ECSs, via the Group 2 intermediate (green) ECSs.}
    \label{fig:distance_X}
\end{figure}

These ten ECSs serve as a reduced representation that captures the essential skeleton of the attractor \( \mathcal{X} \)’s dynamics. Six of these (`Group 2', shown in green) are natural candidates for mediating transitions between the red and blue regions, i.e., for enabling reversals. Figure~\ref{fig:distance_X} shows the distance plot between the trajectory and these ECSs, highlighting the paths of reversals. The remaining four ECSs (`Group 1', shown in red/blue), along with those from the green category, are closely approximated by the trajectory. For example, in the interval \( [29000\tau, 29200\tau] \), during the first reversal of the series (see also Fig.~\ref{fig:time_series_x}), the distance plot shows that the trajectory first approaches the blue ECS group, with the mean velocity oscillating around \( \langle u \rangle \approx -2 \). It then nears some ECSs of the green group, where \( \langle u \rangle \approx 0 \), before finally transitioning to a state where the mean velocity oscillates around \( \langle u \rangle \approx 2 \) and closely follows the red ECS group.

\subsubsection{Shadowing in $\mathcal{X}$}
Among the above chosen subset of ECSs, we identify the closest ECS at each time step within the interval \([29000\tau, 30000\tau]\), and assess whether these close encounters represent genuine shadowing events. To make this determination, we compare the dynamics of the trajectory with those of the ECSs. Specifically, we utilize the time parameter \(s\) associated with each ECS, along with the streamwise translation \(\ell\), which is applied to obtain the closest translated version of the ECS as defined in Equation~\ref{eq:distance}. A shadowing event is characterized by synchronized evolution between the ECS and the turbulent trajectory.
Thus, during a shadowing event, we expect \(\dfrac{ds}{dt} \approx 1\) and \(\dfrac{d\ell}{dt} \approx 0\), indicating that the ECS evolves at nearly the same rate as the trajectory and requires minimal translation to remain aligned with the turbulent trajectory.

\begin{figure}[htbp]
    \centering
    \includegraphics[width=1.0\textwidth]{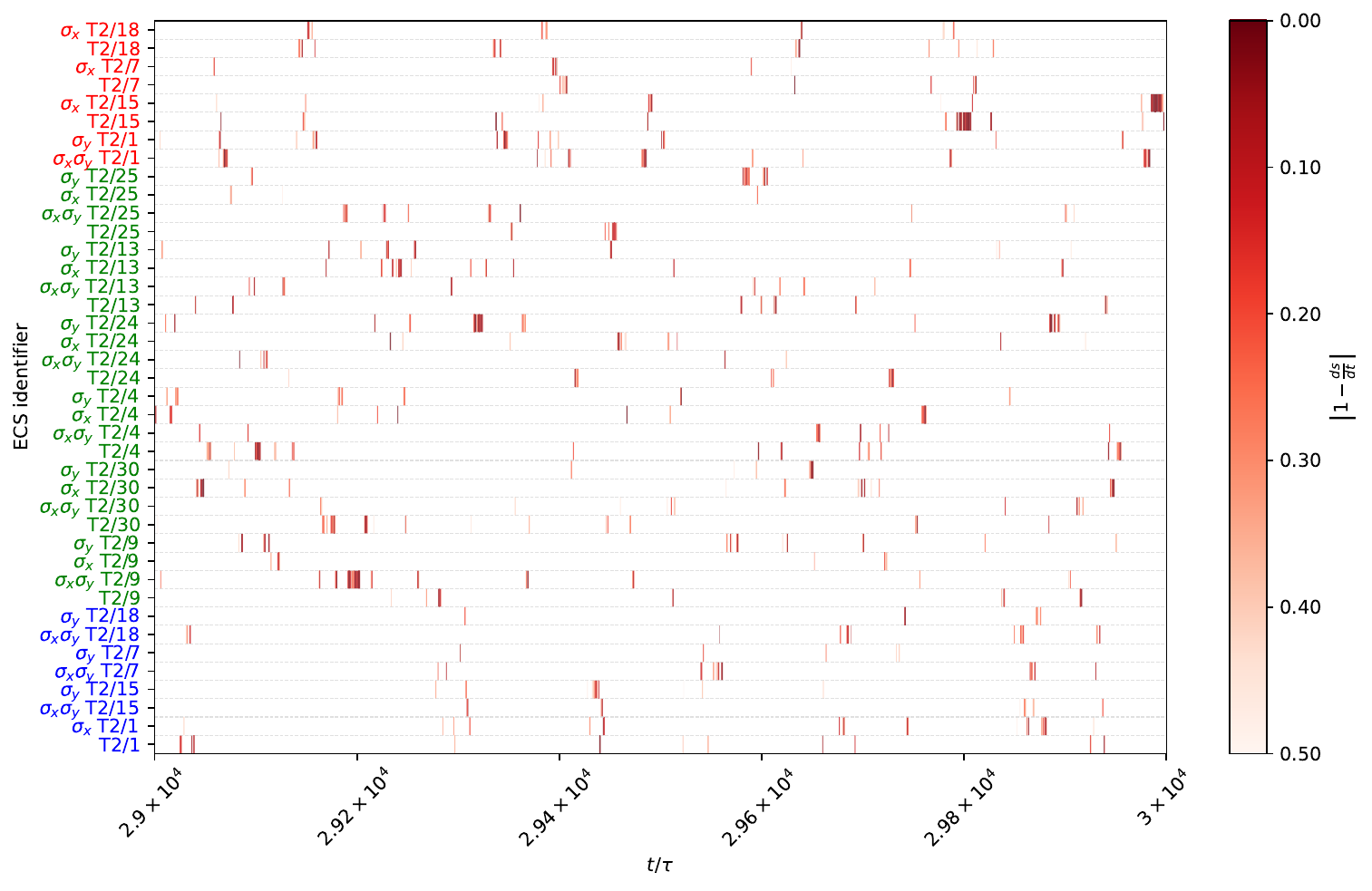}
    \caption{\footnotesize The dynamic shadowing criterion evaluated along the trajectory \(X(t)\) relative to the closest ECS \(X_E(s)\) at discrete times (with $\delta t=0.4\tau$) within the interval \([29000 \tau, 30000 \tau]\). The plot marks instances where the dynamical similarity measure \(\dfrac{d s}{dt}\) lies in the interval \([0.5, 1.5]\), thereby satisfying the criterion. The closer its value is to \(1\), the stronger the shadowing event.}
    \label{fig:shadowing_events_X}
\end{figure}

Since turbulent trajectories inevitably depart from the ECS along its unstable manifold, we adopt a relaxed criterion to identify approximate shadowing. Specifically, the criterion is considered satisfied when
\begin{equation}
0.5 \leq \frac{ds}{dt} \leq 1.5,
\end{equation}
as illustrated in Fig.~\ref{fig:shadowing_events_X}. In such cases, we typically observe
\begin{equation}
\left|\frac{d\ell}{dt}\right| < 1,
\end{equation}
which corresponds to a shift of less than 4\% of the channel length. Time derivatives are computed using central differences for interior points and forward/backward differences at endpoints.  \an{Moreover, shadowing events are expected to occur over time intervals comparable to the inverse of the escape rate $\kappa$~\cite{Krygier2021} of the corresponding shadowed ECS,
\begin{equation}
    \kappa=\sum_{\operatorname{Re}(\lambda_k)>0}\operatorname{Re}(\lambda_k),
\end{equation}
where the sum is taken over the unstable directions, with $\operatorname{Re}(\lambda_k)$ denoting the real parts of the corresponding Floquet exponents. }

For relative periodic orbit ECSs, we additionally account for the discrete jumps in $s$ and $\ell$ at the orbit period $T$:
\begin{equation}
\tau_x(\ell_0) X_E(s=T) = X_E(s=0), \quad \text{so that} \quad s \rightarrow s + T, \;\; \ell \rightarrow \ell + \ell_0,
\end{equation}
which preserves continuity in the symmetry-reduced representation of the RPO and allows consistent computation of the derivatives.

\begin{figure}[htbp]
    \centering

    \includegraphics[width=0.755\textwidth]{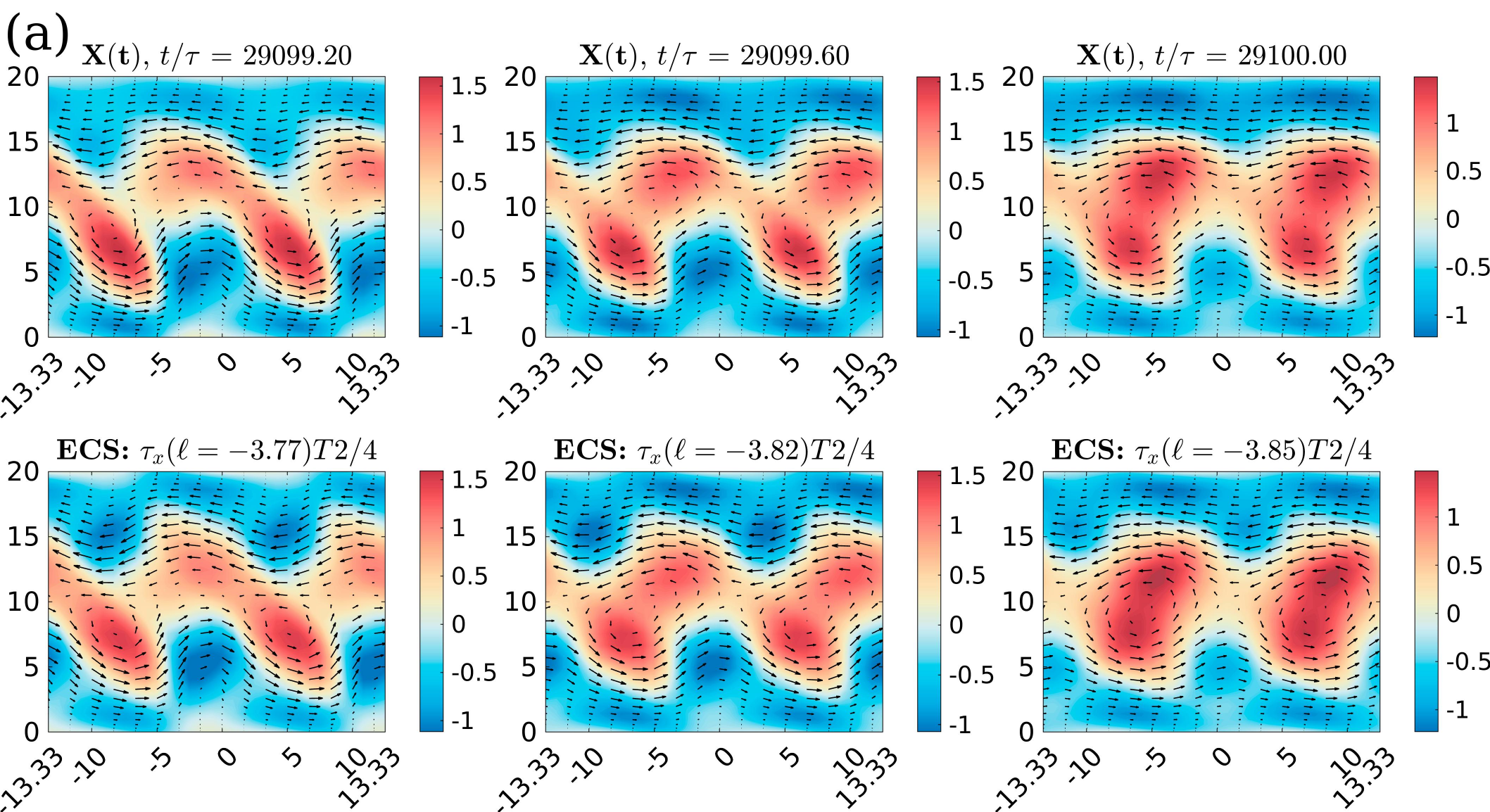}

    \vspace{0.5em}

    \includegraphics[width=0.775\textwidth]{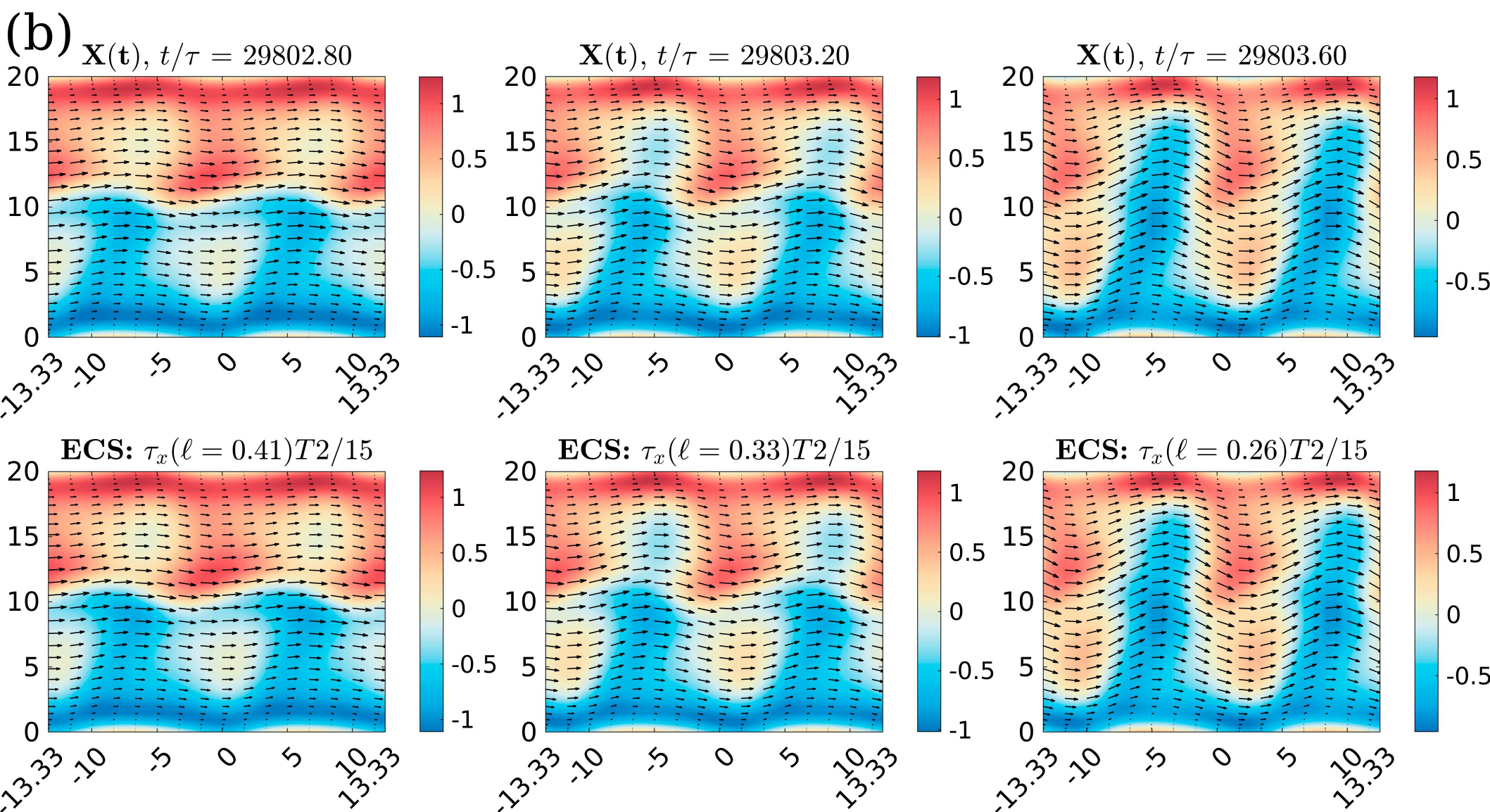}

    \vspace{0.5em}

    \includegraphics[width=0.775\textwidth]{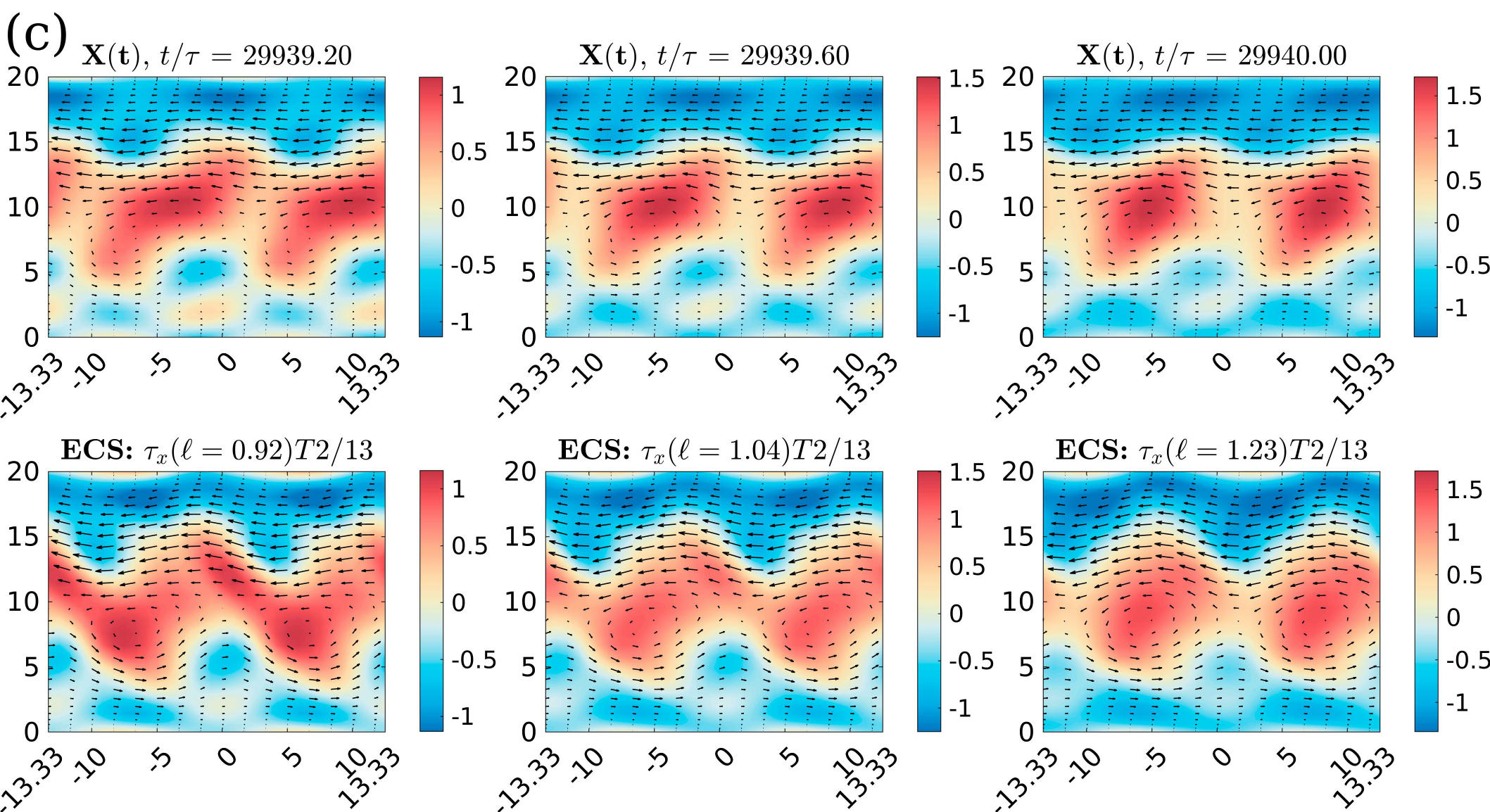}

    \caption{\footnotesize
    \an{Three snapshot sequences extracted from three different confirmed shadowing events of trajectory $X(t)$. 
    The first row of each sequence shows sequential snapshots of the velocity field along the turbulent trajectory, separated by $\delta t=0.4\tau$, while the second row shows the corresponding snapshots of the shadowed ECS at the same times. 
    Notice that the translation $l$ is almost constant during each event, which implies that the \emph{same translated copy of} an ECS is being shadowed throughout the time interval. 
    Sequences (a) and (b) correspond to clear shadowing events of T2/4 and T2/15, respectively. 
    Sequence (c) corresponds to a weaker shadowing event of T2/13 due to the larger approximation distance.}
    }
    \label{fig:shadowing_examples_X}
\end{figure}

\an{We now examine three representative shadowing events to illustrate how the shadowing criteria characterize the proximity and evolution of the turbulent trajectory relative to an ECS. The examples include clear shadowing events, in which the trajectory remains close to an ECS and the corresponding rates remain near their optimal values, as well as a case in which the criteria are satisfied at a larger distance and the resemblance is correspondingly weaker.} Shadowing is typically most evident when the distance between the turbulent trajectory and an ECS remains within a small range (typically between $0.07$ and $0.3$). In the interval $[29099.2\tau, 29104.4\tau]$ (see Fig.~\ref{fig:shadowing_examples_X}(a)), the ECS closest to $X(t)$ is $\mathrm{T}2/4$, with an average distance $\langle d \rangle \approx 0.247$ and average rates $\left\langle \left|\dfrac{ds}{dt}\right| \right\rangle \approx 1.055$ and $\left\langle \left|\dfrac{d\ell}{dt}\right| \right\rangle \approx 0.101$. \an{The length of this time interval is $5.2\tau\approx 0.73\tau_{\mathcal{X}}\approx 2\kappa^{-1}$, where $\tau_{\mathcal{X}}$ is the Lyapunov time of the chaotic set $\mathcal{X}$ and $\kappa^{-1}\approx2.580\tau$ is the escape time of the ECS. During this interval, the trajectory approaches the ECS most closely at $t=29102\tau$, reaching a minimum distance of $d\approx 0.2$. At this time, $\dfrac{ds}{dt}\approx 1$ and $\dfrac{d\ell}{dt}\approx -0.005$, corresponding to the optimal values of these indicators over the interval. Taken together, the consistently small distance, near-unity $\dfrac{ds}{dt}$, and small $\dfrac{d\ell}{dt}$ throughout this interval provide strong \an{evidence} of a shadowing event.}

Another clear shadowing event occurs in the interval $[29799.2\tau, 29806.8\tau]$ (see Fig.~\ref{fig:shadowing_examples_X}(b)). \an{In this case, the closest ECS is $\mathrm{T}2/15$, with an associated escape time $\kappa^{-1}\approx7.609\tau$, which is approximately equal to the length of the shadowing interval and corresponds to about $1.06\tau_{\mathcal{X}}$.} The average distance and rates are $\langle d \rangle \approx 0.126$, $\left\langle \left|\dfrac{ds}{dt}\right| \right\rangle \approx 0.989$, and $\left\langle \left|\dfrac{d\ell}{dt}\right| \right\rangle \approx 0.183$. At $t = 29805.6\tau$, the trajectory has $d \approx 0.105$, $\dfrac{ds}{dt} \approx 1.025$, and $\dfrac{d\ell}{dt} \approx 0.017$.

When the shadowing criteria are satisfied at larger distances, the resemblance to the ECS is generally weaker. For example, around $t = 29939.6\tau$ (see Fig.~\ref{fig:shadowing_examples_X}c), when the turbulent trajectory is in the red region, it does not approach the ECS $\mathrm{T}2/13$ closer than $0.453$, yet the temporal evolution remains similar, with $\dfrac{ds}{dt} \approx 0.813$ and $\dfrac{d\ell}{dt} \approx 0.391$. Although the resemblance to the ECS is not as strong as in cases with smaller distances, these examples suggest that, even at relatively large distances, traces of shadowing can still be detected. This could indicate that the trajectory is closely shadowing another, unconsidered but dynamically related ECS associated with the considered one, or possibly shadowing an invariant manifold instead. Therefore, the analysis of shadowing reveals different nuances that must be taken into account due to the complexity of the phase space geometry.

\noindent\textbf{Observable reconstruction:} Additional indirect evidence for shadowing comes from reconstruction of observables along a turbulent trajectory using ECS data. We approximate a physical observable $\mathcal{O}$ in the active nematic system along a turbulent trajectory $X(t)$ by a weighted superposition of ECS contributions:
\begin{align}
\mathcal{O}\big(X(t)\big) \approx \sum_i w_i(t)\, \mathcal{O}\big(X_{E_i}(s_i(t))\big),\nonumber
\end{align}
where $X_{E_i}(s)$ denotes the state along the $i$-th ECS, $s_i(t)$ is the phase that minimizes the distance to $X(t)$ at time $t$, and $w_i(t)$ represents the instantaneous weight assigned to that ECS.
As a minimal realization, we define the weights using a nearest-ECS criterion: at each time \(t\), the ECS closest (in phase space) to the turbulent state is assigned unit weight, while all others are set to zero. 
We use three spatially averaged observables: the kinetic energy, enstrophy, and Frank elastic energy. These quantities have been previously used to characterize scaling behavior and statistical properties of active turbulent flows \citep{alert2020universal, Krajnik2020Spectral}. We define them as follows:
 \pg{
\begin{align}
E_k &= \frac{1}{2\mathcal{A}} \int_{\mathcal{A}} \left(u^2 + v^2\right) \, dx\,dy, \\
\mathcal{E} &= \frac{1}{2\mathcal{A}} \int_{\mathcal{A}} \left(\partial_y u - \partial_x v\right)^2 \, dx\,dy, \\
E_{\text{Frank}} &= \frac{1}{2\mathcal{A}} \int_{\mathcal{A}} \left( |\nabla Q_{11}|^2 + |\nabla Q_{12}|^2 \right) \, dx\,dy,
\end{align}
where $\mathcal{A}$ is the channel area.}

\begin{figure}
\includegraphics[width=.65\textwidth]{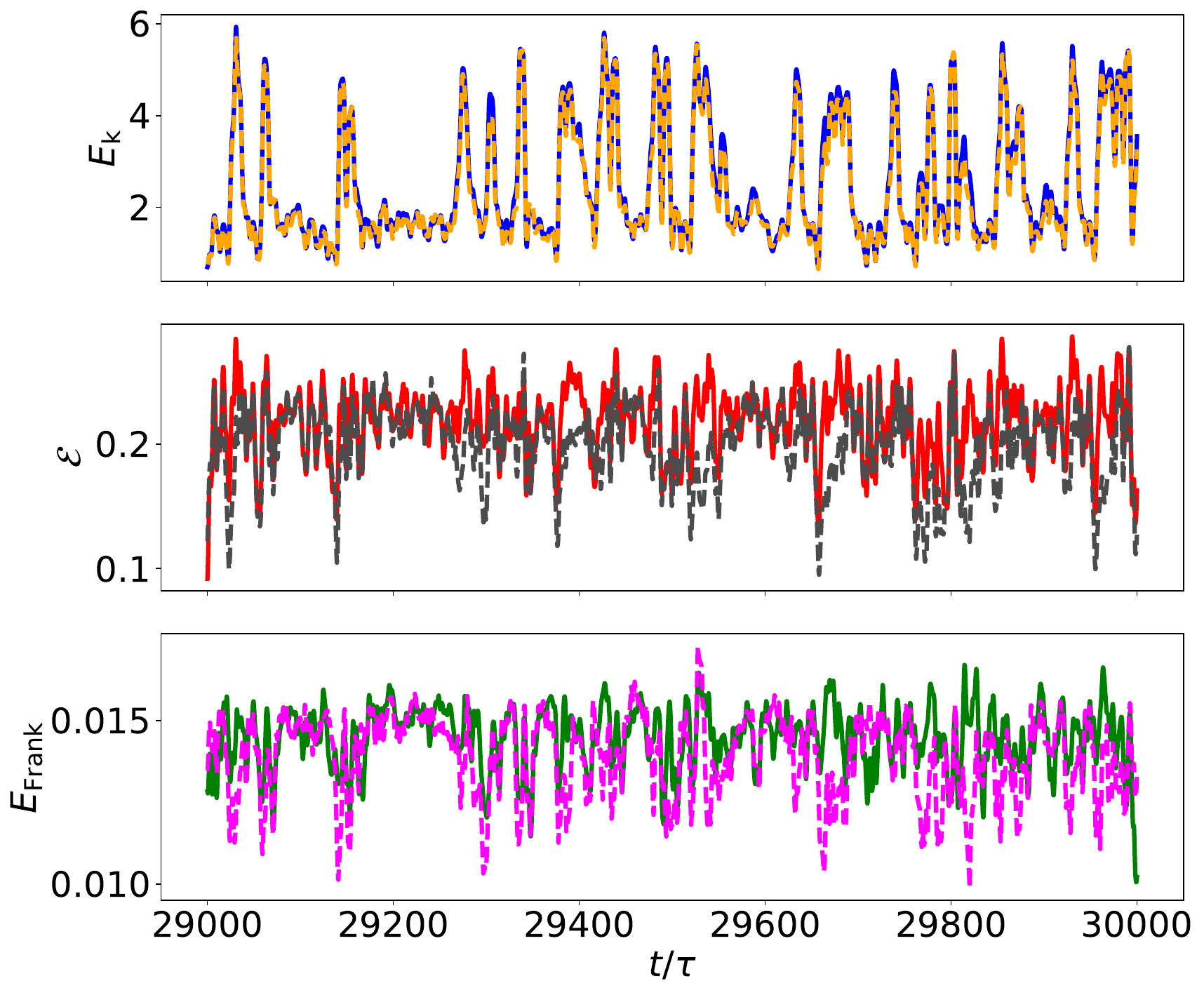}
\caption{\footnotesize
Moving averages (with window $7.2\tau \approx\tau_{\mathcal{X}}$) of each observable, with solid curves showing the turbulent time series and dashed curves the corresponding ECS-based reconstructions.} 
\label{fig:observables}

\end{figure}

Our approximation captures the trends of all three observables \pg{reasonably} well, as illustrated by the  $7.2\tau$ moving average time series in Fig.~\ref{fig:observables}, \an{ where the averaging window size was chosen close to the corresponding Lyapunov time.}  The kinetic energy is reproduced with particularly high fidelity: the correlation between the turbulent signal and its ECS-based approximation is $0.991$, and their time averages are \an{$ \langle E_\mathrm{k} \rangle \approx 2.473$ and $2.373$}, respectively, corresponding to a relative error of about $4.1\%$.

Although enstrophy and Frank elastic energy are harder to reconstruct because they depend on spatial derivatives of the velocity field and nematic director field, the time series in Fig.~\ref{fig:observables} nevertheless reproduces their overall trends reasonably well. Local agreement improves markedly during intervals of stronger shadowing; for instance, around \( t = 29100\tau \) and \( t = 29800\tau \). In general, however, the ECS-based approximation tends to underestimate both enstrophy and Frank energy.

Quantitatively, enstrophy exhibits a correlation of \(0.848\) between the turbulent trajectory and its ECS approximation, with temporal averages \an{\(\langle \mathcal{E} \rangle \approx 0.218\) and \(0.200\)}, corresponding to a relative error of \an{\(8.4\%\)}. The Frank energy shows a correlation of \(0.656\), with temporal averages \an{ \(\langle E_{\mathrm{Frank}} \rangle \approx 0.0145\) and \(0.0139\)}, yielding a relative error of approximately \(4.0\%\). 

\begin{figure}[t]
    \centering
    \includegraphics[width=\textwidth]{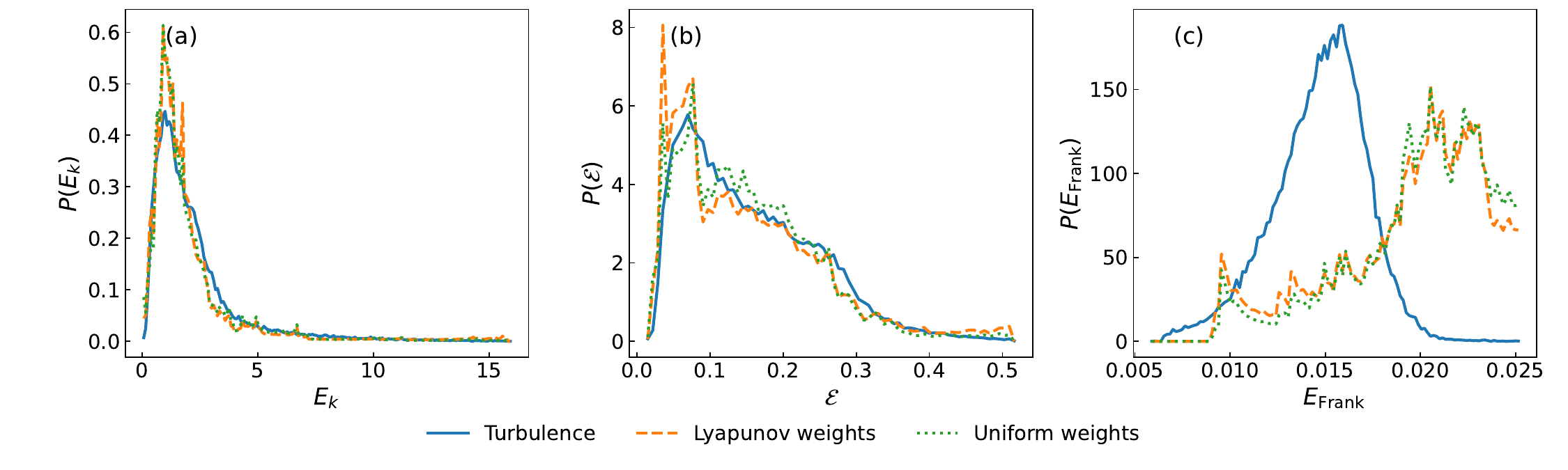}
   \caption{\footnotesize{Reconstruction of the observable distributions using the $32$ ECSs invariant under $\tau_x(L_A/2)$ for the two heuristic weighting schemes providing the best overall reconstruction: uniform and Lyapunov-based weighting. Panels (a) and (b) show reasonable agreement between the reconstructed and actual distributions for the kinetic energy and enstrophy, respectively. In contrast, panel (c) shows that the Frank elastic energy is poorly reconstructed.}}
    \label{fig:observable_distributions_Lyapunov_vs_uniform}
\end{figure}

\an{\noindent\textbf{Reconstruction of Turbulent Statistics}: Motivated by periodic orbit theory \citep{Cvitanovic2020Chaos}, several recent works \citep{Suri2020Capturing, Yalniz2021Coarse, Page2024Recurrent} have proposed approximating inertial turbulent statistics using a collection of ECSs. Here we consider the reconstruction of the empirical probability density functions $P_\mathrm{turb}(O)$ of the observables discussed above using the corresponding empirical distributions obtained from the $32$ ECSs invariant under $\tau_x(L_A/2)$,
\begin{equation}
P_{\mathrm{turb}}(O)\approx\sum_i^{N=32} W_i P_i(O)=P_{\mathrm{ECS}}(O).
\end{equation}
The weights $W_i$ are distinct from those introduced earlier for assessing trajectory-level shadowing. Inspired by the analysis in \citep{Page2024Recurrent,redfern2024dynamically}, three weighting schemes are used here: uniform weighting $W_i=\dfrac{1}{N}$, Lyapunov-based weighting $W_i=\dfrac{{\lambda_i}^{-1}}{\sum_j^N{\lambda_j}^{-1}}$, where $\lambda_i$ is the largest positive Lyapunov exponent of $i$th ECS, and the third scheme consists of weights obtained by minimizing the Kullback--Leibler divergence $D_{\mathrm{KL}}$ between the ECS-based and turbulent distributions,
\begin{equation}
D_{\mathrm{KL}}
=
\int
\left[
P_{\mathrm{turb}}(O)
\log\left(
\frac{P_{\mathrm{turb}}(O)}
{P_{\mathrm{ECS}}(O)}
\right)
+
P_{\mathrm{ECS}}(O)
-
P_{\mathrm{turb}}(O)
\right]
\,\mathrm{d}O.
\end{equation} 
The empirical distributions obtained using the uniform and Lyapunov-based schemes are plotted in Fig.~\ref{fig:observable_distributions_Lyapunov_vs_uniform}, together with that obtained from a long turbulent trajectory of duration \(2.4\times10^4\tau\), yielding \(6\times10^4\) observations sampled at intervals of \(0.4\tau\). To obtain a quantitative comparison, we use $D_{\mathrm{KL}}$ and normalized integrated squared error (NISE),
\begin{equation}
\mathrm{NISE}=\int\frac{\left(P_{\mathrm{turb}}(O)-P_{\mathrm{ECS}}(O)\right)^2}{P_{\mathrm{turb}}(O)}dO,
\end{equation} see Table~\ref{tab:statistical_comparison_ECS}.
 We consider three variants of the $D_\mathrm{KL}$-based scheme, with the weights optimized using the distributions of kinetic energy ($D_\mathrm{KL}$--$E_k$), enstrophy ($D_\mathrm{KL}$--$\mathcal{E}$), and Frank elastic energy ($D_\mathrm{KL}$--$E_{\mathrm{Frank}}$), respectively. When the weights are fitted using either kinetic energy or enstrophy, the other observable is also reconstructed reasonably well, whereas the Frank elastic energy distribution is reproduced poorly. When the Frank elastic energy distribution is used for fitting, its reconstruction improves somewhat, but results in poorer reconstruction of both kinetic energy and enstrophy. The reason for the relatively poor reconstruction of the Frank elastic energy in general is not yet clear and is a topic for future studies.
}

\begin{table}[t]
    \centering
   \pg{ 
    \begin{tabular}{llcc}
        \hline
        Observable & Weighting scheme & $D_{\mathrm{KL}}$ & NISE \\
        \hline

        \multirow{5}{*}{Kinetic energy}
        & Uniform
            & 0.0481
            & 0.0549 \\

        & Lyapunov
            & 0.0536
            & 0.0546 \\

        & $D_{\mathrm{KL}}$--$E_k$
            & 0.0200
            & 0.0159 \\

        & $D_{\mathrm{KL}}$--$\mathcal{E}$
            & 0.0620
            & 0.0454 \\

        & $D_{\mathrm{KL}}$--$E_{\mathrm{Frank}}$
            & 0.137
            & 0.218 \\

        \hline

        \multirow{5}{*}{Enstrophy}
        & Uniform
            & 0.0307
            & 0.0373 \\

        & Lyapunov
            & 0.0452
            & 0.0782 \\

        & $D_{\mathrm{KL}}$--$E_k$
            & 0.0334
            & 0.0647 \\

        & $D_{\mathrm{KL}}$--$\mathcal{E}$
            & 0.00958
            & 0.00789 \\

        & $D_{\mathrm{KL}}$--$E_{\mathrm{Frank}}$
            & 0.222
            & 0.593 \\

        \hline

        \multirow{5}{*}{Frank elastic energy}
        & Uniform
            & 1.89
            & 1.19 \\

        & Lyapunov
            & 1.75
            & 1.10 \\

        & $D_{\mathrm{KL}}$--$E_k$
            & 1.74
            & 1.07 \\

        & $D_{\mathrm{KL}}$--$\mathcal{E}$
            & 1.95
            & 1.16 \\

        & $D_{\mathrm{KL}}$--$E_{\mathrm{Frank}}$
            & 0.947
            & 0.321 \\

        \hline
    \end{tabular}
    \caption{\footnotesize{KL divergences and normalized integrated squared errors for the five weighting methods used to reconstruct turbulent statistics from an ensemble of 32 ECSs invariant under $\tau_x(L_A/2)$.}}\label{tab:statistical_comparison_ECS}}
\end{table}


\subsubsection{Reversal Mechanisms and Heteroclinic connections}
\begin{figure}[t]
    \centering
    \includegraphics[width=1.0\linewidth]{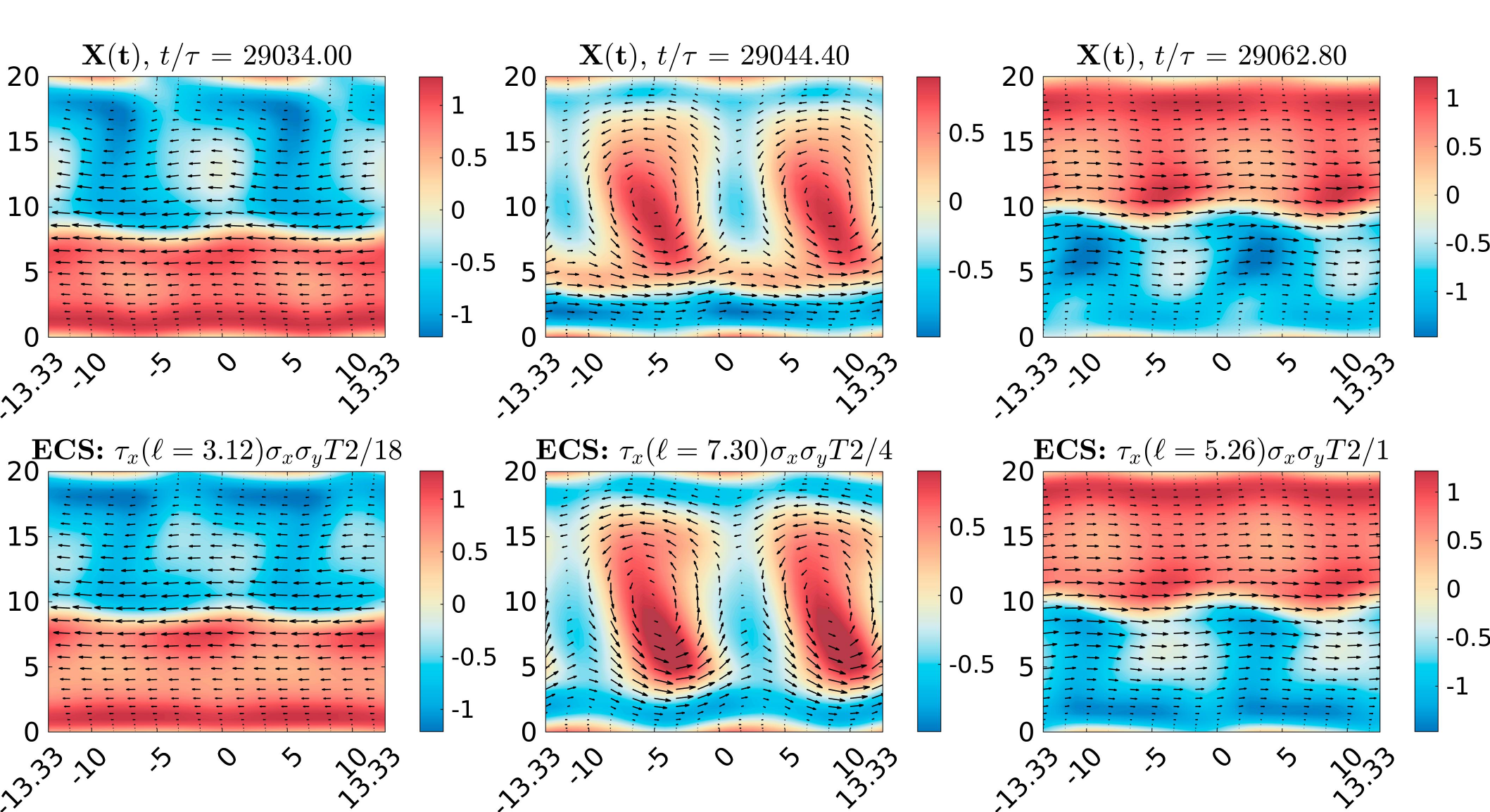}
    \caption{\footnotesize Top row: Velocity field snapshots during the first reversal event for a turbulent trajectory $X(t)$, lying in attractor \( \mathcal{X}(t) \), during the interval \( [29034\tau, 29062.8\tau] \) resembling the left UNI, two-vortex lattice and right UNI solutions, respectively. Bottom row: Velocity field snapshots of the ECSs shadowed during the reversal shown at the respective times.}
    \label{fig:Flow_Reversal_X}
\end{figure}

\begin{figure}[t]
    \centering
    \includegraphics[width=1.0\linewidth]{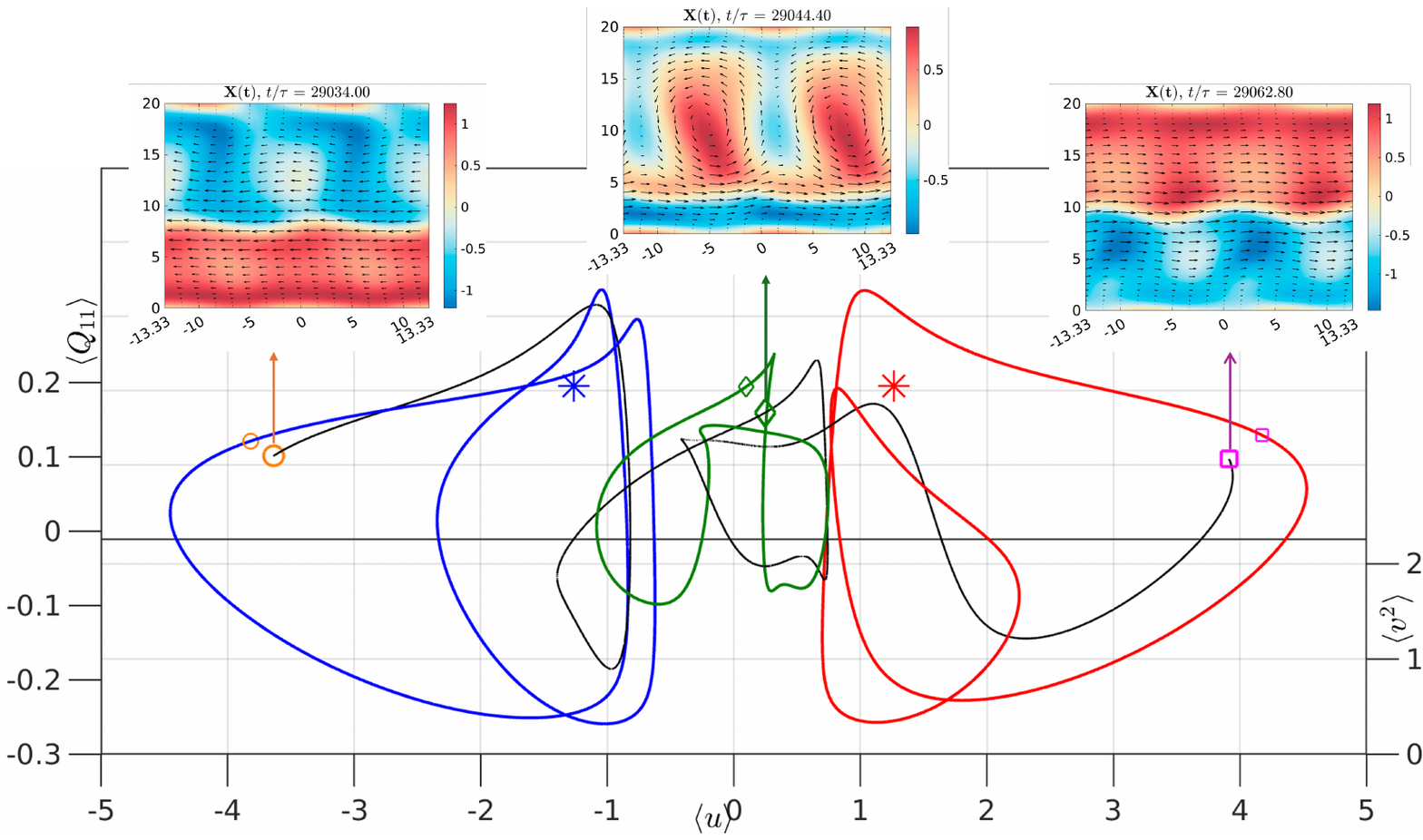}
\caption{\footnotesize 
Shadowing and reversals of Fig. \ref{fig:Flow_Reversal_X} in the reduced phase space. A segment of the turbulent trajectory $X(t)$ is shown in black, while the shadowed 
left-flowing ECS $\sigma_x\sigma_y$T2/18 is shown in blue, the two-vortex  ECS $\sigma_x\sigma_y$T2/4 is shown in green, and the right-flowing $\sigma_x\sigma_y$T2/1 is shown in red. The larger circle, rhombus, and square mark the locations of the snapshots (inset) along the turbulent trajectory, while the smaller circle, rhombus, and square indicate the respective locations along the shadowed ECSs. Additionally, the red and blue asterisks show the UNI and $\sigma_y$UNI, respectively. An animation of this reversal event is provided in the Supplemental Material.} 
    \label{fig:Flow_Reversal_X_diagram}
\end{figure}

The shadowing events confirmed above hint at mechanisms similar to those discussed for the preturbulent regime in Section \ref{sec:preturb}. Consider, for instance, the first clear reversal event in \( X(t) \), occurring within the interval \( [29034\tau, 29062.8\tau] \) (see Fig.~\ref{fig:time_series_x}). The system transitions from a leftward to a rightward unidirectional flow, while passing close to a two-vortex like ECS. Figure~\ref{fig:Flow_Reversal_X} illustrates three representative snapshots capturing the key stages of this reversal event, along with their corresponding shadowed ECSs, highlighting how the trajectory temporarily aligns with these coherent structures. Figure~\ref{fig:Flow_Reversal_X_diagram} complements this view by showing the reduced phase space representation of the same event. 

To obtain more insight into the reversal dynamics, the network of heteroclinic connections among the ECSs needs to be studied. However, even in the reduced representation involving a set of 40 ECSs (modulo symmetries), the transition pathways between the blue, green, and red regions are too complicated to be explicitly traced. But some aspects of the phase space geometry become clearer when we focus on specific pairs of ECSs. 

First, the Group 2 ECSs play a role analogous to that of the `intermediary' POs in the preturbulent regime. These ECSs are likely originated from solutions invariant under symmetries conjugate to \(\sigma_x\sigma_y\) and $\tau_x(L_A/2)$ in the preturbulent regime. These ECSs in the turbulent regime have broken \(\sigma_x\sigma_y\) symmetry, but retain the two-fold $\tau_x(L_A/2)$ translational symmetry. Supporting evidence includes their similarity to PO$_{T2a}$ in the reduced phase space. Second, the Group 1 ECSs play a role analogous to that of HRPOs, and form the structural backbone for generating left- and right-flowing, nearly uniaxial states. Their phase space structures suggests that these ECSs are most likely the result of symmetry-breaking bifurcations of the HRPOs. The Group 1 ECS retain the homoclinic-like structure in phase space, have broken the shift-reflect symmetries of their preturbulent counterparts, but still possess the two-fold $\tau_x(L_A/2)$ translation symmetry. 

\begin{figure}
\includegraphics[width=0.495\columnwidth,height=1.6in]{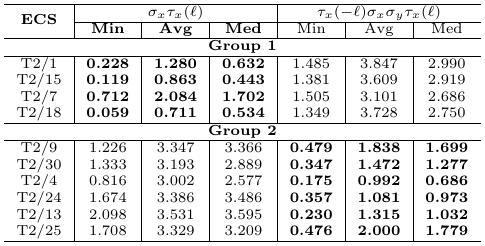}
\includegraphics[width=0.495\columnwidth,height=1.6in]{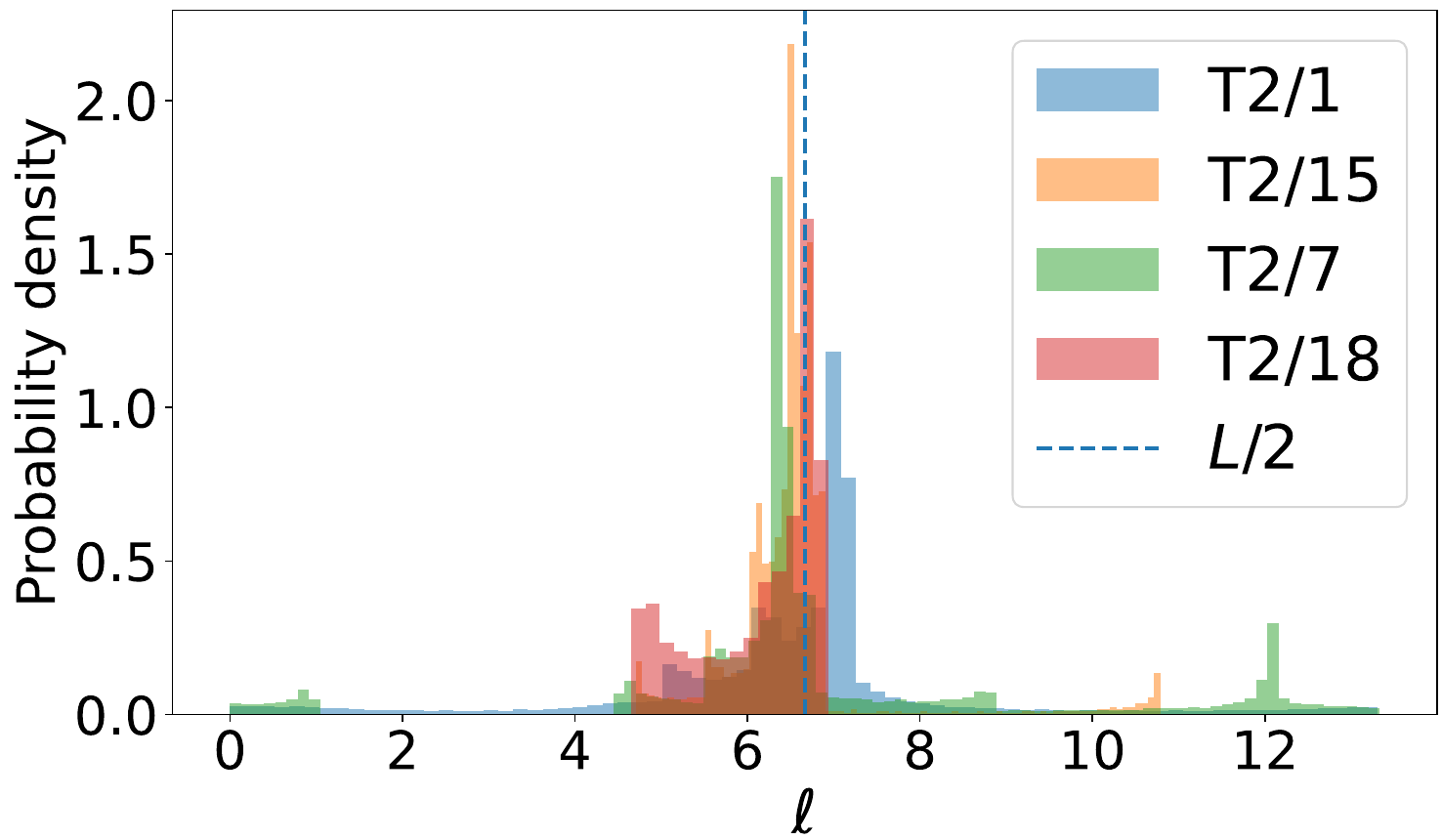}
\caption{\footnotesize
Remanent symmetry analysis based on turbulent ECS orbit data in \an{SPU} B.
Left: Minimum, mean, and median distances between snapshots along each ECS orbit and their symmetry-related images under the two symmetry types. The results indicate that the dominant approximate (remanent) symmetry is shift–reflect ($\sigma_x\tau_x(L_B/2)$) for Group 1, whereas for Group 2 it is a symmetry conjugate to the reflection $\sigma_x\sigma_y$. Right: Distributions of the optimal shifts $\ell$ for Group 1 ECSs, showing strong peaks near $L_B/2\approx 6.667$.
}
\label{fig:symmetry_breaking}
\end{figure}

To move beyond qualitative resemblance, we quantify residual symmetry in the turbulent ECSs by measuring, along each ECS orbit $X_E(t)$, distances to the appropriate symmetry-related images $\tau_x(\ell)\sigma_xX_E(t)$ and $\tau_x(-\ell)\sigma_x\sigma_y\tau_x(\ell)X_E(t)$, for $\ell\in[0,L]$. Since both group 1 and 2 ECSs have two-fold translation symmetry, we work in \an{SPU} B. For each snapshot we minimize over $l$, and use the minimizing shift and the corresponding distance as a remanent symmetry signature, see Fig. \ref{fig:symmetry_breaking}. We find that Group 1 ECSs come markedly closer to their $l=L/2$ shift-reflect images, consistent with their descent from the unstable UNI eigenspace. Group 2 ECSs instead remain closer to images under the conjugates of $\sigma_x\sigma_y$, supporting their interpretation as descendants of LAN-related branches.

\begin{figure}[htbp]
    \centering
    \includegraphics[width=1.0\textwidth]{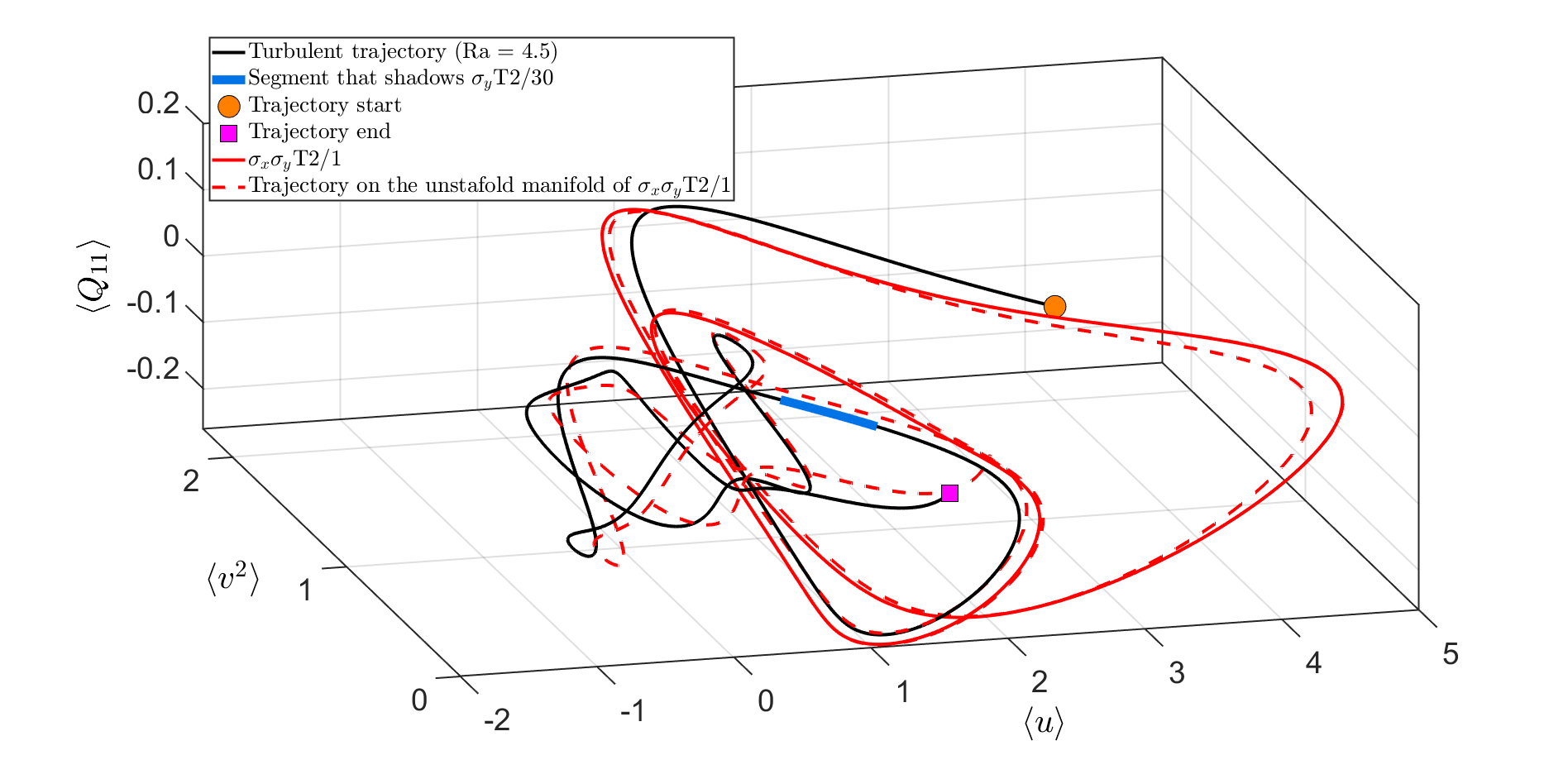}
    \caption{\footnotesize Shadowing of the unstable manifold of a ECS by a turbulent trajectory in $\mathcal{X}$. The turbulent trajectory (black), shadows a trajectory (red dashed) on the unstable manifold of a ECS, $\sigma_x\sigma_y$T2/1, during $[29064\tau, 29094.4\tau]$. The ECS itself is shown in solid red. The later segment of the turbulent trajectory where it shadows a second ECS, $\sigma_y$T2/30, is shown in blue. This latter ECS omitted from the plot for clarity.
    An animation of this shadowing event is provided in the Supplemental Material.}
    \label{fig:Manifold_shadowing}
\end{figure}

The heteroclinic connections between these groups are the likely mechanisms through which turbulence exhibits flow reversals. Indeed, we observe that heteroclinic connections may exist between pairs of ECSs from different regions that are shadowed sequentially. For example, the red ECS $\sigma_x\sigma_y$T2/1 is shadowed during the interval $[29067.20\tau,\,29072\tau]$, and immediately afterward, the green ECS $\sigma_y$T2/30 is shadowed during $[29072.80\tau,\,29073.6\tau]$. This suggests a potential heteroclinic connection between these two ECSs. However, within the green region, the turbulent trajectory sequentially (and sometimes simultaneously) shadows several ECSs, including $\sigma_y$T2/30, $\sigma_x$T2/25, T2/13, $\sigma_y$T2/9, $\sigma_x$T2/30, $\sigma_x\sigma_y$T2/4, and $\sigma_x\sigma_y$T2/13. Consequently, identifying a heteroclinic connection to a target green ECS is challenging, as shadowing events at the red/green and blue/green boundaries involve multiple ECSs and are not sharply defined. Although we cannot conclusively verify a heteroclinic connection between $\sigma_x\sigma_y$T2/1 and $\sigma_y$T2/30, the turbulent trajectory shadows the unstable manifold of the former ECS and eventually approaches the latter. Throughout the subsequent exploration of the green region, the trajectory continues to follow the unstable manifold, as illustrated by the segment of $X(t)$ and a nearby trajectory in Fig.~\ref{fig:Manifold_shadowing} within the interval $[29064.0\tau,\,29094.4\tau]$.

The previous example illustrates how the ECSs and their invariant manifolds shape the dynamics of transitions. To obtain evidence of connections between the green and blue/red ECSs, we perform our computations in \an{SPU} B. Recall that in \an{SPU} B, $\mathcal{X}$ is the only attractor. Additionally, the ECSs and connections that exist in \an{SPU} B and determine the attractor $\mathcal{X}$ dynamics are necessarily those that possess $\tau_x(L_A/2)$ symmetry in \an{SPU} A.

\begin{figure}[htbp]
    \centering
    \subfloat[\footnotesize \an{T2/9 $\rightarrow \sigma_x\sigma_yT2/1$}]{%
        \includegraphics[width=\textwidth]{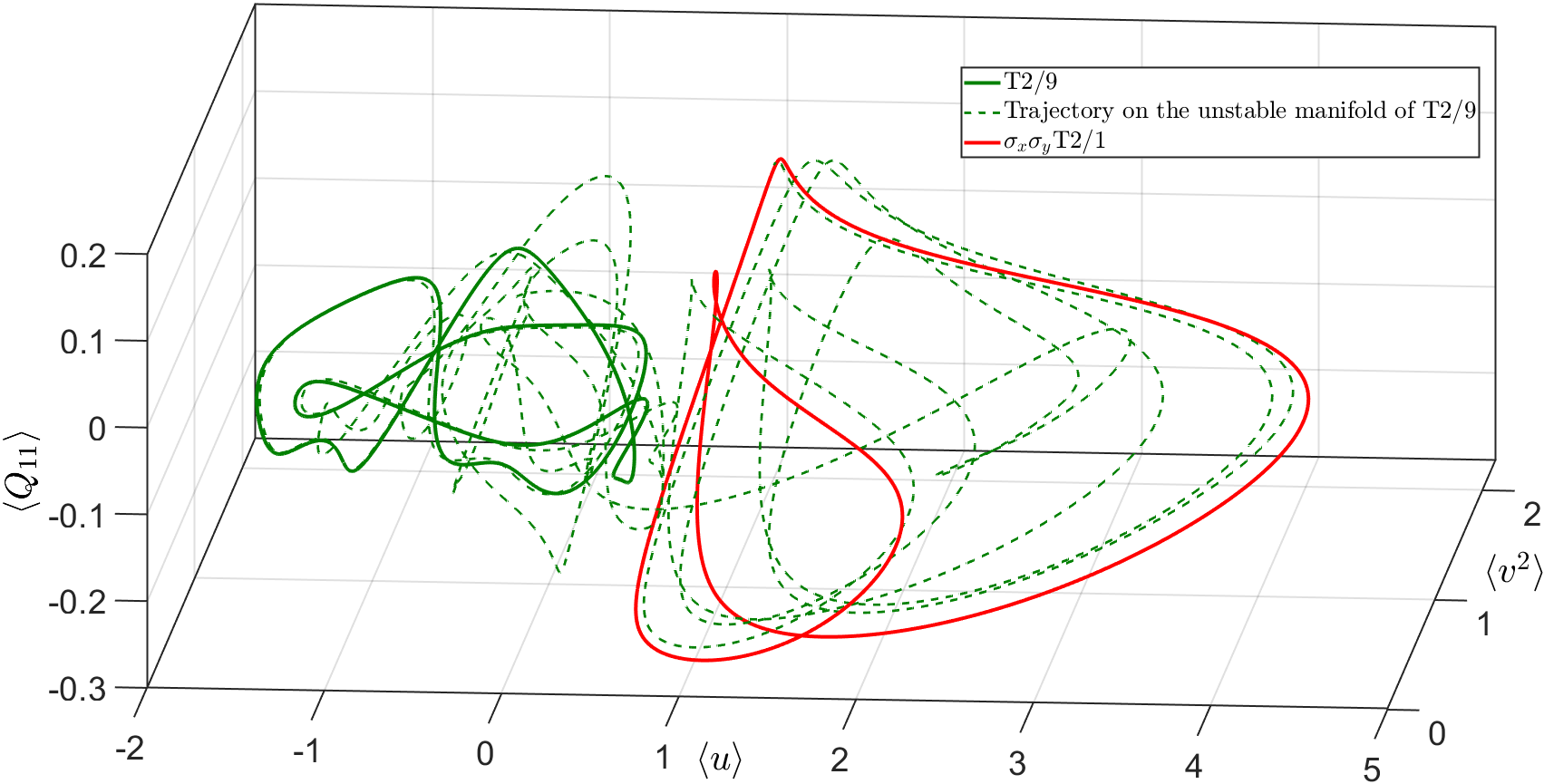}%
        \label{fig:Connection_1}%
    }\\[0.5em]
    \subfloat[\footnotesize \an{T2/30 $\rightarrow \sigma_x\sigma_yT2/15$}]{%
        \includegraphics[width=\textwidth]{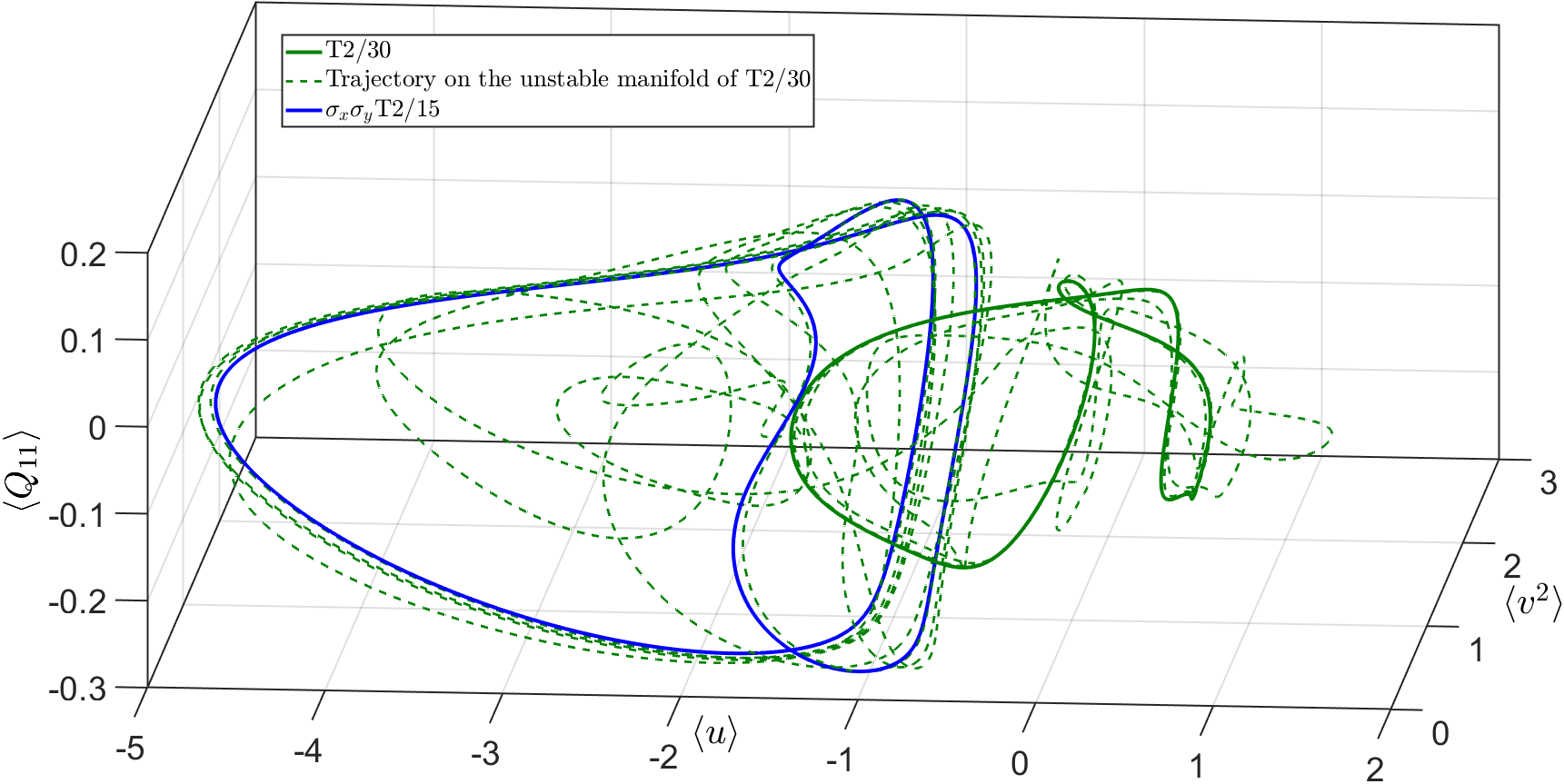}%
        \label{fig:Connection_2}%
    }

    \caption{\footnotesize Examples of trajectories on the unstable manifolds of intermediary (green) ECSs T2/9 and T2/30 that eventually shadow the right-flowing $\sigma_x\sigma_yT2/1$ (red) and left-flowing $\sigma_x\sigma_yT2/15$ (blue) ECSs, respectively. \an{The minimum distance between the trajectory and the target ECS are $0.043$ and $0.009$}, respectively. These results provide strong evidence for the existence of the corresponding heteroclinic connections.}
    \label{fig:Connections}
\end{figure}

In this \an{SPU} B based calculation, we focus on the two most relevant Group 1 (blue/red) ECSs and the three most relevant Group 2 (green) ECSs (and their symmetry counterparts). In this smaller domain, the RPOs T2/1 and T2/15 each have two unstable directions, while the RPOs T2/9, T2/30, and T2/4 have three. We consider the dominant unstable direction to generate a sample of trajectories within each unstable manifold. Several of these trajectories pass nearby and shadow other ECSs, providing evidence of heteroclinic connections. Table~\ref{tab:possible_heteroclinic} lists few of the pairs of ECSs connected by such trajectories, together with the corresponding minimum distances and shadowing time intervals. The Figs.~\ref{fig:Connection_1} and~\ref{fig:Connection_2} highlight two such examples.

\begin{table}[h!]
\centering
\caption{\footnotesize Examples of trajectories on the unstable manifolds of source ECSs that eventually shadow target ECSs. These trajectories connect the red/blue (right-flowing/left-flowing) group of ECSs to the green (intermediate) group of ECSs. Some of these trajectories appear to approach heteroclinic connections between the source and target pairs.}
\begin{tabular}{lllcc}
\hline
\textbf{Source ECS} & \textbf{Target ECS} & \textbf{Type} & \textbf{Min. distance} & \textbf{Shadowing interval} \\
\hline
T2/9   & $\sigma_x\sigma_y$T2/1   & green $\rightarrow$ red  & 0.043 & $10.4\tau$ \\
T2/9   & $\sigma_y$T2/15  & green $\rightarrow$ blue   & 0.053  & $22.4\tau$  \\
T2/30  & $\sigma_y$T2/1   & green $\rightarrow$ red  & 0.061  & $12.8\tau$ \\
T2/30  & $\sigma_x\sigma_y$T2/15  & green $\rightarrow$ blue   & 0.009  & $19.6\tau$ \\
T2/4   & $\sigma_x\sigma_y$ T2/1   & green $\rightarrow$ red  & 0.062  & $12.4\tau$ \\
T2/4   & $\sigma_y$T2/15  & green $\rightarrow$ blue   & 0.010 & $22.4\tau$  \\
\hline
T2/15  & $\sigma_y$T2/9   & red  $\rightarrow$ green  &0.106  & $13.2\tau$  \\
T2/15  & $\sigma_x$T2/30  & red  $\rightarrow$ green  &  0.15&  $13.6\tau$\\
T2/15  & T2/4   & red  $\rightarrow$ green  & 0.115 & $9.6\tau$ \\
T2/1   & $\sigma_x\sigma_y$ T2/9   & blue $\rightarrow$ green  & 0.157 & $14.8\tau$ \\
T2/1   & T2/30  & blue $\rightarrow$ green  & 0.106 & $ 14 \tau$ \\
T2/1   & $\sigma_x$T2/4   & blue $\rightarrow$ green  & 0.167  & $16\tau$  \\

\hline
\end{tabular}
\label{tab:possible_heteroclinic}

\end{table}

Notice that the paths from the green to the red/blue ECSs are clearer than those from the red/blue to the green ECSs. This is related to the fact that the green region is populated by a larger number of unstable ECSs that interact with each other in a more complex manner. For example, the minimum distances between the trajectory on the unstable manifold of T2/30 that shadows T2/15 and the trajectory from T2/9 to T2/1 are on the order of $10^{-3}$ and $10^{-2}$, respectively (see Table~\ref{tab:possible_heteroclinic}), which is remarkable given the level of simplification used to identify them. In contrast, the trajectories originating from the unstable manifolds of T2/1 and T2/15 do not get sufficiently close to a heteroclinic trajectory to confirm a connection; nevertheless, they clearly illustrate the dynamic paths from the red/blue to the green regions by exhibiting a clear shadowing of the latter. These results demonstrate that the unstable manifold of the source ECS eventually shadows the target ECS. This finding \an{supports the view that} the ECSs and their invariant manifolds constitute the scaffold underlying reversals in the turbulent set $X$. 

Since $\mathcal{X}$ is the only attractor in \an{SPU} B, the above discussion implies that (a) arbitrary trajectories in \an{SPU} B ultimately converge to $\mathcal{X}$, and (b) trajectories on this attractor repeatedly shadow various ECSs and portions of their invariant manifolds.



\subsection{Chaotic Set $\mathcal{Y}$}
\subsubsection{Dominant ECSs in $\mathcal{Y}$}
\begin{figure}[htbp]
  \centering
    \includegraphics[width=\textwidth]{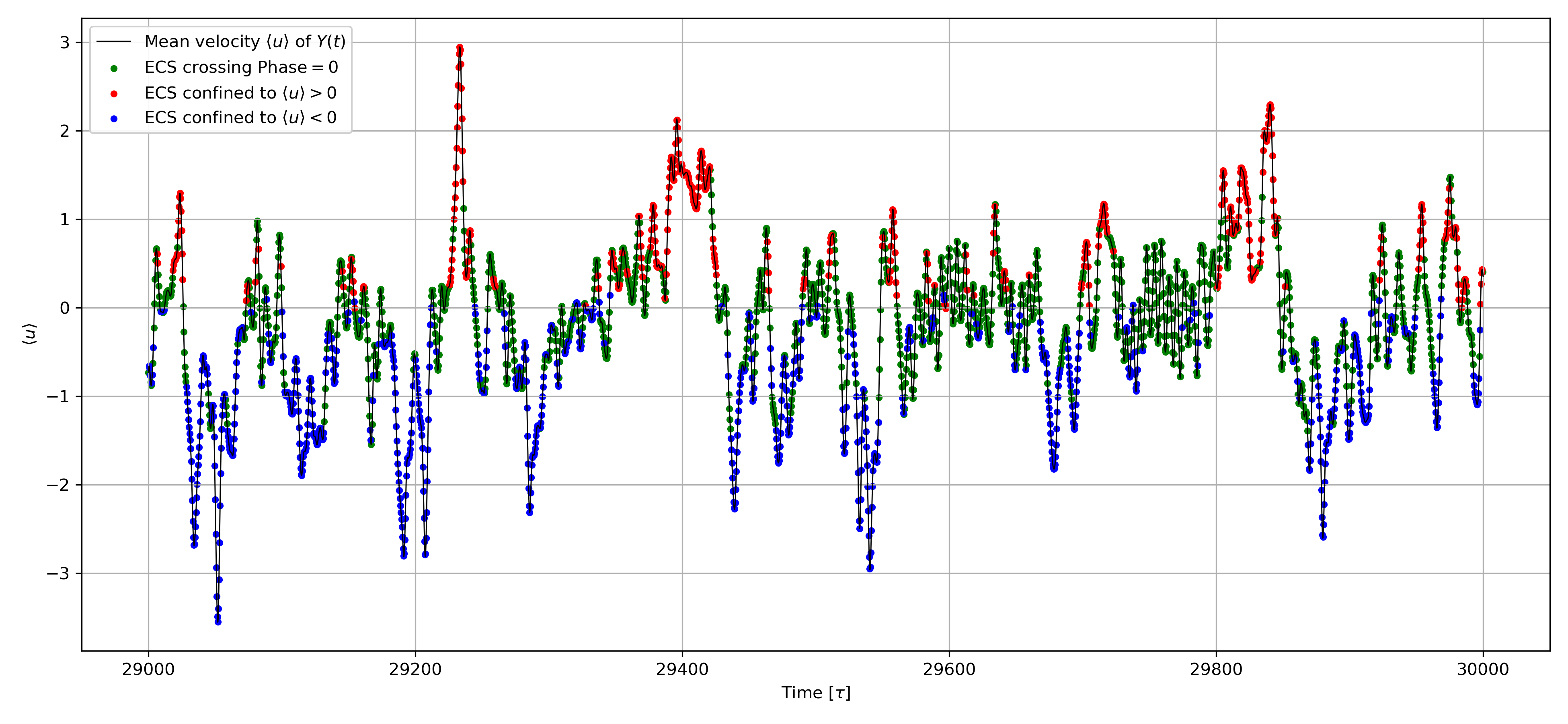}
    \caption{\footnotesize Time series of \( \langle u \rangle \) along the turbulent trajectory $Y(t)$. Various shadowing events are also marked, colored by the region occupied by the corresponding shadowed ECS, using the same color scheme as in Fig. \ref{fig:ecs_analysis}(a).}
    \label{fig:time_series_y}
\end{figure}

We now consider a trajectory \( Y(t) \) within the chaotic set \( \mathcal{Y} \) over the interval \( [29000\tau, 30000\tau] \) in \an{SPU} A. Several key differences are evident when compared to a typical trajectory \( X(t) \) in \( \mathcal{X} \). First,  \an{the time series of  Fig.~\ref{fig:time_series_y} shows that} the oscillations in \( Y(t) \) are shorter, and the maximum value of \( |\langle u \rangle| \) is about 3, compared to about 4 for \( X(t) \). This suggests a less pronounced reversal scenario, with the separation between the blue/red and green regions being less distinct than in \( X(t) \). Second, while \( Y(t) \) occasionally \an{approaches} the blue/red ECSs \an{in the vocinity of \( \langle u \rangle \approx 0 \), such approaches are  largely absent in \( X(t) \)}. Thus, the overlap between the green ECSs and blue/red ECSs is more significant in \( Y(t) \), preventing a clear distinction of ECS groups that mediate the reversals. Third, and most importantly, the minimum distances between the trajectory and the ECSs are larger for \( Y(t) \) than for \( X(t) \). 

In general, the shadowing of the ECSs is less pronounced in \( Y(t) \), as shown in Table~\ref{tab:ecs_top15_T1} in the Appendix for the fifteen most frequent visited ECS groups. For example, the minimum distance from the trajectory to the most frequently visited ECS is 0.308, whereas the minimum distance to the most frequent visited ECSs for \( X(t) \) is only 0.119. Notice that we also needed to include more ECSs to cover about $62\%$ of the closest approaches in $Y(t)$ than we did for $X(t)$.

\begin{figure}[htbp]
    \centering
    \includegraphics[width=0.98\textwidth]{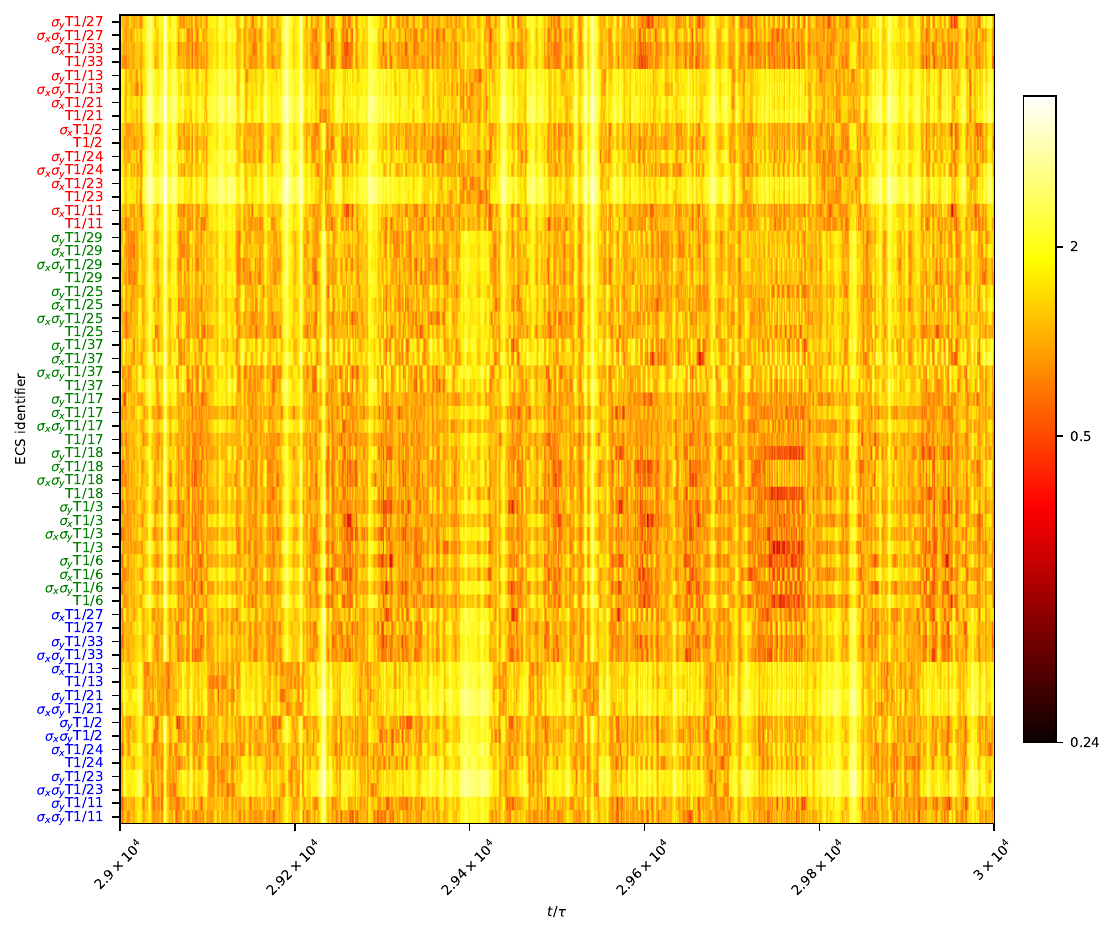}
    \caption{\footnotesize
        Symmetry-reduced distance plot showing the closest approaches between the turbulent trajectory \( Y(t) \) and 15 dynamically relevant ECSs (and their symmetry counterparts). Together, these ECSs account for more than 60\% of all shadowing events. In contrast to the trajectory \( X(t) \) (see Fig. \ref{fig:distance_X}) where the shadowing events were well separated in time, here the shadowing of the green ECSs and the red/blue ECSs appears to occur concurrently. The minimum distance attained is 0.24, roughly three times larger than the smallest distance observed for \( X(t) \).
    }
    \label{fig:distance_Y}
\end{figure}

The fact that the set of 75 ECSs is not exhaustive may explain why we do not observe a clearer picture of the dynamical reversals for \( \mathcal{Y} \). Indeed, the distance plot in Fig.~\ref{fig:distance_Y}, for the fifteen most visited ECS groups, shows no clear pattern or contrast between regions, unlike the plot in Fig.~\ref{fig:distance_X}, which may be attributed to the absence of ECSs that have not yet been computed. The attractor \( \mathcal{Y} \) is not constrained by the $\tau_x(L/2)$ symmetry of \( \mathcal{X} \). The additional complexity arises in part from this doubling of the effective \an{SPU} length. This is consistent with the principal component analysis (PCA) of the corresponding trajectories (see Table \ref{tab:pca-results}): for \( X(t) \), nearly 85\% of the variance is captured by just the first two principal components (PCs), and 90\% is reached with four; in contrast, \( Y(t) \) requires eight components to explain 90\% of the variance, indicating that it explores a significantly higher-dimensional region of phase space.

\subsubsection{Shadowing in $\mathcal{Y}$}
\begin{figure}[htbp]
    \centering
    \includegraphics[width=0.9\textwidth]{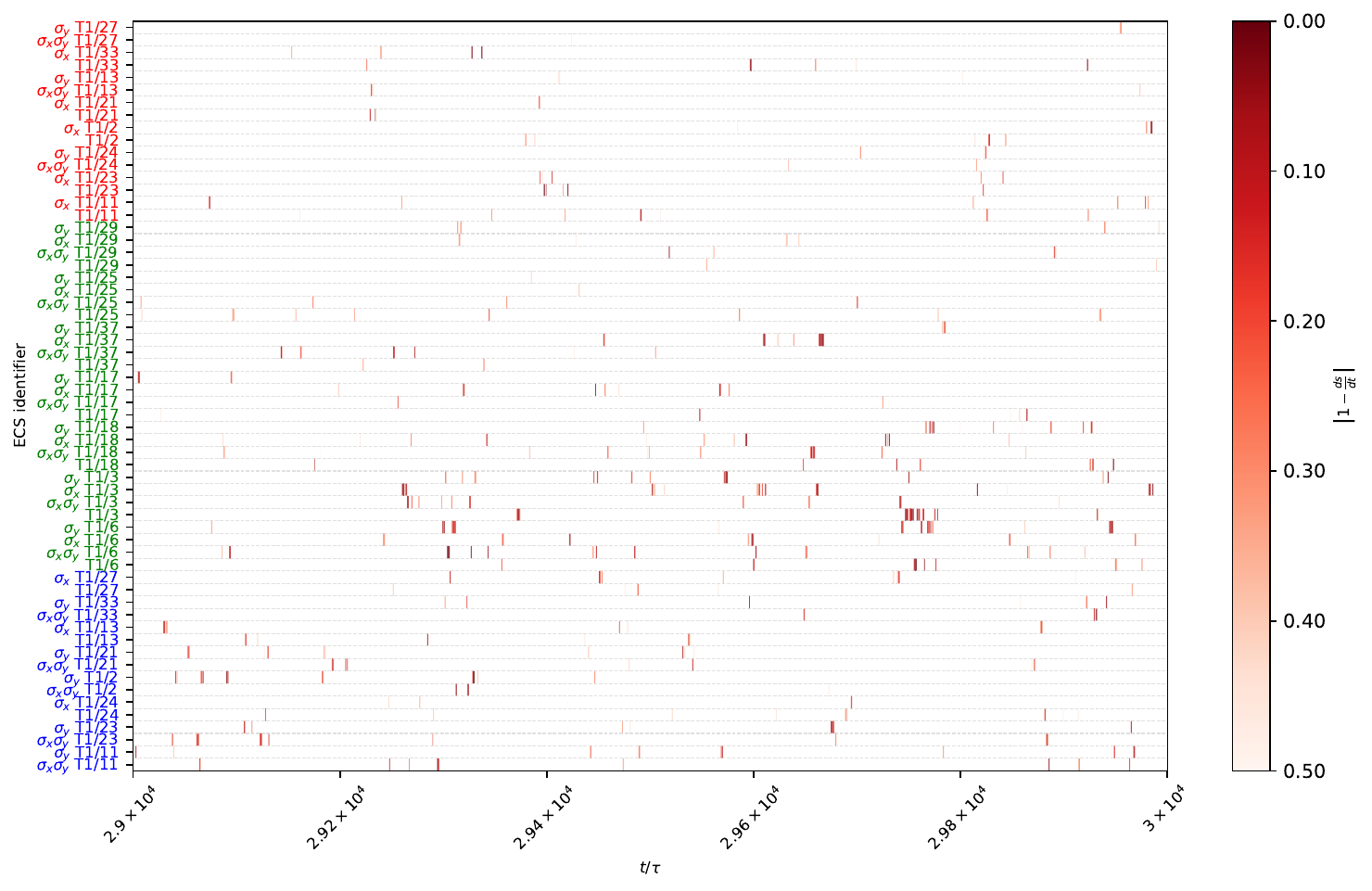}
    \caption{\footnotesize The dynamic shadowing criterion evaluated along the trajectory \(Y(t)\) at discrete times (with $\delta t=0.4\tau$) within the interval \([29000 \tau, 30000 \tau]\). The plot marks instances where the dynamical similarity measure \(\dfrac{d s}{dt}\) lies in the interval \([0.5, 1.5]\), thereby satisfying the criterion. Although fewer shadowing events are detected than for \(X(t)\), their presence points to similar underlying dynamics and motivates the search for additional ECSs to account for the remaining gaps.}
    \label{fig:shadowing_events_Y}
\end{figure}

\begin{figure}[htbp]
    \centering

    \includegraphics[width=0.70\textwidth]{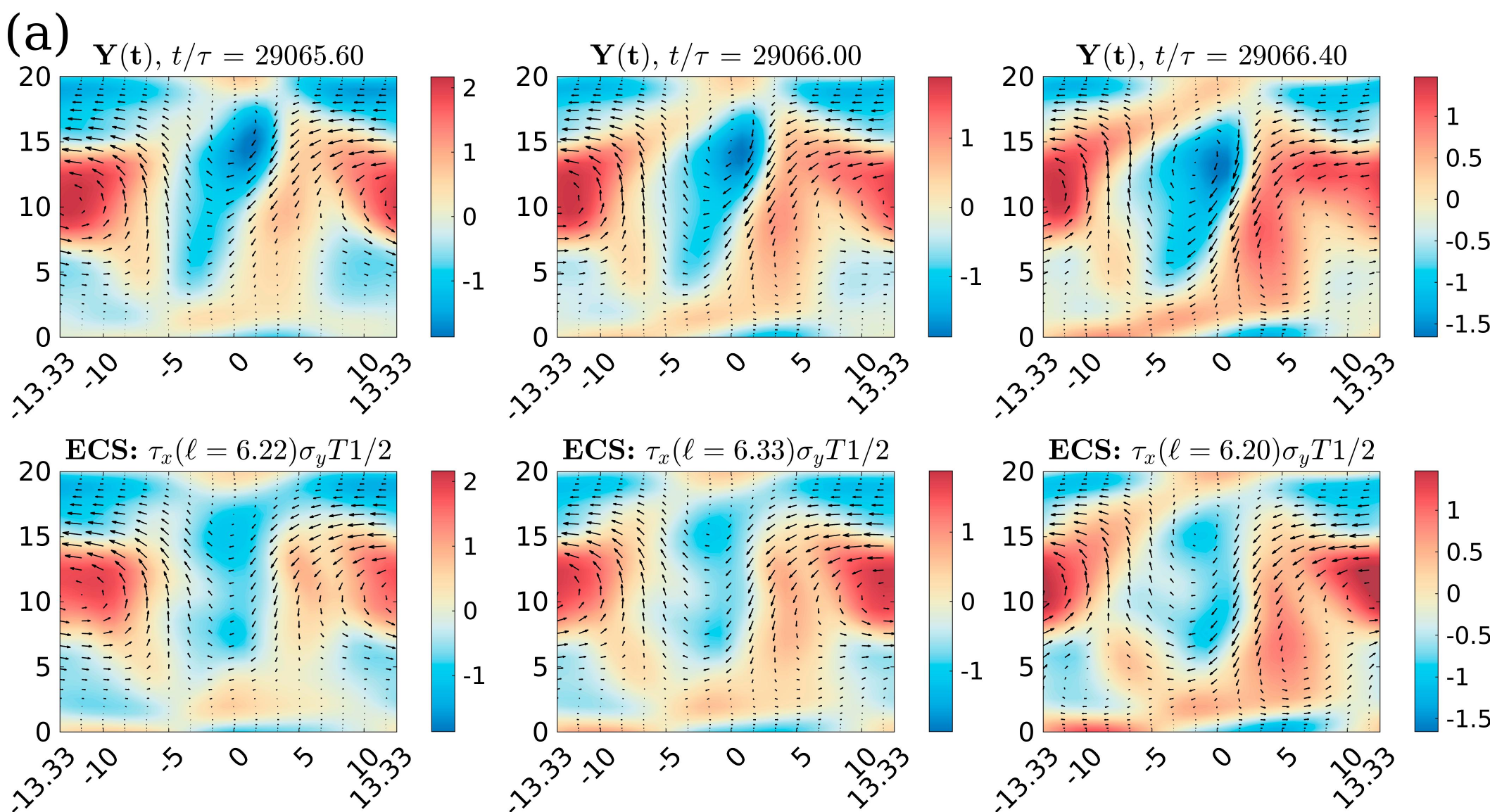}

    \vspace{0.5em}

    \includegraphics[width=0.70\textwidth]{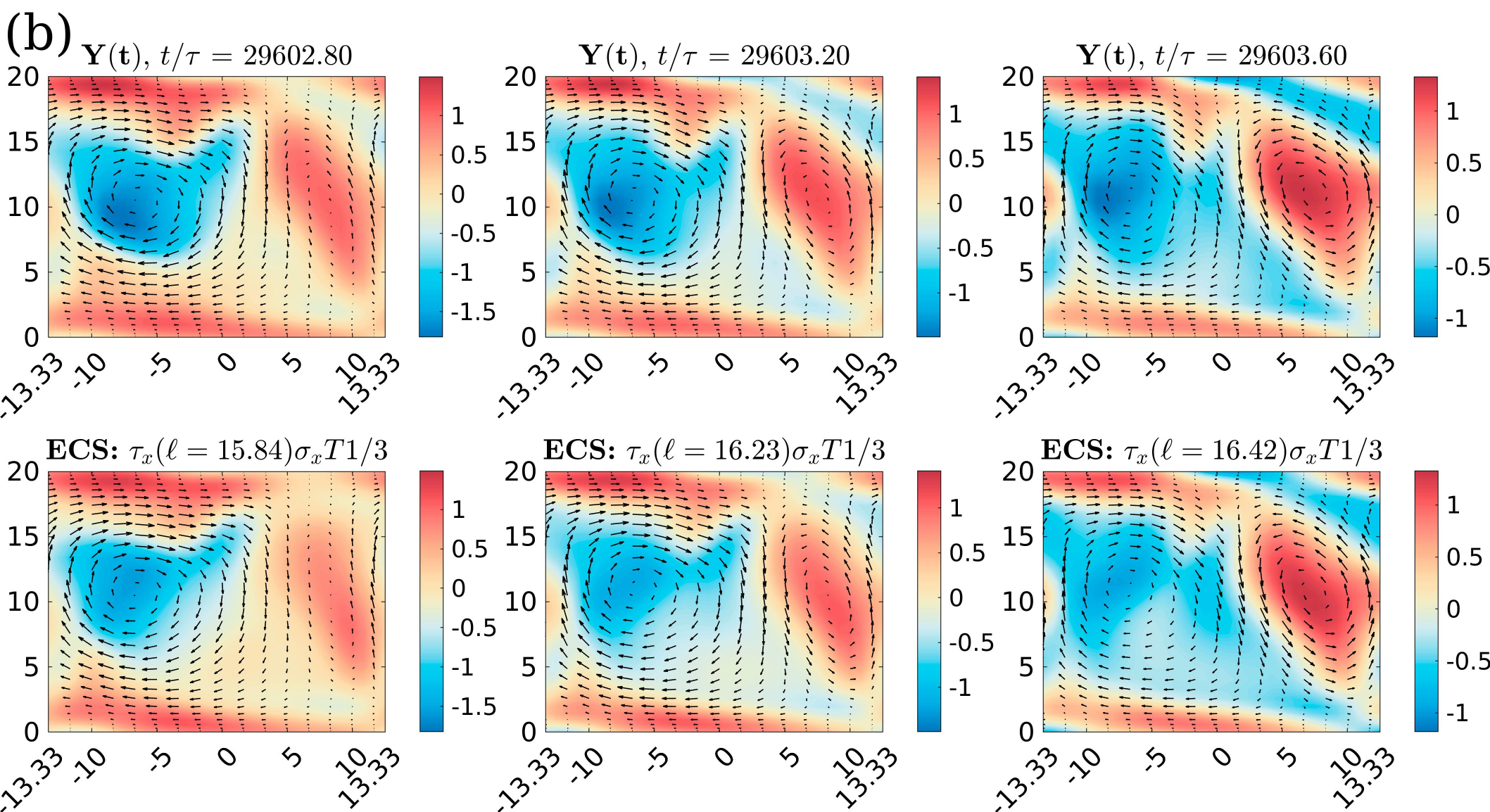}

    \vspace{0.5em}

    \includegraphics[width=0.70\textwidth]{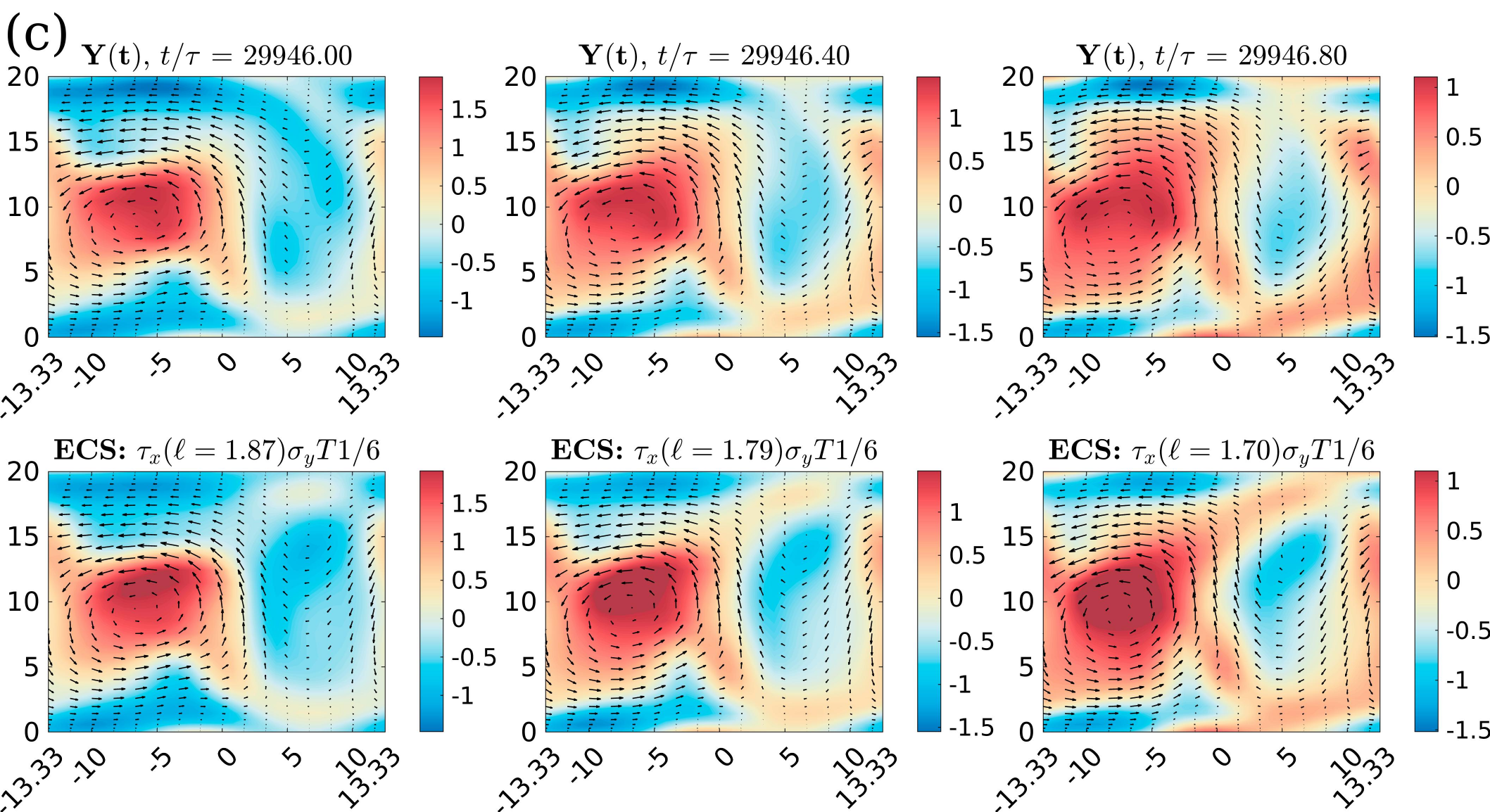}

    \caption{\footnotesize
\an{Velocity field snapshots of three confirmed shadowing events for the trajectory $Y(t)$: subplots a), b), and c) show short snapshot sequences extracted from the shadowing events of $\sigma_y$T1/2, $\sigma_x$T1/3, and $\sigma_y$T1/6, respectively. The first row of each sequence shows three consecutive snapshots of the turbulent trajectory, separated by $\delta t=0.4\tau$, while the second row shows the corresponding snapshots of the shadowed ECS at the same times.}
    }
    \label{fig:shadowing_examples_Y}
\end{figure}

Although the reversal dynamics are not fully captured due to the reduced set of relevant ECSs and the inherent complexity of the chaotic set \( \mathcal{Y} \), our analysis still partly uncovers the underlying deterministic skeleton. Figure~\ref{fig:shadowing_events_Y} shows that shadowing events are less frequent than those of the \( X \)-trajectory. Nevertheless, some of these events exhibit dynamics that closely resemble the behavior of the corresponding ECSs, as illustrated in the examples presented in Fig.~\ref{fig:shadowing_examples_Y}.

The first example occurs in the interval $[29065.2\tau, 29066.4\tau]$, where $Y(t)$ shadows $\sigma_y\mathrm{T1/2}$ \an{with associated escape time $\kappa^{-1}\approx 0.989\tau$. The length of the shadowing interval is comparable to the escape time, and corresponds about $0.28\tau_{\mathcal{Y}}$, where $\tau_{\mathcal{Y}}$ is the Lyapunov time of the chaotic set $\mathcal{Y}$, as defined earlier. The average distance and rates are} $\langle d\rangle \approx 0.443$, $\left\langle \left|\dfrac{ds}{dt}\right| \right\rangle \approx 0.921$, and $\left\langle \left|\dfrac{d\ell}{dt}\right| \right\rangle \approx 0.392$. In Fig.~\ref{fig:shadowing_examples_Y}a, we show three consecutive snapshots from this interval, corresponding to the moment when the distance reaches its minimum value $d \approx 0.393$. The second example in Fig.~\ref{fig:shadowing_examples_Y} is a shadowing of $\sigma_x\mathrm{T1/3}$ within the interval $[29602.8\tau, 29603.6\tau]$ \an{. The interval length is also comparable to the associated escape time $\kappa^{-1}\approx 1.233\tau$ and corresponds to about $0.18\tau_{\mathcal{Y}}$.} In this case, $\langle d\rangle \approx 0.304$, $\left\langle \left|\dfrac{ds}{dt}\right| \right\rangle \approx 1.363$, and $\left\langle \left|\dfrac{d\ell}{dt}\right| \right\rangle \approx 0.724$. Interestingly, although the trajectory approaches this ECS more closely than in the previous example, the dynamics are less synchronized with the ECS. Finally, we discuss a shadowing event in the interval $[29944\tau, 29947.2\tau]$, \an{whose length is approximately $0.36\tau$ longer than the corresponding escape time and about $0.74\tau_{\mathcal{Y}}$. For this shadowing event,} $\langle d\rangle \approx 0.365$, $\left\langle \left|\dfrac{ds}{dt}\right| \right\rangle \approx 0.927$, and $\left\langle \left|\dfrac{d\ell}{dt}\right| \right\rangle \approx 0.267$, \an{and the} trajectory reaches within $d \approx 0.308$ of $\sigma_y\mathrm{T1/6}$ at $t=29946.4$, which corresponds to the snapshots shown in \an{ Fig.~\ref{fig:shadowing_examples_Y}c}. Notice, however, that these shadowing confirmations are less pronounced than our findings in the more constrained attractor~$\mathcal{X}$.

\begin{figure}[htbp]
    \centering
    \includegraphics[width=0.99\linewidth]{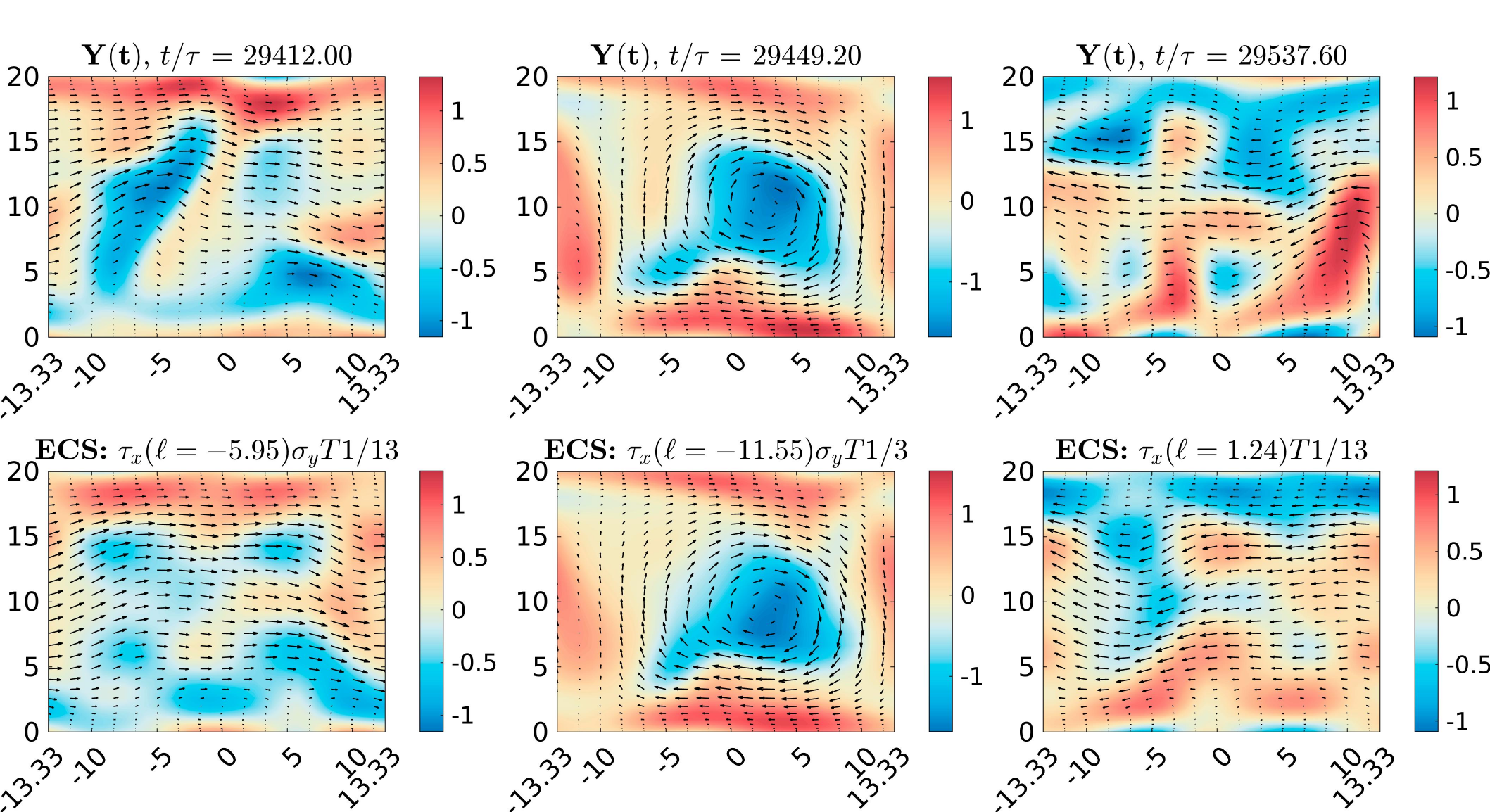}
    \caption{\footnotesize Top row: Velocity field snapshots during the first reversal event for a turbulent trajectory $Y(t)$, evolving in the attractor \( \mathcal{Y}(t) \), during the interval $[29412\tau, 29537.6\tau]$. Bottom row: Velocity field snapshots of the ECSs shadowed during this reversal, shown at the corresponding times. The resemblance between the turbulent flow and the shadowed ECS is noticeably weaker in snapshots where the flow is nearly unidirectional}
    \label{fig:Flow_Reversal_Y}
\end{figure}

The confirmed shadowing events suggest that the turbulence could follow paths that are broadly similar to those observed in the previous trajectory case. Specifically, we expect that the blue and red ECSs shape the nearly uniaxial flow, while the green ECSs mediate the transitions. However, as mentioned, the description with the ECS in hand remains incomplete. Consider, for example, the interval $[29400\tau, 29600\tau]$ (see Fig.~\ref{fig:Flow_Reversal_Y} for a detailed view and Fig.~\ref{fig:time_series_y} for the broader time series). Around $t = 29412\tau$, the mean velocity $\langle u\rangle$ of the trajectory $Y(t)$ oscillates around $\langle u \rangle \approx 1.5$, showing an unclear shadowing of $\sigma_y$T1/13 with a distance $d \approx 0.585$. Following this, brief oscillations occur with successive approximations to the red, green, and blue ECSs. During this interval, there is a clear brief shadowing of the green ECS $\sigma_x$T1/3 at $t=29449.20$ with a minimum distance $d \approx 0.295$. The trajectory eventually reaches the longest and clearest oscillation in the blue region, where at $t=29537.6$, it satisfies the shadowing criterion for T1/13 with a distance $d \approx 0.787$. These three instants highlight the incomplete nature of our description: the large distances to the hypothetically shadowed ECSs when the trajectory is nearly unidirectional, along with the multiple ill-defined oscillations in between, illustrate the gaps in our current understanding.

Overall, the ECS based description of reversals is less clear here than in the case of $X(t)$. Nevertheless, the shadowing evidence indicates that ECSs and their invariant manifolds may still organize the dynamics within this attractor in a manner similar to what was observed for $X(t)$. A more comprehensive search for additional and potentially more intricate ECSs with longer time periods is therefore needed to resolve the gaps in our current understanding. Resolving these questions will be the subject of future work.

\section{Discussion and Conclusions}
Taking an infinite-dimensional dynamical systems perspective on active turbulence, we have studied a hydrodynamic model of active nematics in small 2D periodic channels. \pg{The dynamics in the smallest domain we consider is complex enough to sustain active turbulence, yet,} as we have shown, simple enough that the phase space transport is organized around $\mathcal{O}(10)$ ECSs and a \pg{network of potential heteroclinic connections.}

\pg{Through an equivariant bifurcation analysis, we clarified the origin of several key solutions reported in earlier works and uncovered their role in shaping the global dynamics of the preturbulent regime. In the low activity preturbulent regime, three mechanisms for transient reversals are identified, involving either one of the two $y-$reflection-related copies of RPOs that are nearly homoclinic to the two uniaxial states, or the symmetric PO that is nearly heteroclinic to the two uniaxial states. These ECSs possess the shift-reflect symmetry, and some of them additionally have discrete translation symmetries. Typical reversals involve a potential heteroclinic connection between one of these ECSs (all unstable) and an attractor. Some reversals involve vortex-lattice POs and/or POs that cycle between LAN and vortex-lattice states. These ECSs have $x$-$y$ reflection symmetry, with some again possessing discrete translation symmetries. 

In the active turbulent regime in the original domain, we find that a subset of persistent flow reversals involve homoclinic-like RPOs that have broken shift-reflect symmetry, but retain the discrete translation symmetry of their preturbulent counterparts. These reversals are mediated by RPOs cycling between the LAN-like and vortex-lattice states. These latter RPOs have broken $x$-$y$ reflection symmetry but retain the discrete translation symmetries of the preturbulent POs from which they most likely originate. Building on this, we identify a smaller domain in which the lone turbulent attractor is organized by ECSs that no longer possess any residual spatial symmetries. In this domain, we provide evidence of robust shadowing of the dominating ECSs by typical turbulent trajectories via the use of symmetry-reduced distance and dynamical similarity measures.}

Although our concrete computations are performed for a Beris–Edwards Q-tensor model, much of the structure we uncover relies only on symmetries that are shared across a broad class of continuum descriptions of active nematics. In particular, the local bifurcation analysis depends primarily on translational and reflection symmetries, and the nematic head–tail symmetry, and not on the precise forms of the equations \citep{Marzorati2023}. The catalog of symmetry types of steady and time-periodic solutions and their bifurcations depends only on the total symmetry group $G$ and the representation on the critical eigenspaces, and therefore should be robust across all modeling choices that are expected to have these same symmetries \citep{Golubitsky2002}. We therefore expect the ECS families and reversal pathways identified here to be robust under variations in constitutive parameters and even under modest changes of the modeling framework. Repeating the presented analyses using different models will shift bifurcation points and stability boundaries, but the underlying symmetry-based organization and the existence of low-dimensional ECS skeletons mediating reversals should persist.

While shadowing of coherent structures is clearly observed and quantified in the smallest domain, conclusively establishing the same phenomenon in the original domain proved more challenging, owing to its richer dynamics, the larger dimension of the turbulent attractor, and our inability to compute a more complete set of ECSs. This difficulty is only expected to worsen in longer channels. Future work will focus on closing this gap and clarifying the role of invariant solutions in active nematic turbulence across a wider range of spatial and temporal scales. To do this, we will make use of recent advances in machine learning approaches \citep{linot2023dynamics,beck2025data,page2025computation} that have successfully employed autoencoders to compute ECSs in challenging inertial turbulent flow settings.
\clearpage

\section*{Acknowledgments}
This material is based upon work supported by the U.S. Department of Energy, Office of Science, Office of Basic Energy Sciences under award number DE-SC0022280. This work was completed utilizing the Holland Computing Center of the University of Nebraska, which receives support from the UNL Office of Research and Innovation, and the Nebraska Research Initiative.

\appendix
\section*{Appendix}
\pg{
\subsection{Group theory and equivariance basics}
In this section, we review the group-theoretic definitions, notation, and results from equivariant bifurcation theory used throughout the paper, following \citep{Hoyle2006,Golubitsky2002,krupa1990bifurcations}.
A dynamical system of the form given by Eq.~\eqref{Dynamical_System} is said be equivariant under a group action $\sigma$ if the time evolution and the group operation commute, i.e. for every state $X$ and time $T$, the relationship $\sigma(f^T(X))=f^T(\sigma X)$ is satisfied \citep{Hoyle2006}. For an equivariant system, if $X$ is a solution, then $\sigma X$ is also a solution. Now, let the system Eq.~\eqref{Dynamical_System} be equivariant under a group $G$, which implies that the above two properties are valid for all $\sigma\in G$.

We denote $GL(\mathbb{R}^n)$ as the group of all real invertible $n\times n$ matrices. The subgroup $O(n)\subset GL(\mathbb{R}^n)$ is the set of matrices $A$ satisfying the orthogonality condition $A^TA=1$. We are primarily interested in groups that are isomorphic to $O(n)$ or to any closed subgroup of $O(n)$, i.e., compact Lie groups.  
Strictly speaking, the linear action of the symmetry $\sigma$ on the state $X$ is given by a matrix $\rho(\sigma)$, where the function $\rho$ is called a representation of $G$: 

\begin{definition}[Group Representation]
Let \( G \) be a group and \( V \) a vector space. A \emph{representation} of \( G \) on \( V \) is a homomorphism
\[
\rho: G \to \mathrm{GL}(V),
\]
where \( \mathrm{GL}(V) \) is the group of invertible linear transformations on \( V \). For \( \sigma \in G \) and \( X \in V \), we will often write \( \sigma  X := \rho(\sigma)X \).
\end{definition}

If it is clear in the context, we will use the same symbol to denote the group element and the corresponding matrix.

We will say that a vector $X$ has $\sigma$ symmetry if it is \emph{invariant} under the action of a group element $\sigma$, i.e., $\sigma(X)=X$. 
A solution of an equivariant system can have all, some or none of the symmetries $G$. Usually, we are interested in the group elements that fix a specific vector $X$, motivating the following definition.

\begin{definition}[Isotropy Group]
Let \( G \) act on a set \( V \). For an \( X \in V \), the \emph{isotropy group} (or \emph{stabilizer}) of \( X \) is the subgroup
\[
\Sigma_X=\operatorname{Stab}_G(X) = \{ \sigma \in G \mid \sigma X = X \}.
\]
\end{definition}
On the other hand, it is also relevant to know which vectors are invariant under the action of a specific subgroup. 

\begin{definition}[Fixed Point Subspace of a Subgroup]
Let \( \Sigma \subset G \) be a subgroup. The \emph{fixed point subspace} of \( \Sigma \), denoted \( \operatorname{Fix}(\Sigma) \), is defined as
\[
\operatorname{Fix}(\Sigma) = \{ X \in V \mid \sigma X = X \text{ for all } \sigma \in \Sigma \}.
\]
\end{definition}
An important property of fixed point subspaces is that they are \emph{flow invariant}, i.e., if $X(0)\in Fix(\Sigma)$, then $X(t)\in Fix(\Sigma)$, for all $t>0$.
The final definition we introduce in this section is that of the group orbit. Other useful definitions will be introduced as needed in the subsequent sections.
\begin{definition}[Group Orbit]
For a point $X \in V$, the \textbf{orbit} of $X$ under the action of $G$ is the set
\[
GX = \{ \sigma X \mid \sigma \in G \}.
\]
That is, the orbit of $X$ is the set of all points in $V$ to which $X$ can be moved by the action of elements of $G$. 
\end{definition}

\begin{theorem}[Generalized Equivariant Branching Lemma, double zero eigenvalue]

Consider a 2D $\Gamma-$equivariant dynamical system $\dfrac{dY}{dt}=F_c(Y,\mu)$, with $F_c(0,0)=0$, $D_YF_c(0,0)=0$, and let $\Sigma$ be a isotropy subgroup of $\Gamma$. Assume that $D_{\mu}D_Y F_c(0, 0) w \neq 0$ for nonzero $w \in \operatorname{Fix}(\Sigma)$, and $Fix(\Gamma)=0$. Then, if $\dim(\operatorname{Fix}(\Sigma)) = 1$ (\textbf{axial} subgroup), there exists a smooth branch of fixed points of $F_c$ emerging from the origin with symmetry $\Sigma$.

\end{theorem}
\begin{theorem}[Equivariant Hopf Theorem, a pair of purely imaginary eigenvalues]
   Consider a 2D $\Gamma-$equivariant dynamical system $\dfrac{dY}{dt}=F_c(Y,\mu)$, $F_c(0,\mu)=0$, with Jacobian $DF_c(0,\mu)$ eigenvalues $\lambda(\mu)=\sigma(\mu)\pm i \rho(\mu)$, and $\sigma(0)=0$. Let $\Gamma$ act simply on the phase space, and assume both $\dfrac{d\sigma}{d\mu}|_{(\mu=0)}$ and $\rho(0)$ are different from zero. Suppose there exists an isotropy subgroup $\Sigma\subset\Gamma\times SO(2)$ such that $dim(Fix(\Sigma))=2$. Then there exists a unique branch of time-periodic solutions with period $\approx 2\pi/\omega$ emerging from the origin with  spatio-temporal symmetry  $\Sigma$, where $\omega=\rho(0)$.   
\end{theorem} 

\begin{proposition}[Trivial action case]\label{prop:trivial_action_bif}
Suppose $X_0$ is an equilibrium of dynamical system $\dfrac{dY}{dt}=F_c(Y,\mu)$ at $\mu=0$, where the phase space is either 1D or 2D. Suppose that the action of the isotropy subgroup \( \Sigma_{X_0} \) on the center subspace (at $X_0$) is trivial.
\begin{itemize}
\item 1D Case: If \( D_XF_c(X_0, 0) \) has a single zero eigenvalue, then generically \( F_c \) undergoes a limit point (saddle-node) bifurcation.
\item 2D Case: If \( D_XF_c(X_0, 0) \) has a pair of purely imaginary eigenvalues \(\pm  i\omega \), then \( F_c \) undergoes a Hopf bifurcation to a unique periodic solution.
 \end{itemize}
\end{proposition}

\begin{theorem}[Maximal number of independent frequencies in reconstructed dynamics]\label{thm:relative_equilibrium} ~\citep{krupa1990bifurcations}
Consider a $\Gamma-$ equivariant dynamical system. 
\begin{itemize}
    \item Let \(X \) be an equilibrium of the normal vector field, with isotropy group let \( \Sigma_{X} \). Then the reconstructed trajectory on $\Gamma X$ is generically a quasiperiodic solution with $k$ basic frequencies, where
$k = \operatorname{rank}\left( N(\Sigma_{X}) / \Sigma_{X} \right).$
\item Let $Y$ be a periodic orbit of the normal vector field, and $\Sigma_{Y}$ be the group of symmetries mapping $Y$ to itself. Then the reconstructed trajectory on $\Gamma Y$ is generically a quasiperiodic solution with $k$ basic frequencies, where
$k = \operatorname{rank}\left( N(\Sigma_{Y}) / \Sigma_{Y} \right)+1$. 

Here, \( N(\Sigma_{X}) \) denotes the normalizer of \( \Sigma_{X} \) in \( \Gamma \), and rank refers to the maximum dimension of the resulting torus.

\end{itemize}

\end{theorem}
}

\subsection{Isotropy subgroups of unstable eigenspaces of UNI and LAN}

\begin{table}[ht]
\centering
\caption{Isotropy groups of the unstable eigenspaces of UNI in $L=L_A$. All eigenvalues are complex. The \(\mathrm{R}_{\mathrm{a}}\) column indicates the approximate activity values at which each two-dimensional unstable eigenspace emerges\an{, determined by sampling \(\mathrm{R}{\mathrm{a}}\) in increments of \(\Delta\mathrm{R}{\mathrm{a}}=0.01\)}. From \(\mathrm{R}_{\mathrm{a}} \approx 3.7\) to \(4.5\), no additional local bifurcations from UNI are observed.}
\label{tab:isotropy_groups_UNI}
\resizebox{\textwidth}{!}{%
\begin{tabular}{c l p{7.5cm} l c}
\hline
\textbf{Complex pair} & \textbf{Generators} & \textbf{Group elements} & \textbf{Group structure} & \(\mathbf{R_a}\) \\ \hline

1st & $\{\sigma_x \, \tau_x(L/2)\}$ 
  & $\{ e, \sigma_x \, \tau_x(L/2) \}$ 
  & $\mathbb{Z}_2$ 
  & 0.702 \\[8pt]

2nd & $\{\sigma_x \, \tau_x(L/4)\}$ 
  & $\{ e, \sigma_x \, \tau_x(L/4), \tau_x(L/2), \sigma_x \, \tau_x(3L/4) \}$ 
  & $\mathbb{Z}_4$ 
  & 0.97 \\[8pt]

3rd & $\{\sigma_x\tau_x(L/2) , \tau_x(L/3)\}$ 
  & \begin{tabular}[t]{@{}l@{}}
    $\{ e, \tau_x(L/3), \tau_x(2L/3), \sigma_x \, \tau_x(L/6),$\\
    $\sigma_x \, \tau_x(L/2), \sigma_x \, \tau_x(5L/6) \}$ 
    \end{tabular}
  & $\mathbb{Z}_6\cong \mathbb{Z}_2 \times \mathbb{Z}_3$ 
  & 1.79 \\[8pt]

4th & $\{\sigma_x, \tau_x(L/2)\}$ 
  & $\{ e, \sigma_x, \tau_x(L/2), \sigma_x \, \tau_x(L/2) \}$ 
  & $D_2 \cong \mathbb{Z}_2 \times \mathbb{Z}_2$ 
  & 2.05 \\[8pt]

5th & $\{\sigma_x, \tau_x(L/3)\}$ 
  & \begin{tabular}[t]{@{}l@{}}
   $\{ e, \tau_x(L/3), \tau_x(2L/3), \sigma_x, \sigma_x \, \tau_x(L/3),$\\
  $ \sigma_x \, \tau_x(2L/3) \}$ 
  \end{tabular}
  & $\mathbb{Z}_6 \cong \mathbb{Z}_2 \times \mathbb{Z}_3$ 
  & 2.5 \\[8pt]

6th & $\{\sigma_x \, \tau_x(L/8)\}$ 
  & \begin{tabular}[t]{@{}l@{}}
    $\{ e,\ \sigma_x \, \tau_x(L/8),\ \tau_x(L/4),\ \sigma_x \, \tau_x(3L/8),$ \\
    $\tau_x(L/2),\ \sigma_x \, \tau_x(5L/8),\ \tau_x(3L/4),\ \sigma_x \, \tau_x(7L/8) \}$
  \end{tabular}
  & $\mathbb{Z}_8$ 
  & 3.01 \\[12pt]

7th & $\{\sigma_x, \tau_x(L/4)\}$ 
  & \begin{tabular}[t]{@{}l@{}}
    $\{ e, \tau_x(L/4), \tau_x(L/2), \tau_x(3L/4),$ \\
    $\sigma_x, \sigma_x \, \tau_x(L/4), \sigma_x \, \tau_x(L/2), \sigma_x \, \tau_x(3L/4) \}$
  \end{tabular}
  & $\mathbb{Z}_2 \times \mathbb{Z}_4$ 
  & 3.7 \\
\hline
\end{tabular}%
}
\end{table}

\begin{table}[ht]
\centering
\caption{Isotropy groups of the unstable eigenspaces of LAN in $L=L_A$. Each real eigenvalue has multiplicity two. The \(\mathrm{R}_{\mathrm{a}}\) column indicates the approximate activity values at which each two-dimensional unstable eigenspace emerges\an{, determined by sampling \(\mathrm{R}{\mathrm{a}}\) in increments of \(\Delta\mathrm{R}{\mathrm{a}}=0.01.\)}}
\label{tab:isotropy_groups_LAN}
\resizebox{\textwidth}{!}{
\begin{tabular}{c l p{7.5cm} l c}
\hline
\textbf{Real pair} & \textbf{Generators} & \textbf{Group elements} & \textbf{Group structure} & \(\mathbf{R_a}\) \\ \hline

1st & $\{e\}$ 
  & $\{e\}$ 
  & Trivial group 
  & 0.85 \\[8pt]

2nd & $\{\tau_x(L/2)\}$ 
  & $\{e,\, \tau_x(L/2)\}$
  & $\mathbb{Z}_2$ 
  & 1.25 \\[8pt]

3rd & $\{\tau_x(L/3)\}$ 
  & $\{e, \tau_x(L/3), \tau_x(2L/3)\}$ 
  & $\mathbb{Z}_3$ 
  & 2.11 \\[8pt]
  
4th & $\{\tau_x(L/2)\}$ 
  & $\{e, \tau_x(L/2)\}$ 
  & $\mathbb{Z}_2$ 
  & 2.51 \\[8pt]

5th & $\{\tau_x(L/3)\}$ 
  & $\{e, \tau_x(L/3), \tau_x(2L/3)\}$ 
  & $\mathbb{Z}_3$ 
  & 3.10 \\[8pt]

6th & $\{\tau_x(L/4)\}$ 
  & $\{e, \tau_x(L/4), \tau_x(L/2), \tau_x(3L/4)\}$ 
  & $\mathbb{Z}_4$ 
  & 3.30 \\[8pt]

7th & $\{\tau_x(L/4)\}$
  & $\{e, \tau_x(L/4), \tau_x(L/2), \tau_x(3L/4)\}$ 
  & $\mathbb{Z}_4$ 
  & 4.21 \\[8pt]

8th & $\{e\}$ 
  & $\{e\}$ 
  & Trivial group 
  & 4.36 \\

\hline
\end{tabular}
}
\end{table}

\subsection{Symmetry classification: isotropy subgroups of the main ECS solutions}
For ECSs in the preturbulent regime, the subscript is of the form $Tmx$, where the number $m$ labels the originating unstable eigenspace, while the letter x (=a, b, c, …) distinguishes different RPOs arising from the same eigenpair.

Each ECS in the turbulent regime is denoted $Tk/n$, where $k$ specifies the discrete translation symmetry $\tau_x(L_A/k)$, and $n$ indexes the different ECSs within that symmetry class.

\begin{table}[H]
\centering
\caption{\footnotesize Summary of the main ECSs identified in SPU A. All discrete translation symmetries and the number/value of unstable eigenvalues $N_u$, reported at the representative value of $Ra$ shown in the table, refer to \an{SPU} A, with $L=L_A=\tilde{L}/3$.
Here, $T$ denotes the temporal period, and $d$ the spatial drift over one period. SPU B contains the half-domain representations of the SPU A ECSs with two-fold translational symmetry. }
\label{tab:ECS}

\renewcommand{\arraystretch}{0.9}
\setlength{\tabcolsep}{3pt}

\begin{tabular}{
>{\raggedright\arraybackslash}p{7.5cm}
>{\centering\arraybackslash}p{1.25cm}
ccccc
}
\hline
Description & Name & $Ra$ & $N_u$ & $T$ & $d$ & Symmetry Generators \\
\hline

Unstable traveling wave associated to the
first SNIPER bifurcation
& UTW & 0.65 & 1 & 51.37630 & 0 & $\sigma_x \tau_x(L/2)$\\

Homoclinic-like RPO from 1st unstable
pair of UNI
& HRPO$_{T1a}$ & 1.24 & 1 & 74.78663 & -7.54619 & $\sigma_x\tau_x(L/2)$ \\

Homoclinic-like RPO from 2nd unstable
pair of UNI
& HRPO$_{T2a}$ & 1.5 & 2 & 55.21704 & 8.119028 & $\sigma_x\tau_x(L/4)$  \\

Heteroclinic-like PO from 1st unstable
pair of UNI
& HTPO$_{T1a}$ & 1.5 & 2 & 79.63479 & $1.06\times10^{-12}$ & $\sigma_x\tau_x(L/2)$ \\

Heteroclinic-like PO from 1st unstable
pair of UNI
& HTPO$_{T1b}$ & 1.5 & 6 & 210.6514 & $1.00\times10^{-12}$ & $\sigma_x\tau_x(L/2)$ \\

One-vortex steady solution from LAN
& SV$_{T1a}$ & 0.851 & 1 & -- & -- & $\sigma_x\sigma_y$ \\

One-vortex traveling wave from LAN
& VTW & 0.851 & 0 & 95.82117 & 0 & $\mathbb{I}$ \\

One-vortex periodic-orbit family of standing waves
from LAN
& SPO$_{T1a}$ & 0.93 & 2 & 149.1892 & $1.00\times10^{-12}$ & $\sigma_x\sigma_y$ \\

Periodic orbit cycling between LAN-like state
and one-vortex lattice state
& PO$_{T1a}$ & 0.98 & 1 & 81.60422 & $9.07\times10^{-13}$ & $\sigma_x\sigma_y$ \\

One-vortex periodic orbit from LAN
(Dancing disclination)
& DPO$_{T1a}$ & 1.5 & 1 & 17.70984 & $-2.04\times10^{-23}$ & $\sigma_x\sigma_y$ \\

Second family of One-vortex periodic orbit
(Dancing disclination)
& DPO$_{T1b}$ & 1.5 & 2 & 17.28738 & $-1.55\times10^{-13}$ & $\sigma_x\sigma_y$ \\

RPO possibly bifurcated from DPO family
& DRPO$_{T1a}$ & 1.5 & 1 & 18.4127 & 12.77133 & $\mathbb{I}$ \\

RPO possibly bifurcated from DPO family
& DRPO$_{T1b}$ & 1.5 & 1 & 20.74712 & 7.251976 & $\mathbb{I}$ \\

RPO possibly bifurcated from DPO family
& DRPO$_{T1c}$ & 1.5 & 0 & 62.98886 & -3.09302 & $\mathbb{I}$ \\

Two-vortex steady solution from LAN
& SV$_{T2a}$ & 1.48 & 2 & -- & -- & $\sigma_x\sigma_y, \tau_x(L/2)$ \\

Two-vortex periodic-orbit family of standing waves from LAN
& SPO$_{T2a}$ & 1.5 & 4 & 53.59093 & $1.00\times10^{-12}$ & $\sigma_x\sigma_y, \tau_x(L/2)$ \\

Periodic orbit cycling between LAN-like 
and two-Vortex lattice state
& PO$_{T2a}$ & 2.25 & 2 & 17.63302 & $1.00\times10^{-12}$ & $\sigma_x\sigma_y, \tau_x(L/2)$ \\

Homoclinic-like RPO (turbulent regime)
& T2/1 & 4.5 & 5 & 18.86329 & -6.23718 & $\tau_x(L/2)$ \\

Homoclinic-like RPO (turbulent regime)
& T2/15 & 4.5 & 6 & 22.1869 & -4.30037 & $\tau_x(L/2)$ \\

Homoclinic-like RPO (turbulent regime)
& T2/18 & 4.5 & 3 & 20.66953 & 3.47867 & $\tau_x(L/2)$ \\

Homoclinic-like RPO (turbulent regime)
& T2/7 & 4.5 & 7 & 18.99337 & 8.646005 & $\tau_x(L/2)$\\

RPO cycling between LAN-like and two-vortex lattice states (turbulent regime)
& T2/9 & 4.5 & 8 & 28.02853 & 2.168216 & $\tau_x(L/2)$ \\

RPO cycling between LAN-like and two-vortex lattice states (turbulent regime)
& T2/30 & 4.5 & 5 & 16.76355 & -1.28716 & $\tau_x(L/2)$ \\

RPO cycling between LAN-like and two-vortex lattice states (turbulent regime)
& T2/4 & 4.5 & 6 & 18.00231 & 11.80098 & $\tau_x(L/2)$ \\

RPO cycling between LAN-like and two-vortex lattice states (turbulent regime)
& T2/24 & 4.5 & 11 & 21.71903 & 5.999602 & $\tau_x(L/2)$ \\

RPO cycling between LAN-like and two-vortex lattice states (turbulent regime)
& T2/13 & 4.5 & 12 & 18.35934 & -2.10171 & $\tau_x(L/2)$\\

RPO cycling between LAN-like and two-vortex lattice states (turbulent regime)
& T2/25 & 4.5 & 6 & 13.06355 & -5.12427 & $\tau_x(L/2)$ \\

\hline
\end{tabular}
\end{table}

    

\subsection{Shadowing in Turbulent Regime: Additional information}
\begin{table}[htbp]
\centering
\caption{\footnotesize \textbf{Top 10 ECS groups ranked by frequency of minimum distance occurrences of trajectory $X(t)$. These groups cover approximately 62\% of all minimum distance instances.}}
\label{tab:ecs_top10}
\resizebox{\textwidth}{!}{%
\begin{tabular}{>{\huge}l>{\huge}r>{\huge}r>{\huge}r>{\huge}r>{\huge}r>{\huge}r>{\huge}r>{\huge}l}
\hline
\textbf{\huge ECS Group} & \textbf{\huge Freq.} & \textbf{\huge Rel. Freq.} & \textbf{\huge Cum. Freq.} & \textbf{\huge Cum. Frac.} & \textbf{\huge Min Min Dist.} & \textbf{\huge Max Min Dist.} & \textbf{\huge Avg Min Dist.} & \textbf{\huge Region} \\
\hline
T2/1   & 257 & 0.103 & 257  & 0.103 & 0.119 & 0.940 & 0.465 & blue/red \\
T2/15  & 204 & 0.082 & 461  & 0.184 & 0.070 & 0.809 & 0.331 & blue/red \\
T2/9   & 181 & 0.072 & 642  & 0.257 & 0.113 & 0.770 & 0.396 & green \\
T2/7   & 161 & 0.064 & 803  & 0.321 & 0.256 & 1.132 & 0.561 & blue/red \\
T2/18  & 134 & 0.054 & 937  & 0.375 & 0.139 & 1.056 & 0.513 & blue/red \\
T2/30  & 134 & 0.054 & 1071 & 0.428 & 0.203 & 0.607 & 0.341 & green \\
T2/4   & 133 & 0.053 & 1204 & 0.481 & 0.154 & 0.761 & 0.329 & green \\
T2/24  & 118 & 0.047 & 1322 & 0.529 & 0.244 & 0.648 & 0.394 & green \\
T2/13  & 116 & 0.046 & 1438 & 0.575 & 0.207 & 0.835 & 0.396 & green \\
T2/25  & 110 & 0.044 & 1548 & 0.619 & 0.153 & 0.807 & 0.412 & green \\
\hline
\end{tabular}%
}
\end{table}

\begin{table}[htbp]
\centering

\caption{\footnotesize \textbf{Top 15 ECS groups ranked by frequency of minimum distance occurrences of trajectory Y(t). These groups cover approximately 62\% of all `closest approach' instances.}}
\label{tab:ecs_top15_T1}
\resizebox{\textwidth}{!}{%
\begin{tabular}{>{\huge}l>{\huge}r>{\huge}r>{\huge}r>{\huge}r>{\huge}r>{\huge}r>{\huge}r>{\huge}l}
\hline
\textbf{ ECS Group} & \textbf{ Freq.} & \textbf{ Rel. Freq.} & \textbf{ Cum. Freq.} & \textbf{ Cum. Frac.} & \textbf{ Min Min Dist.} & \textbf{ Max Min Dist.} & \textbf{ Avg Min Dist.} & \textbf{Region} \\
\hline
T1/6   & 167 & 0.067 & 167  & 0.067 & 0.308 & 1.046 & 0.593 & green \\
T1/3   & 158 & 0.063 & 325  & 0.130 & 0.243 & 1.011 & 0.611 & green \\
T1/11  & 158 & 0.063 & 483  & 0.193 & 0.306 & 0.996 & 0.678 & blue/red \\
T1/18  & 144 & 0.058 & 627  & 0.251 & 0.399 & 1.027 & 0.662 & green \\
T1/23  & 113 & 0.045 & 740  & 0.296 & 0.572 & 1.039 & 0.790 & blue/red \\
T1/24  & 108 & 0.043 & 848  & 0.339 & 0.533 & 1.010 & 0.742 & blue/red \\
T1/17  & 101 & 0.040 & 949  & 0.379 & 0.465 & 0.965 & 0.694 & green \\
T1/2   & 99  & 0.040 & 1048 & 0.419 & 0.393 & 1.009 & 0.660 & blue/red \\
T1/37  & 85  & 0.034 & 1133 & 0.453 & 0.299 & 0.955 & 0.652 & green\\
T1/21  & 78  & 0.031 & 1211 & 0.484 & 0.556 & 1.011 & 0.782 & blue/red \\
T1/25  & 77  & 0.031 & 1288 & 0.515 & 0.489 & 1.036 & 0.724 & green \\
T1/13  & 74  & 0.030 & 1362 & 0.545 & 0.539 & 1.104 & 0.772 & blue/red \\
T1/33  & 71  & 0.028 & 1433 & 0.573 & 0.424 & 0.879 & 0.624 & blue/red \\
T1/29  & 61  & 0.024 & 1494 & 0.597 & 0.596 & 0.927 & 0.742 & green \\
T1/27  & 59  & 0.024 & 1553 & 0.621 & 0.320 & 0.870 & 0.610 & blue/red \\
\hline
\end{tabular}%
}
\end{table}

\begin{table}[ht]
\centering
\caption{\footnotesize PCA results comparing the two trajectories embedded in chaotic sets $\mathcal{X}$ and $\mathcal{Y}$, respectively}
\label{tab:pca-results}
\begin{tabular}{lcc}
\hline
\textbf{PCA Metric}                                & $X(t)$ & $Y(t)$ \\
\hline
Number of time steps analyzed                       & 2501                                & 2501                            \\
Number of components explaining 90\% variance       & 4                                   & 8                               \\
Number of components explaining 95\% variance       & 6                                   & 12                              \\
Total number of components (time points)             & 2501                                & 2501                            \\
Variance explained by PC1 (\%)                        & 60.37                              & 28.41                          \\
Variance explained by PC2 (\%)                        & 24.40                              & 18.11                          \\
Variance explained by PC3 (\%)                        & 4.28                               & 16.01                          \\
Variance explained by PC4 (\%)                        & 4.13                               & 15.32                          \\
Variance explained by PC5 (\%)                        & 1.61                               & 3.62                           \\
Variance explained by PC6 (\%)                        & 1.25                               & 3.05                           \\
\hline
\end{tabular}
\end{table}
\newpage
\clearpage
\bibliography{References_AN}

@article{shendruk2017dancing,
   author = {Tyler N. Shendruk and Amin Doostmohammadi and Kristian Thijssen and Julia M. Yeomans},
   issn = {17446848},
   issue = {21},
   journal = {Soft Matter},
   pages = {3853-3862},
   pmid = {28345089},
   publisher = {Royal Society of Chemistry},
   title = {Dancing disclinations in confined active nematics},
   volume = {13},
   year = {2017}
}

@article{smith2005low,
  title={Low-dimensional models for turbulent plane Couette flow in a minimal flow unit},
  author={Smith, TR and Moehlis, J and Holmes, Ph},
  journal={Journal of Fluid Mechanics},
  volume={538},
  pages={71--110},
  year={2005},
  publisher={Cambridge University Press}
}

@article{linot2023dynamics,
  title={Dynamics of a data-driven low-dimensional model of turbulent minimal Couette flow},
  author={Linot, Alec J and Graham, Michael D},
  journal={Journal of Fluid Mechanics},
  volume={973},
  pages={A42},
  year={2023},
  publisher={Cambridge University Press}
}

@article{nishiguchi2025vortex,
  title={Vortex reversal is a precursor of confined bacterial turbulence},
  author={Nishiguchi, Daiki and Shiratani, Sora and Takeuchi, Kazumasa A and Aranson, Igor S},
  journal={Proceedings of the National Academy of Sciences},
  volume={122},
  number={11},
  pages={e2414446122},
  year={2025},
  publisher={National Academy of Sciences}
}

@phdthesis{dore2022active,
  author    = {Dor{\'e}, Claire},
  title     = { Active nematic films under confinement: harnessing topological defects, shaping active flows and designing autonomous microfluidic machines},
  school    = {Universit{\'e} Paris sciences et lettres},
  year      = {2022},
  type      = {PhD thesis},
  address   = {Paris, France},
  note      = {NNT: 2022UPSLS066, tel-04080803},
  language  = {English}
}

@article{burns2020dedalus,
  title={Dedalus: A flexible framework for numerical simulations with spectral methods},
  author={Burns, Keaton J and Vasil, Geoffrey M and Oishi, Jeffrey S and Lecoanet, Daniel and Brown, Benjamin P},
  journal={Phys. Rev. Res.},
  volume={2},
  number={2},
  pages={023068},
  year={2020},
  publisher={APS}
}

@article{koon2000heteroclinic,
  title={Heteroclinic connections between periodic orbits and resonance transitions in celestial mechanics},
  author={Koon, Wang Sang and Lo, Martin W and Marsden, Jerrold E and Ross, Shane D},
  journal={Chaos: An Interdisciplinary Journal of Nonlinear Science},
  volume={10},
  number={2},
  pages={427--469},
  year={2000},
  publisher={American Institute of Physics}
}

@article{viswanath2007recurrent,
  title={Recurrent motions within plane {C}ouette turbulence},
  author={Viswanath, D},
  journal={J. Fluid. Mech.},
  volume={580},
  pages={339--358},
  year={2007},
  publisher={Cambridge University Press}
}

@incollection{holmes1993symmetries,
  title={Symmetries, heteroclinic cycles and intermittency in fluid flow},
  author={Holmes, Philip},
  booktitle={Turbulence in Fluid Flows: A Dynamical Systems Approach},
  pages={49--58},
  year={1993},
  publisher={Springer}
}

@article{armbruster1988heteroclinic,
  title={Heteroclinic cycles and modulated travelling waves in systems with O (2) symmetry},
  author={Armbruster, Dieter and Guckenheimer, John and Holmes, Philip},
  journal={Physica D: Nonlinear Phenomena},
  volume={29},
  number={3},
  pages={257--282},
  year={1988},
  publisher={Elsevier}
}

@article{kreilos2013edge,
  title={Edge states for the turbulence transition in the asymptotic suction boundary layer},
  author={Kreilos, Tobias and Veble, Gregor and Schneider, Tobias M and Eckhardt, Bruno},
  journal={Journal of Fluid Mechanics},
  volume={726},
  pages={100--122},
  year={2013},
  publisher={Cambridge University Press}
}

@article{krupa1997robust,
  title={Robust heteroclinic cycles},
  author={Krupa, Martin},
  journal={Journal of Nonlinear Science},
  volume={7},
  number={2},
  pages={129--176},
  year={1997},
  publisher={Springer}
}

@article{ahlers2009heat,
  title={Heat transfer and large scale dynamics in turbulent Rayleigh-B{\'e}nard convection},
  author={Ahlers, Guenter and Grossmann, Siegfried and Lohse, Detlef},
  journal={Reviews of modern physics},
  volume={81},
  number={2},
  pages={503--537},
  year={2009},
  publisher={APS}
}

@article{bai2016ability,
  title={Ability of a low-dimensional model to predict geometry-dependent dynamics of large-scale coherent structures in turbulence},
  author={Bai, Kunlun and Ji, Dandan and Brown, Eric},
  journal={Physical Review E},
  volume={93},
  number={2},
  pages={023117},
  year={2016},
  publisher={APS}
}

@misc{ecsact,
howpublished={\url{https://github.com/DynamicalSystemsLab-UNL/ECSAct}},
author={Wagner, Caleb and Grover, Piyush},
year={2022}
}

@article{mori2023viscoelastic,
  title={Viscoelastic confinement induces periodic flow reversals in active nematics},
  author={Mori, Francesco and Bhattacharyya, Saraswat and Yeomans, Julia M and Thampi, Sumesh P},
  journal={Physical Review E},
  volume={108},
  number={6},
  pages={064611},
  year={2023},
  publisher={APS}
}

@article{ohm2022weakly,
  title={Weakly nonlinear analysis of pattern formation in active suspensions},
  author={Ohm, Laurel and Shelley, Michael J},
  journal={Journal of Fluid Mechanics},
  volume={942},
  pages={A53},
  year={2022},
  publisher={Cambridge University Press}
}

@article{araujo2005wind,
  title={Wind reversals in turbulent Rayleigh-B{\'e}nard convection},
  author={Araujo, Francisco Fontenele and Grossmann, Siegfried and Lohse, Detlef},
  journal={Physical Review Letters},
  volume={95},
  number={8},
  pages={084502},
  year={2005},
  publisher={APS}
}

@article{sommeria1986experimental,
  title={Experimental study of the two-dimensional inverse energy cascade in a square box},
  author={Sommeria, Joel},
  journal={Journal of Fluid Mechanics},
  volume={170},
  pages={139--168},
  year={1986},
  publisher={Cambridge University Press}
}

@article{glatzmaier1999role,
  title={The role of the Earth's mantle in controlling the frequency of geomagnetic reversals},
  author={Glatzmaier, Gary A and Coe, Robert S and Hongre, Lionel and Roberts, Paul H},
  journal={Nature},
  volume={401},
  number={6756},
  pages={885--890},
  year={1999},
  publisher={Nature Publishing Group UK London}
}

@article{lavi2025nonlinear,
  title={Nonlinear spontaneous flow instability in active nematics},
  author={Lavi, Ido and Alert, Ricard and Joanny, Jean-Fran{\c{c}}ois and Casademunt, Jaume},
  journal={Physical Review Letters},
  volume={134},
  number={23},
  pages={238301},
  year={2025},
  publisher={APS}
}

@article{mishra2011dynamics,
  title={Dynamics of reorientations and reversals of large-scale flow in Rayleigh--B{\'e}nard convection},
  author={Mishra, Pankaj Kumar and De, Arnab K and Verma, Mahendra K and Eswaran, Vinayak},
  journal={Journal of Fluid Mechanics},
  volume={668},
  pages={480--499},
  year={2011},
  publisher={Cambridge University Press}
}

@article{dellnitz1995admissible,
  title={Admissible symmetry increasing bifurcations},
  author={Dellnitz, Michael and Heinrich, Cordula},
  journal={Nonlinearity},
  volume={8},
  number={6},
  pages={1039},
  year={1995},
  publisher={IOP Publishing}
}

@article{chossat1988symmetry,
  title={Symmetry-increasing bifurcation of chaotic attractors},
  author={Chossat, Pascal and Golubitsky, Martin},
  journal={Physica D: Nonlinear Phenomena},
  volume={32},
  number={3},
  pages={423--436},
  year={1988},
  publisher={Elsevier}
}

@book{strogatz2024nonlinear,
  title={Nonlinear dynamics and chaos: with applications to physics, biology, chemistry, and engineering},
  author={Strogatz, Steven H},
  year={2024},
  publisher={Chapman and Hall/CRC}
}

@book{guillemin2025differential,
  title={Differential topology},
  author={Guillemin, Victor and Pollack, Alan},
  volume={370},
  year={2025},
  publisher={American Mathematical Society}
}

@article{brauns2024nonreciprocal,
  title={Nonreciprocal pattern formation of conserved fields},
  author={Brauns, Fridtjof and Marchetti, M Cristina},
  journal={Physical Review X},
  volume={14},
  number={2},
  pages={021014},
  year={2024},
  publisher={APS}
}

@article{ramaswamy2019active,
  title={Active fluids},
  author={Ramaswamy, Sriram},
  journal={Nature Reviews Physics},
  volume={1},
  number={11},
  pages={640--642},
  year={2019},
  publisher={Nature Publishing Group UK London}
}

@book{kurzthaler2023out,
  title={Out-of-equilibrium Soft Matter: Active Fluids},
  author={Kurzthaler, Christina and Gentile, Luigi and Stone, Howard A},
  year={2023},
  publisher={Royal Society of Chemistry}
}

@article{kness1992symmetry,
  title={Symmetry-breaking bifurcations in one-dimensional excitable media},
  author={Kness, Mark and Tuckerman, Laurette S and Barkley, Dwight},
  journal={Physical Review A},
  volume={46},
  number={8},
  pages={5054},
  year={1992},
  publisher={APS}
}

@article{Krygier2021,
   author = {Michael C. Krygier and Joshua L. Pughe-Sanford and Roman O. Grigoriev},
   doi = {10.1017/jfm.2021.522},
   issn = {14697645},
   journal = {Journal of Fluid Mechanics},
   publisher = {Cambridge University Press},
   title = {Exact coherent structures and shadowing in turbulent Taylor-Couette flow},
   volume = {923},
   year = {2021}
}

@article{Wagner2023,
   author = {Caleb G. Wagner and Rumayel H. Pallock and Jae Sung Park and Michael M. Norton and Piyush Grover},
   doi = {10.1103/PhysRevFluids.8.124401},
   issn = {2469990X},
   issue = {12},
   journal = {Physical Review Fluids},
   month = {12},
   publisher = {American Physical Society},
   title = {Exploring regular and turbulent flow states in active nematic channel flow via Exact Coherent Structures and their invariant manifolds},
   volume = {8},
   year = {2023}
}

@article{Crowley2022,
  author  = {Christopher J. Crowley and Joshua L. Pughe-Sanford and Wesley Toler and Michael C. Krygier and Roman O. Grigoriev and Michael F. Schatz},
  title   = {Turbulence tracks recurrent solutions},
  journal = {Proceedings of the National Academy of Sciences of the United States of America},
  volume  = {119},
  number  = {34},
  pages   = {e2120665119},
  year    = {2022}
}

@article{Crowley2023,
   author = {C. J. Crowley and J. L. Pughe-Sanford and W. Toler and R. O. Grigoriev and M. F. Schatz},
   doi = {10.1098/rsta.2022.0137},
   issn = {1364503X},
   issue = {2243},
   journal = {Philosophical Transactions of the Royal Society A: Mathematical, Physical and Engineering Sciences},
   month = {3},
   pmid = {36709779},
   publisher = {Royal Society Publishing},
   title = {Observing a dynamical skeleton of turbulence in Taylor-Couette flow experiments},
   volume = {381},
   year = {2023}
}

@article{alert2025active,
  title={Active turbulence},
  author={Alert, Ricard and Casademunt, Jaume and Joanny, Jean-Fran{\c{c}}ois},
  journal={Annual Review of Condensed Matter Physics},
  volume={13},
  number={1},
  pages={143--170},
  year={2022},
  publisher={Annual Reviews}
}

@article{gabern2006application,
  title={Application of tube dynamics to non-statistical reaction processes},
  author={Gabern, F and Koon, WS and Marsden, Jerrold E and Ross, SD and Yanao, T},
  journal={Few-Body Systems},
  volume={38},
  number={2},
  pages={167--172},
  year={2006},
  publisher={Springer}
}

@article{joshi2022data,
  title={Data-driven discovery of active nematic hydrodynamics},
  author={Joshi, Chaitanya and Ray, Sattvic and Lemma, Linnea M and Varghese, Minu and Sharp, Graham and Dogic, Zvonimir and Baskaran, Aparna and Hagan, Michael F},
  journal={Physical Review Letters},
  volume={129},
  number={25},
  pages={258001},
  year={2022},
  publisher={APS}
}

@article{page2025computation,
  title={Computation of simple invariant solutions in fluid turbulence with the aid of deep learning},
  author={Page, Jacob},
  journal={Nonlinear dynamics},
  volume={113},
  year={2025},
  publisher={Springer}
}

@article{kuehn2009scaling,
  title={Scaling of saddle-node bifurcations: degeneracies and rapid quantitative changes},
  author={Kuehn, Christian},
  journal={Journal of Physics A: Mathematical and Theoretical},
  volume={42},
  number={4},
  pages={045101},
  year={2009}
}

@article{beck2025data,
  title={Data-driven guessing and gluing of unstable periodic orbits},
  author={Beck, Pierre and Parker, Jeremy P and Schneider, Tobias M},
  journal={Physical Review E},
  volume={112},
  number={2},
  pages={024203},
  year={2025},
  publisher={APS}
}

@book{Cvitanovic2020Chaos,
  author    = {Cvitanovi{\'c}, Predrag and Artuso, Roberto and Mainieri, Ronnie and Tanner, Gregor and Vattay, G{\'a}bor},
  title     = {Chaos: Classical and Quantum},
  publisher = {Niels Bohr Institute},
  address   = {Copenhagen},
  year      = {2020},
  url       = {https://ChaosBook.org}
}

@article{Walton2020,
   author = {Joshua Walton and Geoffrey McKay and Michael Grinfeld and Nigel J. Mottram},
   doi = {10.1140/epje/i2020-11973-8},
   issn = {1292895X},
   issue = {8},
   journal = {European Physical Journal E},
   month = {8},
   pmid = {32743686},
   publisher = {Springer},
   title = {Pressure-driven changes to spontaneous flow in active nematic liquid crystals},
   volume = {43},
   year = {2020}
}

@article{Marzorati2023,
   author = {Arsène Marzorati and Stefano Turzi},
   doi = {10.1088/1751-8121/ace408},
   issn = {17518121},
   issue = {31},
   journal = {Journal of Physics A: Mathematical and Theoretical},
   month = {8},
   publisher = {Institute of Physics},
   title = {Bifurcation analysis of spontaneous flows in active nematic fluids},
   volume = {56},
   year = {2023}
}

@article{redfern2024dynamically,
  title={Dynamically relevant recurrent flows obtained via a nonlinear recurrence function from two-dimensional turbulence},
  author={Redfern, Edward M and Lazer, Andrei L and Lucas, Dan},
  journal={Physical Review Fluids},
  volume={9},
  number={12},
  pages={124401},
  year={2024},
  publisher={APS}
}

@article{krupa1990bifurcations,
  title={Bifurcations of relative equilibria},
  author={Krupa, Martin},
  journal={SIAM journal on mathematical analysis},
  volume={21},
  number={6},
  pages={1453--1486},
  year={1990},
  publisher={SIAM}
}

@article{Wagner2022,
   author = {Caleb G. Wagner and Michael M. Norton and Jae Sung Park and Piyush Grover},
   doi = {10.1103/PhysRevLett.128.028003},
   issn = {10797114},
   issue = {2},
   journal = {Physical Review Letters},
   month = {1},
   pmid = {35089772},
   publisher = {American Physical Society},
   title = {Exact Coherent Structures and Phase Space Geometry of Preturbulent 2D Active Nematic Channel Flow},
   volume = {128},
   year = {2022}
}

@book{Chossat2000,
   author = {Pascal Chossat and Reiner Lauterbach},
   city = {Singapore},
   publisher = {World Scientific Publishing Co. Pte. Ltd.},
   title = {Methods in Equivariant Bifurcations and Dynamical Systems},
   volume = {15},
   year = {2000}
}

@article{tan2019topological,
  title={Topological chaos in active nematics},
  author={Tan, Amanda J and Roberts, Eric and Smith, Spencer A and Olvera, Ulyses Alvarado and Arteaga, Jorge and Fortini, Sam and Mitchell, Kevin A and Hirst, Linda S},
  journal={Nature Physics},
  volume={15},
  number={10},
  pages={1033--1039},
  year={2019},
  publisher={Nature Publishing Group UK London}
}

@article{klein2025spontaneous,
  title={Spontaneous Optimal Mixing via Defect-Vortex Coupling in Confined Active Nematics},
  author={Klein, Brandon and Franco, Alejandro J Soto and Sabbir, Md Mainul Hasan and Deutsch, Matthew J and Kliegman, Ross and Selinger, Robin LB and Mitchell, Kevin A and Beller, Daniel A},
  journal={arXiv preprint arXiv:2503.10880},
  year={2025}
}

@book{Hoyle2006,
   author = {Rebecca Hoyle},
   city = {New York},
   publisher = {University Press, Cambridge},
   title = {Pattern Formation: An introduction to methods},
   year = {2006}
}

@book{Golubitsky2002,
   author = {Martin Golubitsky and Ian Stewart},
   city = {Basel; Boston; Berlin},
   publisher = {Birkhäuser Verlag},
   title = {The Symmetry Perspective: From Equilibrium to Chaos in Phase Space and Physical Space},
   volume = {200},
   year = {2002}
}

@article{Thampi2014,
   author = {Sumesh P. Thampi and Ramin Golestanian and Julia M. Yeomans},
   doi = {10.1098/rsta.2013.0366},
   issn = {1364503X},
   issue = {2029},
   journal = {Philosophical Transactions of the Royal Society A: Mathematical, Physical and Engineering Sciences},
   month = {11},
   publisher = {Royal Society of London},
   title = {Vorticity, defects and correlations in active turbulence},
   volume = {372},
   year = {2014}
}

@article{Opathalage2019,
   author = {Achini Opathalage and Michael M. Norton and Michael P.N. Juniper and Blake Langeslay and S. Ali Aghvami and Seth Fraden and Zvonimir Dogic},
   issn = {10916490},
   issue = {11},
   journal = {Proceedings of the National Academy of Sciences of the United States of America},
   pages = {4788-4797},
   pmid = {30804207},
   publisher = {National Academy of Sciences},
   title = {Self-organized dynamics and the transition to turbulence of confined active nematics},
   volume = {116},
   year = {2019}
}

@article{Giomi2011,
   author = {L. Giomi and L. Mahadevan and B. Chakraborty and M. F. Hagan},
   doi = {10.1103/PhysRevLett.106.218101},
   issn = {00319007},
   issue = {21},
   journal = {Physical Review Letters},
   month = {5},
   title = {Excitable patterns in active nematics},
   volume = {106},
   year = {2011}
}

@article{Doostmohammadi2017,
   author = {Amin Doostmohammadi and Tyler N. Shendruk and Kristian Thijssen and Julia M. Yeomans},
   doi = {10.1038/ncomms15326},
   issn = {20411723},
   journal = {Nature Communications},
   month = {5},
   pmid = {28508858},
   publisher = {Nature Publishing Group},
   title = {Onset of meso-scale turbulence in active nematics},
   volume = {8},
   year = {2017}
}

@article{Suri2024,
   author = {Balachandra Suri},
   doi = {10.1103/PhysRevLett.133.154002},
   issn = {10797114},
   issue = {15},
   journal = {Physical Review Letters},
   month = {10},
   pmid = {39454162},
   publisher = {American Physical Society},
   title = {Predictive Framework for Flow Reversals and Excursions in Turbulence},
   volume = {133},
   year = {2024}
}

@article{Yang2025,
   author = {Rui Yang and Peter J. Schmid},
   doi = {10.1017/jfm.2025.371},
   issn = {14697645},
   journal = {Journal of Fluid Mechanics},
   month = {5},
   publisher = {Cambridge University Press},
   title = {Complex-network modeling of reversal events in two-dimensional turbulent thermal convection},
   volume = {1011},
   year = {2025}
}

@article{Ramaswamy2016,
   author = {Rajesh Ramaswamy and Frank Jülicher},
   doi = {10.1038/srep20838},
   issn = {20452322},
   journal = {Scientific Reports},
   month = {2},
   pmid = {26877263},
   publisher = {Nature Publishing Group},
   title = {Activity induces traveling waves, vortices and spatiotemporal chaos in a model actomyosin layer},
   volume = {6},
   year = {2016}
}

@article{Mitchell2024,
   author = {Kevin A. Mitchell and Md Mainul Hasan Sabbir and Kevin Geumhan and Spencer A. Smith and Brandon Klein and Daniel A. Beller},
   doi = {10.1103/PhysRevE.109.014606},
   issn = {24700053},
   issue = {1},
   journal = {Physical Review E},
   month = {1},
   pmid = {38366395},
   publisher = {American Physical Society},
   title = {Maximally mixing active nematics},
   volume = {109},
   year = {2024}
}

@article{graham2020exact,
  title={Exact coherent states and the nonlinear dynamics of wall-bounded turbulent flows},
  author={Graham, Michael D and Floryan, Daniel},
  journal={Annu. Rev. Fluid Mech.},
  volume={53},
  year={2020},
  publisher={Annual Reviews}
}

@article{hopf1948mathematical,
  title={A mathematical example displaying features of turbulence},
  author={Hopf, Eberhard},
  journal={Commun. Pure Appl. Math.},
  volume={1},
  number={4},
  pages={303--322},
  year={1948},
  publisher={Wiley Online Library}
}

@article{ruelle1971nature,
  title={On the nature of turbulence},
  author={Ruelle, David and Takens, Floris},
  journal={Les rencontres physiciens-math{\'e}maticiens de Strasbourg-RCP25},
  volume={12},
  pages={1--44},
  year={1971}
}

@article{henshaw2023dynamic,
  title={Dynamic mode structure of active turbulence},
  author={Henshaw, Richard J and Martin, Olivia G and Guasto, Jeffrey S},
  journal={Physical Review Fluids},
  volume={8},
  number={2},
  pages={023101},
  year={2023},
  publisher={APS}
}

@article{thampi2022channel,
  title={Channel Confined Active Nematics},
  author={Thampi, Sumesh P},
  journal={Current Opinion in Colloid \& Interface Science},
  pages={101613},
  year={2022},
  publisher={Elsevier}
}

@article{alert2020universal,
  title={Universal scaling of active nematic turbulence},
  author={Alert, Ricard and Joanny, Jean-Fran{\c{c}}ois and Casademunt, Jaume},
  journal={Nature Physics},
  volume={16},
  number={6},
  pages={682--688},
  year={2020},
  publisher={Nature Publishing Group}
}

@article{Giomi2015,
  title = {Geometry and Topology of Turbulence in Active Nematics},
  author = {Giomi, Luca},
  journal = {Phys. Rev. X},
  volume = {5},
  issue = {3},
  pages = {031003},
  numpages = {11},
  year = {2015},
  month = {Jul},
  publisher = {American Physical Society}
 
}

@article{linkmann2020condensate,
  title={Condensate formation and multiscale dynamics in two-dimensional active suspensions},
  author={Linkmann, Moritz and Marchetti, M Cristina and Boffetta, Guido and Eckhardt, Bruno},
  journal={Phys. Rev. E},
  volume={101},
  number={2},
  pages={022609},
  year={2020},
  publisher={APS}
}

@article{martinez2019selection,
  title={Selection mechanism at the onset of active turbulence},
  author={Mart{\'\i}nez-Prat, Berta and Ign{\'e}s-Mullol, Jordi and Casademunt, Jaume and Sagu{\'e}s, Francesc},
  journal={Nature physics},
  volume={15},
  number={4},
  pages={362--366},
  year={2019},
  publisher={Nature Publishing Group UK London}
}

@article{you2020nonreciprocity,
  title={Nonreciprocity as a generic route to traveling states},
  author={You, Zhihong and Baskaran, Aparna and Marchetti, M Cristina},
  journal={Proceedings of the National Academy of Sciences},
  volume={117},
  number={33},
  pages={19767--19772},
  year={2020},
  publisher={National Acad Sciences}
}

@article{hillebrand2025discontinuous,
  title={Discontinuous transition to active nematic turbulence},
  author={Hillebrand, Malcolm and Alert, Ricard},
  journal={Nature Communications},
  volume={16},
  number={1},
  pages={11169},
  year={2025},
  publisher={Nature Publishing Group UK London}
}

@article{giomi2012banding,
  title={Banding, excitability and chaos in active nematic suspensions},
  author={Giomi, L and Mahadevan, L and Chakraborty, B and Hagan, MF},
  journal={Nonlinearity},
  volume={25},
  number={8},
  pages={2245},
  year={2012},
  publisher={IOP Publishing}
}

@article{duclos2018spontaneous,
  title={Spontaneous shear flow in confined cellular nematics},
  author={Duclos, G and Blanch-Mercader, C and Yashunsky, V and Salbreux, G and Joanny, J-F and Prost, J and Silberzan, Pascal},
  journal={Nature Physics},
  volume={14},
  number={7},
  pages={728--732},
  year={2018},
  publisher={Nature Publishing Group UK London}
}

@article{Simha2002,
  title = {Hydrodynamic Fluctuations and Instabilities in Ordered Suspensions of Self-Propelled Particles},
  author = {Aditi Simha, R. and Ramaswamy, Sriram},
  journal = {Phys. Rev. Lett.},
  volume = {89},
  issue = {5},
  pages = {058101},
  numpages = {4},
  year = {2002},
  month = {Jul},
  publisher = {American Physical Society},
}

@phdthesis{varghese2021active,
  title={Active Nematics in Three Dimensions A Dissertation Presented to The Faculty of the Graduate School of Arts and Sciences},
  author={Varghese, Minu},
  year={2021},
  school={Brandeis University}
}

@article{golden2023physically,
  title={Physically informed data-driven modeling of active nematics},
  author={Golden, Matthew and Grigoriev, Roman O and Nambisan, Jyothishraj and Fernandez-Nieves, Alberto},
  journal={Science Advances},
  volume={9},
  number={27},
  year={2023},
  publisher={American Association for the Advancement of Science}
}

@article{Marchetti2013,
  author  = {Marchetti, M. C. and Joanny, J. F. and Ramaswamy, S. and Liverpool, T. B. and Prost, J. and Rao, M. and Simha, R. A.},
  title   = {Hydrodynamics of soft active matter},
  journal = {Reviews of Modern Physics},
  volume  = {85},
  number  = {3},
  pages   = {1143--1189},
  year    = {2013},
  publisher = {American Physical Society},
  month = {7}
}

@article{Jimenezi1991,
   author = {Javier Jimenezi and Parviz Moin},
   journal = {J. Fluid Mech},
   pages = {213-240},
   title = {The minimal flow unit in near-wall turbulence},
   volume = {225},
   year = {1991}
}

@article{Kawahara2011,
   author = {Genta Kawahara and Markus Uhlmann and Lennaert Van Veen},
   issn = {00664189},
   journal = {Annual Review of Fluid Mechanics},
   pages = {203-225},
   title = {The significance of simple invariant solutions in turbulent flows},
   volume = {44},
   year = {2011}
}

@article{suri2020capturing,
  title        = {Capturing turbulent dynamics and statistics in experiments with unstable periodic orbits},
  author       = {Suri, Balachandra and Kageorge, Logan and Grigoriev, Roman O. and Schatz, Michael F.},
  journal      = {Physical Review Letters},
  volume       = {125},
  number       = {6},
  pages        = {064501},
  year         = {2020},
  doi          = {10.1103/PhysRevLett.125.064501}
}

@article{yalniz2021coarse,
  title        = {Coarse graining the state space of a turbulent flow using periodic orbits},
  author       = {Yalniz, Gökhan and Hof, Björn and Budanur, Nazmi Burak},
  journal      = {Physical Review Letters},
  volume       = {126},
  number       = {24},
  pages        = {244502},
  year         = {2021},
  doi          = {10.1103/PhysRevLett.126.244502}
}

@article{page2024recurrent,
  title        = {Recurrent flow patterns as a basis for two-dimensional turbulence: predicting statistics from structures},
  author       = {Page, J. and Norgaard, P. and Brenner, M. P. and Kerswell, R. R.},
  journal      = {Proceedings of the National Academy of Sciences},
  volume       = {121},
  pages        = {e2320007121},
  year         = {2024}
}

@article{Krajnik2020Spectral,
  title   = {Spectral energy analysis of bulk three-dimensional active nematic turbulence},
  author  = {Ziga Krajnik and Ziga Kos and Miha Ravnik},
  journal = {Soft Matter},
  volume  = {16},
  pages   = {9059--9068},
  year    = {2020},
  doi     = {10.1039/C9SM02492A}
}

\end{document}


\begin{center}
{\Large \textbf{Supplementary Information}}\\[0.5em]
{\large \textbf{Symmetry, Invariant Manifolds and Flow Reversals in Active Nematic Turbulence}}\\[1em]

Angel Naranjo$^{1}$, Rumayel Pallock$^{1}$, Caleb G. Wagner$^{2}$, and Piyush Grover$^{1}$\\[0.5em]

{\small
$^1$Mechanical and Materials Engineering, University of Nebraska–Lincoln, Lincoln, NE 68588, USA
$^2$Parsons Corporation, USA}
\end{center}

\vspace{1.5em}

\begin{center}
\textbf{CONTENTS}
\end{center}

\noindent
I.\quad List and description of movies \dotfill 2\\
II.\quad Methods \dotfill 2\\

\newpage

\section{Description of movies}
The movies listed below are available via two sources:\\
1). Downloadable at Zenodo: \url{https://zenodo.org/records/17850230}\\
2). Youtube playlist: \url{https://www.youtube.com/playlist?list=PL2a93ODHI2F6DyQxC-HBLJALmCXOyRxIe}

\begin{table}[h!]
\begin{tabular}{l p{10cm}}
\hline
\textbf{Name (.mp4)} & \textbf{Description} \\
\hline

movie1\_DPO\_T1a\_Ra1.5.mp4 &
Nematic director and velocity field for the periodic orbit DPO$_{T1a}$ at $\mathrm{R}_{\mathrm{a}} = 1.5$ in the channel. \\[0.2cm]

movie2\_PO\_T2a\_Ra2.25.mp4 &
Nematic director and velocity field for the periodic orbit PO$_{T2a}$ at $\mathrm{R}_{\mathrm{a}} = 2.25$ in the channel. \\[0.2cm]

movie3\_HRPO\_T1a\_Ra1.24.mp4 &
Nematic director and velocity field for the homoclinic-like relative periodic orbit HRPO$_{T1a}$ at $\mathrm{R}_{\mathrm{a}} = 1.24$ in the channel. \\[0.2cm]

movie4\_HRPO\_T2a\_Ra1.24.mp4 &
Nematic director and velocity field for the homoclinic-like relative periodic orbit HRPO$_{T2a}$ at $\mathrm{R}_{\mathrm{a}} = 1.24$ in the channel. \\[0.2cm]

movie5\_HTPO\_T1a\_Ra1.5.mp4 &
Nematic director and velocity field for the heteroclinic-like periodic orbit HTPO$_{T1a}$  at $\mathrm{R}_{\mathrm{a}} = 1.5$ in the channel. \\[0.2cm]

movie6\_HTPO\_T1b\_Ra1.5.mp4 &
Nematic director and velocity field for the heteroclinic-like periodic orbit HTPO$_{T1b}$ at $\mathrm{R}_{\mathrm{a}} = 1.5$ in the channel. \\[0.2cm]

movie7\_One\_vortex\_TW.mp4 &
Nematic director and velocity field for the one–vortex traveling wave  at $\mathrm{R}_{\mathrm{a}} = 0.851$ in the channel. \\[0.2cm]

movie8\_One\_vortex\_standing\_wave.mp4 &
Nematic director and velocity field for the one–vortex standing wave at $\mathrm{R}_{\mathrm{a}} = 0.93$ in the channel. \\[0.2cm]

movie9\_Preturbulent\_reversals\_Ra1.5.mp4 &
Nematic director and velocity field for a preturbulent trajectory at $\mathrm{R}_{\mathrm{a}} = 1.5$ in the channel exhibiting reversals. \\[0.2cm]

movie10\_T2\_4\_Ra4.5.mp4 &
Nematic director and velocity field for the green (Vortex lattice RPO) ECS T2/4  at $\mathrm{R}_{\mathrm{a}} = 4.5$ in the channel. \\[0.2cm]

movie11\_T2\_1\_Ra4.5.mp4 &
Nematic director and velocity field for the blue (homoclinic-like RPO) ECS T2/1 at $\mathrm{R}_{\mathrm{a}} = 4.5$ in the channel. \\[0.2cm]

movie12\_Shadowing\_turbulent.mp4 &
Example of shadowing of an ECS and its unstable manifold in turbulent regime. \\[0.2cm]

movie13\_Turbulent\_Three\_reversals\_in\_X.mp4 &
Nematic director and velocity field for a turbulent trajectory with three reversals of $X$ in the channel and in the reduced phase space representation. \\[0.2cm]

\hline
\end{tabular}
\end{table}

\section{Methods}
Our calculation framework uses extensively the open toolkit Exact Coherent Structures in Active Matter (ECSAct) which is based on the open-source pseudospectral code Dedalus. 

\subsection{Symmetry and Equivariant Bifurcation Analysis}
For \emph{local equivariant analysis}, we employed the symmetry tools provided by the \texttt{ECSAct} code to identify the isotropy subgroups associated with the bifurcating ECSs and their corresponding eigenspaces. In several instances, additional analysis required projecting full time-dependent simulations onto the appropriate invariant subspaces in order to get solutions that are stable within those subspaces and predicted by the equivariant bifurcation theory.

For \emph{global bifurcation analysis}, we needed to compute multiple RPOs to confirm the presence of SNIPER, homoclinic, and heteroclinic bifurcations. Long-period RPOs were particularly challenging, since the associated Newton solver often struggled with very large temporal periods nearby the bifurcation point. To mitigate this difficulty, we projected the solutions onto the invariant subspace in which the relevant RPO or PO is more stable. For example, every global-bifurcation branch associated with the first complex unstable pair of \texttt{UNI} required projecting the trajectories that initiated from the bifurcating ECS onto the subspace invariant under $\sigma_x \tau_x(L/2)$ by applying the operator,
\begin{equation}\label{eq:distance}
  P=\frac{\mathbf{I}+\sigma_x\tau_x(L/2)}{2}
\end{equation}


\subsection{Heteroclinic Connections and Shadowing Verification}
\subsubsection{Heteroclinic Connections}

Heteroclinic connections are confirmed using the distance function
\begin{equation}\label{eq:distance}
    d\!\left(X_1(t), X_{E}\right)
    = \min_{\substack{0 \leq \ell \leq L \\[1mm] 0 \leq s \leq T}}
    \left\| \tau_x(\ell)\, X(t) - X_{\text{target}}(s) \right\|_2
    < \epsilon,
\end{equation}
where $X(t)$ is the trajectory originating from a source ECS along its unstable manifold. In practice, the unstable manifolds were approximated using approximately $100$ to $1000$ points along the unstable directions. 

In general, heteroclinic connections were difficult to verify to tolerances much smaller than $10^{-2}$, unless the target ECS lies within an attractor contained in a specific invariant subspace. However, if the unstable manifold of the source ECS is one-dimensional, or dominated by a single unstable eigenvalue, and the target ECS is not too unstable, a shooting method can be applied by minimizing the function
\begin{equation}\label{eq:dist2_func}
\begin{aligned}
    d^{2}(\mu, t, \ell, s)
    = \left\| \phi\!\left(X_0 + \mu v,\, t\right)
        - \tau_x(\ell)\, X_{\text{target}}(s)
      \right\|_2^{2},
\end{aligned}
\end{equation}
where $v$ is the dominant unstable eigenvector. The corresponding minimization problem,
\begin{equation}\label{eq:dist2_opt}
    \min_{\mu,\, t,\, \ell,\, s} \; d^{2}(\mu, t, \ell, s),
\end{equation}
can be solved using Gauss-Newton or Levenberg-Marquardt methods. Using the former, with a good initial guess, we successfully verified the heteroclinic connections 
\[
\text{DPO}_{T1b} \longrightarrow \text{RPO}_{T1a},
\qquad
\text{HTPO}_{T1a} \longrightarrow \text{DRPO}_{T1a},
\]
with tolerance as low as $10^{-8}$ and $10^{-6}$, respectively. Although the shooting formulation can be generalized easily for the case where the unstable manifold is effectively multidimensional, we could only verify such connections with a tolerance of $10^{-2}$. An adjoint-based method could be used to confirm these connections to a lower tolerance value in the future. 

\subsubsection{Shadowing}
Although the distance calculation already gives us direct information of shadowing events, we also verified the correspondence of the dynamics between the trajectory and the target ECSs.  In calculation of the numerical derivatives $\dfrac{ds}{dt}$ and $\dfrac{d\ell}{dt}$, the discontinuities of $s$ and $\ell$ for a relative periodic orbit $u$ with period $T$ and shift $\ell_0$ need to be accounted for. To do so, we use the relation
\[
\tau_x(\ell_o)u(T)=u(0).
\]
The derivative $\dfrac{ds}{dt}$ also allowed us to verify shadowing of unstable manifolds. By considering a shadowed ECS, we looked for trajectories on its unstable manifold that are shadowed for a non-trivial time interval while satisfying  $\dfrac{ds}{dt}\approx 1$. By doing this, we found for example the shadowing of the unstable manifold of $\sigma_x\sigma_y$ T2/1 in the interval $[29064\tau,29094.4\tau]$.